# Ice-free geomorphometry of Enderby Land, East Antarctica: 2. Coastal oases

## I.V. Florinsky[1*], S.O. Zharnova[2]

[1] Institute of Mathematical Problems of Biology, Keldysh Institute of Applied Mathematics, Russian Academy of Sciences, Pushchino, Moscow Region, 142290, Russia

[2] National Research Tomsk State University, 36 Lenin Ave., Tomsk, 634050, Russia

### Abstract

Geomorphometric modeling and mapping of ice-free Antarctic areas can be applied for obtaining new quantitative knowledge about the topography of these unique landscapes and for the further use of morphometric information in Antarctic research. Within the framework of a project of creating a physical geographical thematic scientific reference geomorphometric atlas of ice-free areas of Antarctica, we performed geomorphometric modeling and mapping of five key coastal oases of Enderby Land, East Antarctica. These include, from west to east, the Konovalov Oasis, Thala Hills (Molodezhny and Vecherny Oases), Fyfe Hills, and Howard Hills. As input data, we used five fragments of the Reference Elevation Model of Antarctica (REMA). For the coastal oases and adjacent ice sheet and glaciers, we derived models and maps of eleven, most scientifically important morphometric variables (i.e., slope, aspect, horizontal curvature, vertical curvature, minimal curvature, maximal curvature, catchment area, topographic wetness index, stream power index, total insolation, and wind exposition index). In total, we derived 60 maps in 1:50,000 and 1:75,000 scales. The obtained models and maps describe the coastal oases of Enderby Land in a rigorous, quantitative, and reproducible manner. New morphometric data can be useful for further geological, geomorphological, glaciological, ecological, and hydrological studies of this region.

### Contents

## 1 Introduction

There are three main types of ice-free areas in Antarctica: (1) Antarctic oases, i.e. coastal, shelf, and mountainous ice-free areas; (2) ice-free islands (or areas thereof) situated outside the ice shelves; and (3) ice-free mountain chains (or their portions) and nunataks (Markov et al., 1970; Simonov, 1971; Korotkevich, 1972; Alexandrov, 1985; Pickard, 1986; Beyer and Bölter, 2002; Sokratova, 2010). Topography is obviously a key element of the natural environment of ice-free Antarctic landscapes.

Geomorphometry deals with the mathematical modeling and analysis of topography as well as relationships between topography and other components of geosystems (Evans, 1972;

---

[*] Correspondence to: iflor@mail.ru

Moore et al., 1991; Wilson and Gallant, 2000; Shary et al., 2002; Hengl and Reuter, 2009; Minár et al., 2016; Florinsky, 2017, 2025a). Geomorphometric modeling and mapping of ice-free Antarctic areas (Florinsky, 2023a, 2023b; Florinsky and Zharnova, 2025a, 2025b, 2025c, 2025d) can be used for obtaining new quantitative knowledge about the topography of these terrains and for the further use of morphometric information in solving problems of geomorphology, geology, glaciology, soil science, ecology, and other sciences. A project has recently been launched to create a physical geographical, thematic scientific reference geomorphometric atlas of ice-free areas of Antarctica (Florinsky, 2024, 2025b).

As part of this project, we carried out a geomorphometric modeling and mapping of ice-free areas of Enderby Land, East Antarctica. Previously, we published original series of geomorphometric maps for its ice-free mountainous areas (Florinsky and Zharnova, 2025d). In this paper, we report results for key coastal oases of Enderby Land.

## 2 Study areas

Enderby Land extends from the Shinnan (Carnebreen)[1] Glacier, Prince Olav Coast, Queen Maud Land on the west to Edward VIII Gulf, Kemp Land on the east; from 67.92°S, 44.69°E to 66.86°S, 57.10°E. To the north, Enderby Land is bounded by two seas of the Southern Ocean, the Cosmonauts Sea to the northwest and the Cooperation Sea to the northeast. Its interior is an ice-capped plateau. Along the coast, there are vast, partly ice-free mountainous areas (Florinsky and Zharnova, 2025d), several outlet glaciers, and some small coastal oases located on the shores of Freeth Bay, Alasheev Bight, Lena (Casey) Bay, and Amundsen Bay (Fig. 1).

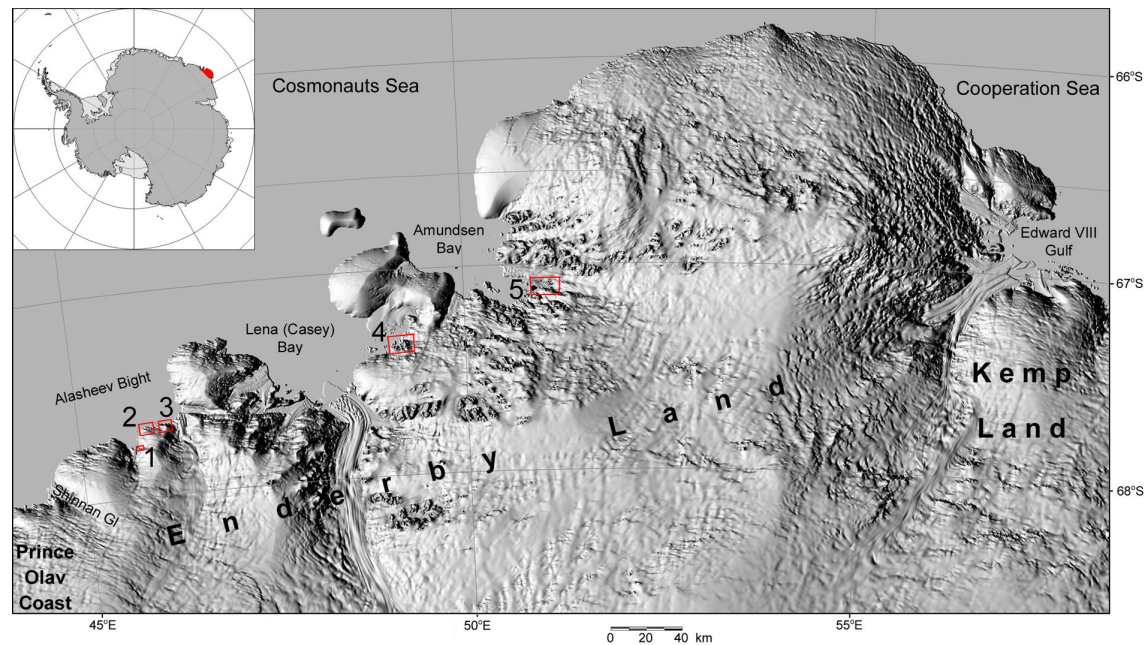

**Fig. 1** Geographical location of the study areas (red frames): 1 – Konovalov Oasis; 2 – Molodezhny Oasis, Thala Hills; 3 – Vecherny Oasis, Thala Hills; 4 – Fyfe Hills; 5 – Howard Hills. Hill shading was performed by REMA Explorer (PGC, 2022–2024).

---

[1] Some geographical objects of Enderby Land have double names assigned independently by the Soviet, Norwegian, and Australian cartographic authorities (Division of National Mapping, 1962–1964; Bakaev and Tolstikov, 1966; Donidze, 1987; Korotkevich et al., 2005; Australian Antarctic Division, 2023). In such cases, we use names of the both origin, with a second name in brackets.

**Table 1** Characteristics of the study areas (Fig. 1) and their DEMs.

| # | Area name | Geographical coordinates of a conditional center | Area,[*] $km^2$ | DEM size, m | DEM size, point | Points with elevation values |
|---|-----------|--------------------------------------------------|-----------------|-------------|-----------------|------------------------------|
| 1 | Konovalov Oasis | 67.75899° S, 45.76434° E | 3 | 5,216 × 2,648 | 652 × 331 | 202,710 |
| 2 | Molodezhny Oasis | 67.66987° S, 45.87278° E | 12 | 6,960 × 3,664 | 870 × 458 | 258,239 |
| 3 | Vecherny Oasis | 67.65804° S, 46.11360° E | 6 | 7,696 × 4,368 | 962 × 546 | 315,708 |
| 4 | Fyfe Hills | 67.36599° S, 49.23295° E | 25 | 10,728 × 7,688 | 1,341 × 961 | 934,624 |
| 5 | Howard Hills | 67.10550° S, 51.12242° E | 43 | 14,080 × 9,384 | 1,760 × 1,173 | 1,891,610 |

[*] Ice-free areas were estimated from satellite images (Fig. 2).

Coastal oases of Enderby Land have intensively been studied by the Soviet, American, Australian, Japanese, and Belorussian researchers (Klimov et al., 1962; Simonov, 1968; Myers and MacNamara, 1970; Alexandrov and Simonov, 1971; Alexandrov, 1971, 1972, 1985; Grew, 1975, 1978; Sobotovich et al., 1976; Kamenev, 1979; Black et al., 1983; Sandiford and Wilson, 1984; Kamenev and Hofmann, 1988; Yoshimura et al., 2000; Ishizuka, 2008; Myasnikov, 2011).

We consider five key coastal oases of Enderby Land (Fig. 1, Table 1), namely: the Konovalov Oasis (Fig. 2a), Thala Hills (Molodezhny and Vecherny Oases) (Figs. 2b and 2c), Fyfe Hills (Fig. 2d), and Howard Hills (Fig. 2e).

### 2.1 Konovalov Oasis

The Konovalov Oasis (also known as the Konovalov Mountains) is a small group of partly ice-free, rocky hills located on the coast of Freeth Bay, east of the Campbell outlet glacier (Figs. 2a and 3a). The oasis is 4.5 km long and 1.5 km wide. Its total area is about 3 $km^2$, including snowfields and intra-oasis glaciers.

The Konovalov Oasis is divided by the Khrustalnoe Lake valley. The area adjacent to Freeth Bay occupies about 20% of the entire oasis area and is located at altitudes of 50–90 m above sea level (ASL). The western low foothill part of the oasis (up to of 160 m) includes hills and valleys. There are snowfields on the northern slopes of the hills and lakes in the valleys. The oasis foothill occupies approximately 55% of the entire oasis area. The eastern elevated part of the oasis lies at elevations of 200–350 m ASL. It is divided into three sections by two intra-oasis glaciers 300–400 m wide. The highest point of the oasis is the Gorodkov Hill reaching 346 m ASL (Alexandrov, 1971).

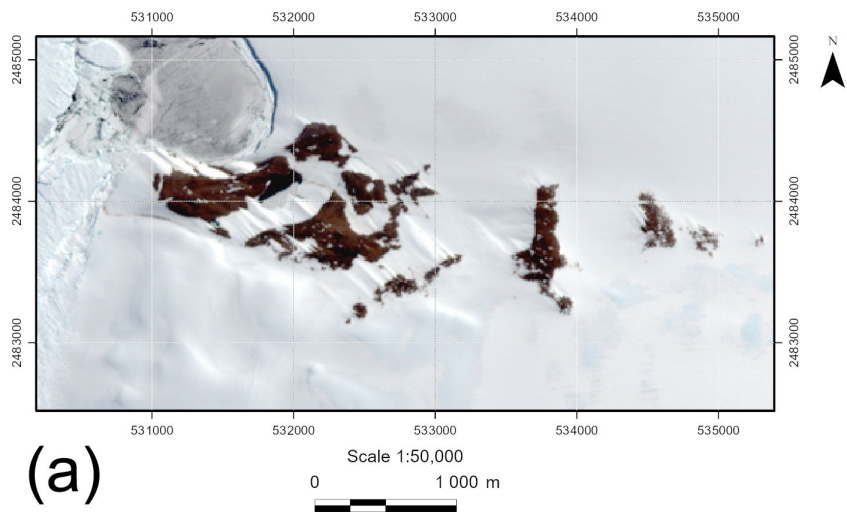

(a)

**Fig. 2** Satellite images of the study areas: (a) Konovalov Oasis.

*(Continued)*

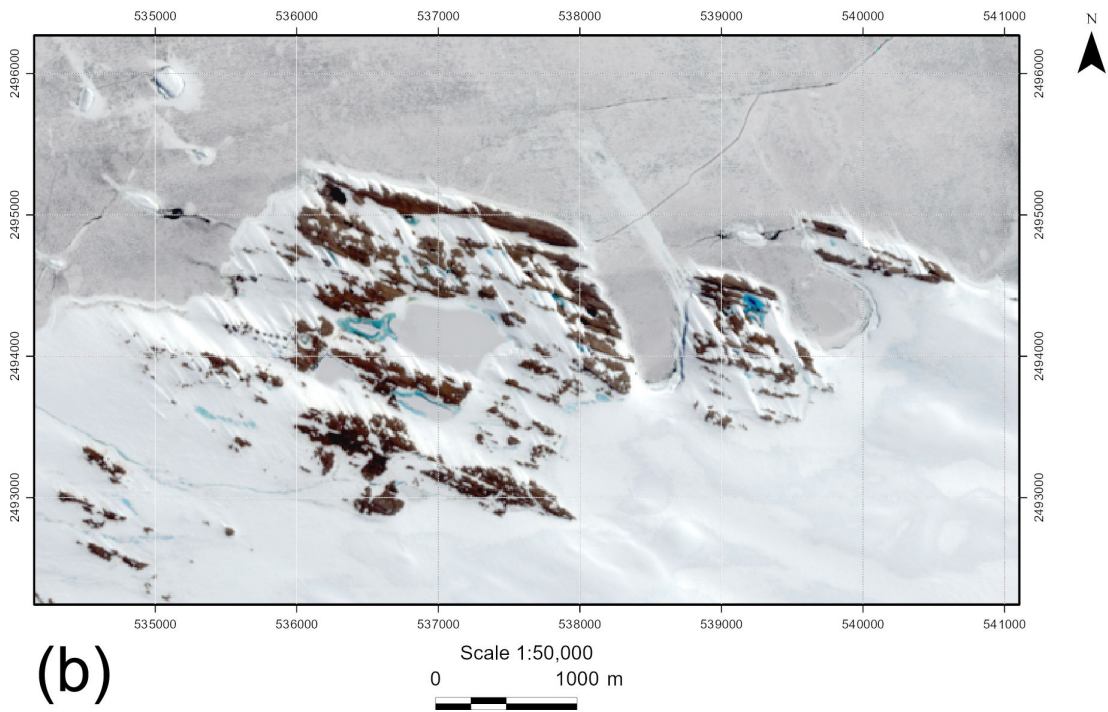

(b)

Scale 1:50,000

0          1000 m

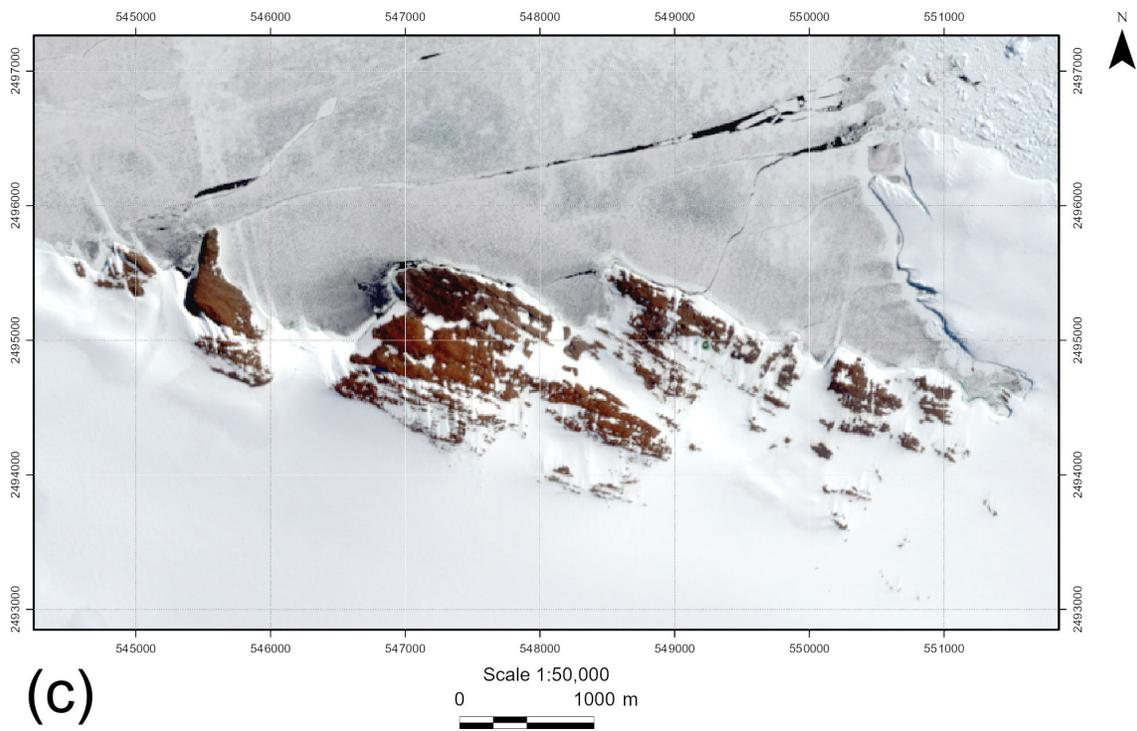

(c)

Scale 1:50,000

0          1000 m

**Fig. 2, cont'd** Satellite images of the study areas: (b) Molodezhny Oasis, Thala Hills. (c) Vecherny Oasis, Thala Hills.

*(Continued)*

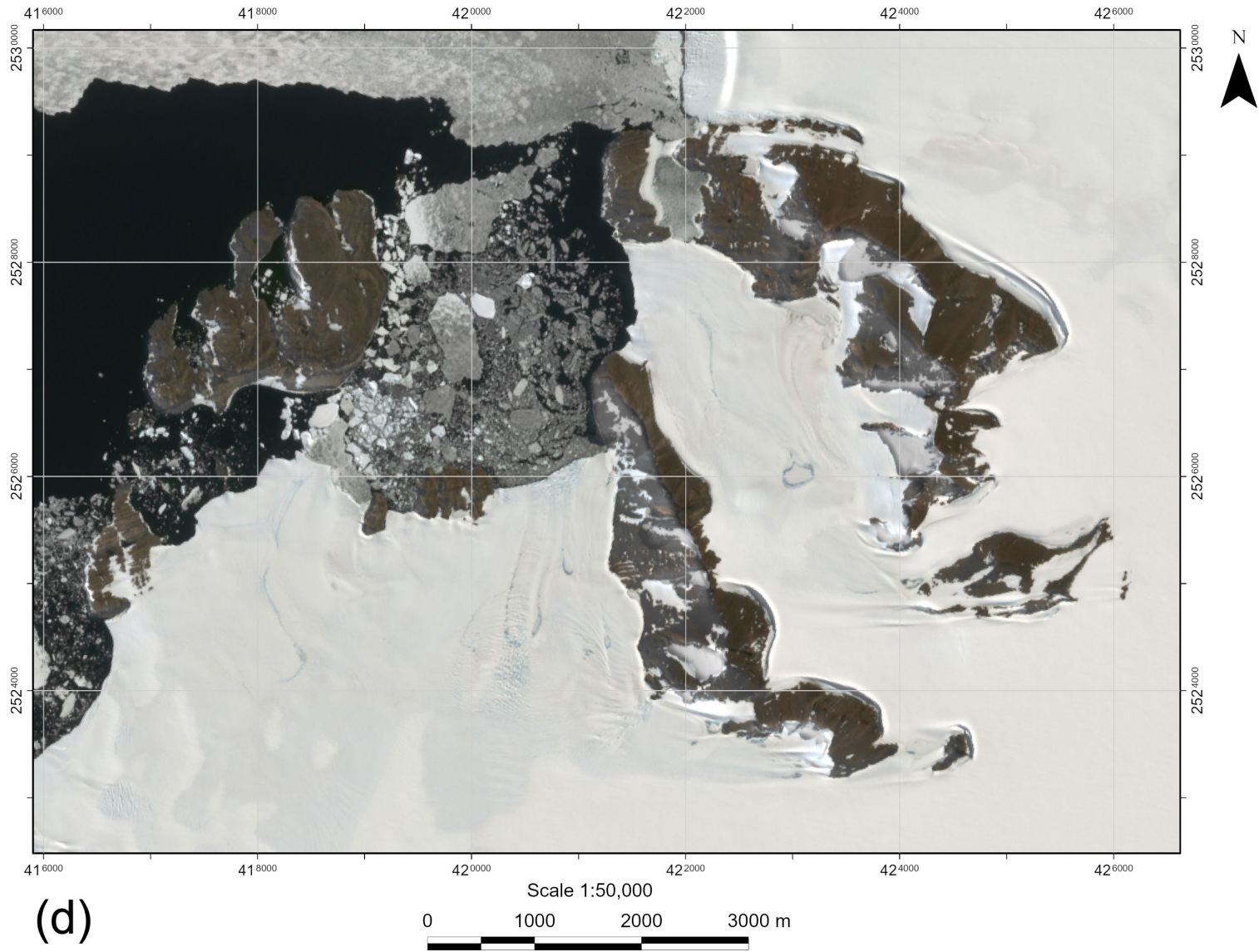

(d)

Scale 1:50,000

0    1000    2000    3000 m

**Fig. 2, cont'd** Satellite images of the study areas: (d) Fyfe Hills.

*(Continued)*

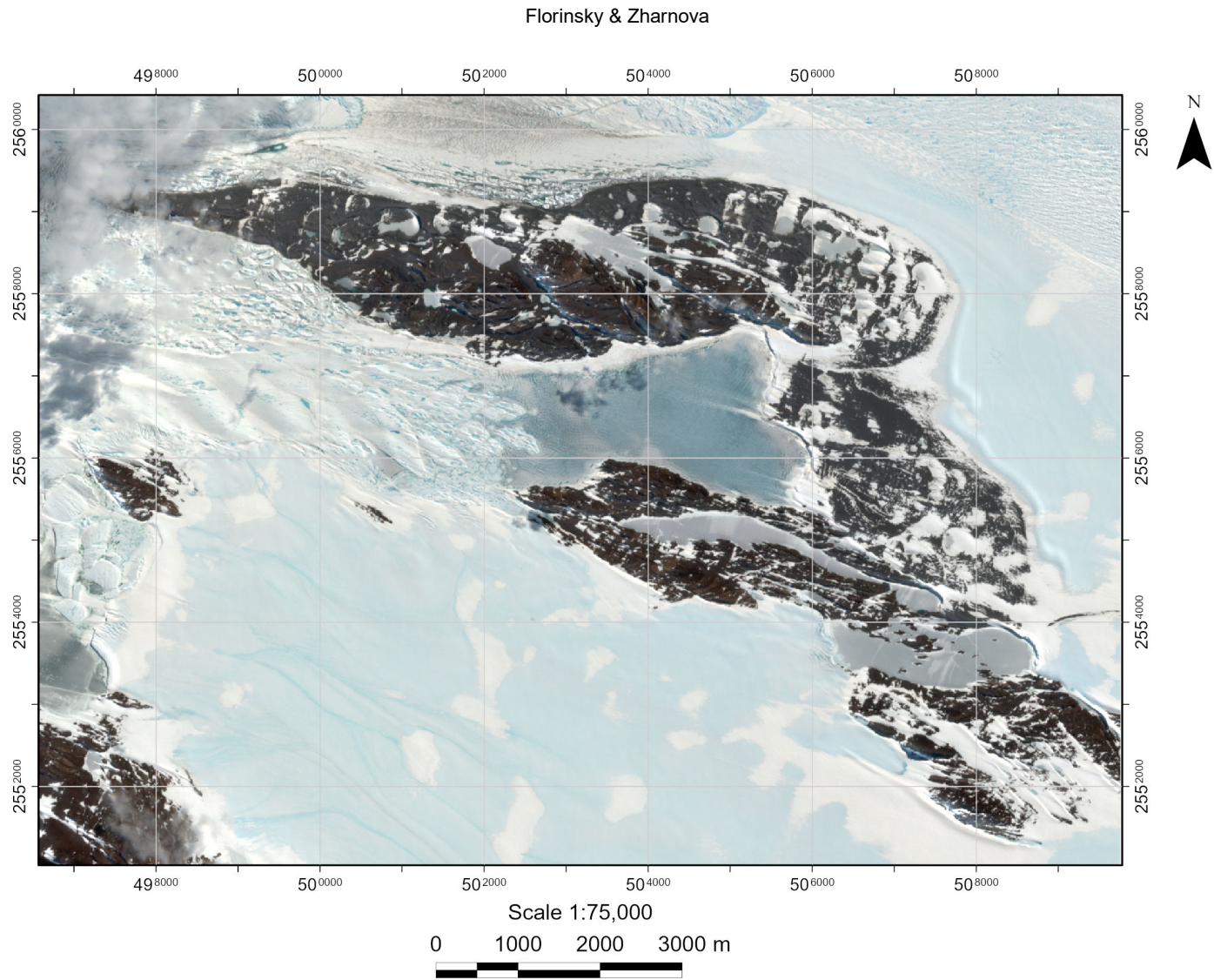

Scale 1:75,000

**Fig. 2, cont'd** Satellite images of the study areas: (e) Howard Hills.
Fragments of Sentinel-2A MSI scenes captured on 15.12.2024 for the Howard Hills and 20.01.2025 for the other study areas (ESA, 2025). True color synthesis with brightness and contrast enhancement. UTM projection (zone 38S for the Konovalov, Molodezhny, and Vecherny Oases; zone 39S for the Fyfe and Howard Hills).

The Konovalov Oasis is generally composed of rocks of the Proterozoic Rayner Complex, one of the ultrahigh-temperature metamorphic complexes of Enderby Land. The rocks of the Konovalov Oasis include charnockitic and enderbitic gneisses with subordinate amounts of two-pyroxene granulites, hornblende and biotite two-feldspar granitic gneisses intruded by pegmatites and granite aplite, as well as migmatites with alternating layers of quartzo-feldspathic and hornblende, garnet–biotite, or hornblende-pyroxene gneisses (Krylov et al., 1998).

### 2.2 Molodezhny Oasis

The Molodezhny Oasis (Figs. 2b and 4a) is the western portion of the Thala Hills located on the shore of Alasheev Bight of the Cosmonauts Sea. The oasis is bounded by the ice sheet in the south, by the sea in the west and north, and by the Groznensky outlet glacier in the east. The oasis stretches from west to east for 8.3 km, and from south to north for 2.7 km; its total area is about 12 km$^2$, including intra-oasis glaciers and snowfields.

The Molodezhny Oasis is an exaratic, rocky hummocky terrain consisting of several ice-free ridges stretched parallel to the coast and separated by snow-covered terraced valleys with glaciers, snowfields, lakes, and channels of temporary watercourses (Simonov, 1968; Alexandrov and Simonov, 1971; Alexandrov, 1971, 1985). The absolute heights of the oasis do not exceed 100 m, while the relative elevations of hills and ridges are 10–40 m. The oasis's elevations decrease from the southeast to the northwest, from the ice sheet edge to the sea coast.

The Molodezhny Oasis is cut by faults formed as a result of tectonic movements of individual sections of the coast. The northwest-striking fault system is best manifested in the topography. These faults control all large depressions, mainly filled with glaciers, snowfields and lakes. Parallel to the depressions, the oasis is crossed by cuesta-like hill ridges 600–1000 m long and up to 150 m wide. The northeastern slopes of the ridges are steep and short, with cliffs; the opposite slopes are gentle, some of them are 200 m long. The ridge surface is heavily damaged by physical weathering. The ridge tops are usually covered with eluvium (fine soil with rubble) and a thin layer of superficial moraine deposits (Simonov, 1968; Alexandrov, 1985).

Traces of glacial activity can be found in many parts of the oasis in the form of deepened valleys, polished rocks, and moraine deposits. However, these traces are masked by traces of physical weathering of rocks such as nivation, cellular weathering, and desquamation (onion-skin weathering). As a result of nivation, significant areas of the oasis are occupied by stone placers, which are most often found on the tops and gentle slopes of hills (Simonov, 1968).

The Molodezhny Oasis is composed of rocks belonging to the Proterozoic Ongul complex of polymetamorphic rocks of granulite and amphibolite facies. The predominant rocks are charnockite gneisses, biotite, biotite-garnet, and amphibole-biotite gneisses and plagiogneisses, amphibolites, as well as amphibole-pyroxene-plagioclase crystalline schists (Kamenev and Hofmann, 1988).

In the western part of the oasis, there is the Molodezhnaya seasonal research station (Russia), which was founded in 1962 as a year-round polar research station, but was mothballed in 1999.

### 2.3 Vecherny Oasis

The Vecherny Oasis (Figs. 2c and 5a) is the eastern portion of the Thala Hills located on the shore of Alasheev Bight of the Cosmonauts Sea. The oasis is bounded by the Hays outlet glacier flowing into Spooner Bay in the east, by the ice sheet in the south, and by the sea in the north. The Groznensky outlet glacier separates the Vecherny Oasis from the Molodezhny

Oasis in the west. The Vecherny Oasis stretches from west to east for 6.7 km, and from south to north for 1.8 km. Its total area is about 6 km$^2$, including intra-oasis glaciers and snowfields.

The landforms of the Vecherny Oasis are stretched in the northwestern direction, as in the neighboring Molodezhny Oasis. Geomorphically, the oasis can be divided into three parts, such as central, eastern, and western.

The central part of the oasis is formed by the Mount Vechernyaya massif (272 m ASL) and the lowland foothills lying between the massif and the sea (Alexandrov, 1971, 1985). Orographically, the central part of the oasis can be divided into three sections. The first section includes relatively flat mountain summits 200–270 m high. They are complicated by small, isometric rocky ridges and hills with relative heights of up to 10 m. The ridges and hills are partly smooth, while the depressions separating them contain debris deposits and snowfields. The second, foothill section is a low, hilly and ridged terrain adjacent to the sea coast. In some places, its heights slightly exceed 100 m. The third section lies between the first two and is represented by a rocky slope up to 500 m long and 2.5 km wide. The generally steep slope has a bend in the form of a gently sloping terrace, reaching 100 m in width. The slope has a smooth surface composed of charnockite (Alexandrov, 1985).

There is a small intra-oasis glacier between the central and eastern parts of the Vecherny Oasis. The eastern part is divided into sections by near-north-south- and near-west-east-striking valleys. The valleys are filled with glaciers and snowfields. There are also small ridges and hills with relative heights of several meters. The highest point of the eastern part, the Rubin Hill, is about 160 m ASL, but elevations less than 100 m predominate there. The western part of the oasis includes two minor rocky capes, the Rog Point and the Tri Valuna Point. They are separated from Mount Vechernyaya by small outlet glaciers 200–300 m wide (Alexandrov, 1971).

In the Vecherny Oasis, tectonic movements have formed a number of narrow and long, mutually perpendicular valley-like depressions. Moraine as well as eluvial and deluvial deposits formed a discontinuous and thin mantle. Erratic boulders can be found throughout the oasis, but they are particularly noticeable on smooth rock surfaces (Alexandrov, 1985).

The Vecherny Oasis can be classified as exaratic, rocky hummocky terrain, with the exception of areas of smooth mountain slope.

The oasis is predominantly composed of the Proterozoic charnockite series, including amphibole-pyroxene-plagioclase-quartz feldspar charnockitized enderbites, amphibole-feldspar-quartz-plagioclase charnockites, and ultrametamorphic feldspar-quartz-plagioclase-amphibole-biotite charnockites as well as the Neoarchean biotite-hornblende-two-pyroxene plagiogneisses (Myasnikov, 2011).

In the central part of the oasis, there is the Gora Vechernyaya seasonal research station (Belarus), which was constructed in 2015.

### 2.4 Fyfe Hills

The Fyfe Hills (Figs. 2d and 6a) are a group of coastal hills lying southeast of Khmara Bay of Lena (Casey) Bay and south of the Molokov Glacier. In the Fyfe Hills, there are three main features: Mounts Enderbitovaya, Zubastaya, and Nikitin. Mount Enderbitovaya is the largest in the height (650 m ASL) and area; it includes several alpine-type peaks. Mount Zubastaya is located to the west; its pointed peaks forming a ridge. Mount Nikitin (also known as the Nikitin Oasis) is the third in the height (about 500 m ASL) and area (1.8 km$^2$). It is located south of Mount Enderbitovaya, 3.5 km from the sea. Mount Nikitin has several flatter summits. In the intermountain valleys, there are glaciers from 350 m to 1.5 km wide (Alexandrov, 1972).

The topography of the Fyfe Hills can be classified as a mountain-glacial dissected periglacial terrain, but, in some places, it has been changed by exaration processes. Most of the Fyfe Hills are covered by eluvial, deluvial, and glacial deposits, among which large-fragment placers and erratic boulders predominate. Permafrost processes led to the formation of polygonal patterned grounds. Along with this, there are rock outcrops in cliffs in the form of steep walls and inclined slabs as well as *roches moutonnee*. In places where glaciers and rocky areas meet, there are dammed lakes usually occupying randkluft bottoms (Alexandrov, 1972).

Ridges, hills, and cirques are common in the Fyfe Hills. The ridges are jagged, while the hills have relatively flat tops and steep slopes. The cirques face northwest, northeast, and south. At the cirque bottoms, there are lakes and glaciers of intermountain valleys. At the cirque tops, there are snowdrift glaciers and persistent snowbanks. The distance between cirque edges does not exceed 700 m; the slope length reaches 300 m (Alexandrov, 1985).

Many rock outcrops have tafoni formations. Different hilltops are covered either by moraine deposits with erratic boulders up to 1–2 m in diameter, or fine eluvium, or large bedrock fragments with coarse-grained eluvium. The hill slopes are predominantly covered with large-scale debris scree. The origin of the detrital material is eluvial, deluvial, and moraine (Alexandrov, 1985).

The Fyfe Hills are composed of the Early Archaean rocks of the Napier Complex, one of the ultrahigh-temperature metamorphic complexes of Enderby Land. These rocks include predominantly enderbites and charnockites, as well as pyroxene-plagioclase crystalline schists, pyroxenites, peridotites, garnet-pyroxene-magnetite-quartz and plagioclase crystalline schists, and garnet or sillimanite-garnet mesoperthite gneisses (Kamenev, 1979).

The enderbite samples, collected on the southwest slopes of Mount Nikitin[2], were isotopically dated to 4000±200 Ma (Sobotovich et al., 1976; Kamenev, 1979). Isotopic and geological data led Sobotovich et al. (1976) to suggest that the metamorphic rocks of the Fyfe Hills represent a relic of the primary crust, formed by accretion in the last stages of the Earth's formation and later metamorphosed under conditions of the two-pyroxene gneiss facies.

### 2.5 Howard Hills

The Howard Hills (Figs. 2e and 7a) are an area of low ice-free hills and meltwater lakes located northeast of the Krzhizhanovsky Mountains, south of the Beaver Glacier, and east of Amundsen Bay (see Fig. 13 in Florinsky and Zharnova, 2025d). The Howard Hills extend in a northwestern direction for 17 km with a maximum width of 5 km. The ice sheet drops steeply towards the eastern edge of the oasis. From the west, a 1.7 km wide epishelf lake-lagoon juts into the oasis for 3 km, dividing the oasis in two, northern and southern, halves. The total area of the oasis is 43 km[2]. Within the oasis, one can observe two types of topography, namely, glacial accumulative (moraine) and denudation (exaratic, rocky hummocky) ones (Alexandrov, 1985).

Denudation topography predominates in the central part of the northern half of the Howard Hills. Elongated convex rocky landforms are controlled by near-northeast-striking geological structures. There are two mutually perpendicular systems of valley-like depressions associated with faults, which together form block structure of the topography. In some places where faults intersect, closed depressions have formed. All depressions are filled with detrital material of eluvial, deluvial, moraine, and fluvioglacial origin. On the rocky surface, there are many erratic boulders of various sizes (Alexandrov, 1985).

Glacial accumulative topography is observed in the west of the Howard Hills. The moraine deposits reach a thickness of 50 m. There are several lake basins measuring up to 200 × 400 m with steep (35° and higher) and long slopes. In some areas, one can see drumlin-like hills (Alexandrov, 1985).

---

[2] In the geological literature, toponymic errors were made: Mount Nikitin, where the geological sampling site is located, was mistakenly called Mount Novogodnyaya (Sobotovich et al., 1976) and Mount Enderbitovaya (Kamenev, 1979). One can compare maps by Kamenev (1979, p. 12) and Alexandrov (1972, p. 19).

The topography of the eastern part of the Howard Hills also has glacial accumulative origin. Orographically, three sections can be distinguished there: two gently sloping sections, which are separated by a steep-sloping one. It stretches from the lagoon shore to the east, to a group of thermokarst lake basins located 0.5 km from the ice sheet edge. In the gently sloping sections, the thick moraine, lying on the bedrocks, has a wavy surface complicated by a number of erratic boulders up to several meters in size. In the steep-sloping section, the landscape is complicated by minor parallel ridges. Deposits of lesser thickness, lying on bedrocks and buried ice, form a gently sloping surface with relatively small thermokarst depressions 10–15 m in diameter and 2–3 m in depth, as well as low and narrow ridges up to 100 m long, located 50–100 m from each other. The lagoon shores are adjoined by moraine deposits lying on buried ice (Alexandrov, 1985).

The Howard Hills area belongs to the Napier complex, one of the ultrahigh-temperature metamorphic complexes of Enderby Land. The oasis is predominately composed of garnet felsic gneiss, orthopyroxene felsic gneiss, and aluminous gneiss containing garnet, sapphirine, and sillimanite (Yoshimura et al., 2000).

### 3 Materials and methods

We used five digital elevation models (DEMs) with a grid spacing of 8 m. The DEMs were extracted from the Reference Elevation Model of Antarctica (REMA) (Howat et al., 2019; REMA, 2018–2022). For data preprocessing and mapping (see below), we applied a procedure described earlier in our papers (Florinsky, 2025b; Florinsky and Zharnova, 2025b, 2025c, 2025d).

#### 3.1 Preprocessing

REMA is presented in the polar stereographic projection with an elevation datum of the WGS84 ellipsoid. There is no bathymetry in REMA; such cells have 'no data' values. The extracted DEMs were reprojected into the UTM projection (zone 38S for the Konovalov, Molodezhny, and Vecherny Oases; zone 39S for the Fyfe and Howard Hills) preserving the original grid spacing (Table 1). Ellipsoidal elevations were transformed into orthometric ones.

#### 3.2 Calculations

Digital models of eleven, most scientifically important morphometric variables were derived from the preprocessed DEMs. The list of morphometric variables (Table 2) includes six local attributes: slope ($G$), aspect ($A$), horizontal curvature ($k_h$), vertical curvature ($k_v$), minimal curvature ($k_{min}$), and maximal curvature ($k_{max}$); one nonlocal variable—catchment area ($CA$); two combined variables: topographic wetness index ($TWI$) and stream power index ($SPI$); as well as two two-field-specific attributes—total insolation ($TIns$) and wind exposition index ($WEx$). Formulas and detailed interpretations of these morphometric variables can be found elsewhere (Shary et al., 2002; Florinsky 2017, 2025a, chap. 2).

To derive digital models of local variables, we used a finite-difference method by Evans (1980). To compute models of $CA$, we applied a maximum-gradient based multiple flow direction algorithm by Qin et al. (2007) to preprocessed sink-filled DEMs. $CA$ digital models were then logarithmized. To derive digital models of combined morphometric variables, we used calculated models of $CA$ and $G$. To compute digital models of two-field-specific attributes, we applied two related methods by Böhner (2004). $TIns$ were estimated for one mid-summer day (1st January) with a temporal step of 0.5 h.

**Table 2** Definitions and interpretations of key morphometric variables (Shary et al., 2002; Florinsky, 2017, 2025a, chap. 2).

| Variable, notation, and unit | Definition and interpretation |
|---|---|
| *Local morphometric variables* | |
| Slope, $G$ (°) | An angle between the tangential and horizontal planes at a given point of the topographic surface. Relates to the velocity of gravity-driven flows. |
| Aspect, $A$ (°) | An angle between the north direction and the horizontal projection of the two-dimensional vector of gradient counted clockwise, from 0° to 360°, at a given point of the surface. Relates to the direction of gravity-driven flows |
| Horizontal (tangential) curvature, $k_h$ (m$^{-1}$) | A curvature of a normal section tangential to a contour line at a given point of the surface. A measure of flow convergence and divergence. Gravity-driven lateral flows converge where $k_h < 0$, and diverge where $k_h > 0$. $k_h$ mapping reveals crest and valley spurs. |
| Vertical (profile) curvature, $k_v$ (m$^{-1}$) | A curvature of a normal section having a common tangent line with a slope line at a given point of the surface. A measure of relative deceleration and acceleration of gravity-driven flows. They are decelerated where $k_v < 0$, and are accelerated where $k_v > 0$. $k_v$ mapping reveals terraces and scarps. |
| Minimal curvature, $k_{min}$ (m$^{-1}$) | A curvature of a principal section with the lowest value of curvature at a given point of the surface. $k_{min} > 0$ corresponds to local convex landforms, while $k_{min} < 0$ relates to elongated concave landforms. |
| Maximal curvature, $k_{max}$ (m$^{-1}$) | A curvature of a principal section with the highest value of curvature at a given point of the surface. $k_{max} > 0$ corresponds to elongated convex landforms, while $k_{max} < 0$ relate to local concave landforms. |
| *Nonlocal morphometric variables* | |
| Catchment area, $CA$ (m$^2$) | An area of a closed figure formed by a contour segment at a given point of the surface and two flow lines coming from upslope to the contour segment ends. A measure of the contributing area. |
| *Combined morphometric variables* | |
| Topographic wetness index, $TWI$ | A ratio of catchment area to slope gradient at a given point of the topographic surface. A measure of the extent of flow accumulation. |
| Stream power index, $SPI$ | A product of catchment area and slope gradient at a given point of the surface. A measure of potential flow erosion and related landscape processes. |
| *Two-field specific morphometric variables* | |
| Total insolation, $TIns$ (kWh/m$^2$) | A measure of the topographic surface illumination by solar light flux. Total potential incoming solar radiation, a sum of direct and diffuse insolations. |
| Wind exposition index, $WEx$ | A measure of an average exposition of slopes to wind flows of all possible directions at a given point of the topographic surface. |

### 3.3 Mapping

First, we created hypsometric maps of the five key coastal oases of Enderby Land from the preprocessed DEMs. To display the elevations of the ice-free topography, we applied the green-yellow part of the standard spectral hypsometric scale of color plasticity (Kovaleva, 2014). To display the elevations of the glacier topography, we utilized a modified hypsometric tint scale for polar regions (Patterson and Jenny, 2011). Then, two hypsometric tintings were combined with achromatic hill shading derived from the DEMs by a standard procedure (Figs. 3a, 4a, 5a, 6a, and 7a). Finally, we put geographical names on the hypsometric maps. As a source of this information, we used available topographic and geological maps of the study areas (Division of National Mapping, 1962–1964; Bakaev and Tolstikov, 1966; Sheraton, 1985; Korotkevich et al., 2005; Australian Antarctic Division, 2023) as well as schemes from the related articles (Alexandrov, 1971, 1972).

Second, from the calculated digital morphometric models, we produced series of morphometric maps for the five key coastal oases of Enderby Land (Figs. 3b–l, 4b–l, 5b–l, 6b–l, and 7b–l). For optimal visual perception of morphometry, we used the following rules for applying gradient tint scales:

1. $G$ and $CA$ can take only positive values. To map $G$ and $CA$, we applied a standard gray tint scale; the minimum and maximum values of $G$ or $CA$ correspond to white and black, respectively (Figs. 3b, 3h, 4b, 4h, 5b, 5h, 6b, 6h, 7b, and 7h).

2. $A$ is circular variable taking values from 0° to 360°. To map $A$, we applied an eight-color, eight-cardinal direction tint scale (Figs. 3c, 4c, 5c, 6c, and 7c).

3. $k_h$, $k_v$, $k_{min}$, and $k_{max}$ can take both negative and positive values, having opposite physical mathematical and physical geographical sense and interpretation. To map curvatures, we used a two-color tint scale consisting of two contrasting parts, blue and orange (negative and positive values, respectively). The most and least saturated shades of blue or orange colors correspond to the absolute maximum and absolute minimum values, respectively, of $k_h$, $k_v$, $k_{min}$, and $k_{max}$ (Figs. 3d–g, 4d–g, 5d–g, 6d–g, and 7d–g).

4. $TWI$ and $SPI$ can take only positive values. To map these indices, we applied a standard spectral tint scale; the minimum and maximum values of $TWI$ or $SPI$ correspond to violet and red, respectively (Figs. 3i, 3j, 4i, 4j, 5i, 5j, 6i, 6j, 7i, and 7j).

5. $TIns$ is a nonnegative variable. To map it, we used an orange tint scale: the minimum and maximum $TIns$ values correspond to the darkest and lightest orange shades depicting the least and most illuminated areas, respectively (Figs. 3k, 4k, 5k, 6k, and 7k).

6. $WEx$ is a positive dimensionless variable, wherein values below and above 1 relate to wind-shadowed and -exposed areas, respectively. To map this index, we applied a two-color tint scale consisting of two contrasting parts, orange and violet (values below and above 1, respectively). The darkest shades of orange and violet colors correspond to the minimum and maximum $WEx$ values, respectively, while the lightest shades of the colors correspond to 1 (Figs. 3l, 4l, 5l, 6l, and 7l).

There is no lake bathymetry data in REMA; lake cells contain interpolated values of lake coastal elevations, that is, artifacts. Also, REMA has multiple island-like artifacts related to icebergs. On the produced maps, the lakes and island-like artifacts were masked.

To create masks, we used maps of three spectral indices, namely, modified normalized difference water index (Xu, 2006), ferrous minerals ratio, and iron oxide ratio (Segal, 1982). We derived the spectral index maps from the Sentinel-2A multispectral imager (MSI) satellite scenes captured on 15.12.2024 for the Howard Hills and 20.01.2025 for the other study areas (ESA, 2025). To map spectral indices, we used composite legends allowing for the most successful classification of objects by moisture level and iron content. A technique description can be found elsewhere (Florinsky and Zharnova, 2025b, 2025c).

For DEM preprocessing and geomorphometric calculations, we used a software SAGA 9.8.1 (Conrad et al., 2015). For morphometric mapping and map masking, we utilized ArcGIS Pro 3.0.1 (ESRI, 2015–2024).

## 4 Results

Geomorphometric modeling and mapping resulted in the series of morphometric maps for the five key coastal oases of Enderby Land including the Konovalov Oasis (Fig. 3), Molodezhny Oasis (Fig. 4), Vecherny Oasis (Fig. 5), Fyfe Hills (Fig. 6), and Howard Hills (Fig. 7). In total, we derived 60 maps in 1:50,000 and 1:75,000 scales.

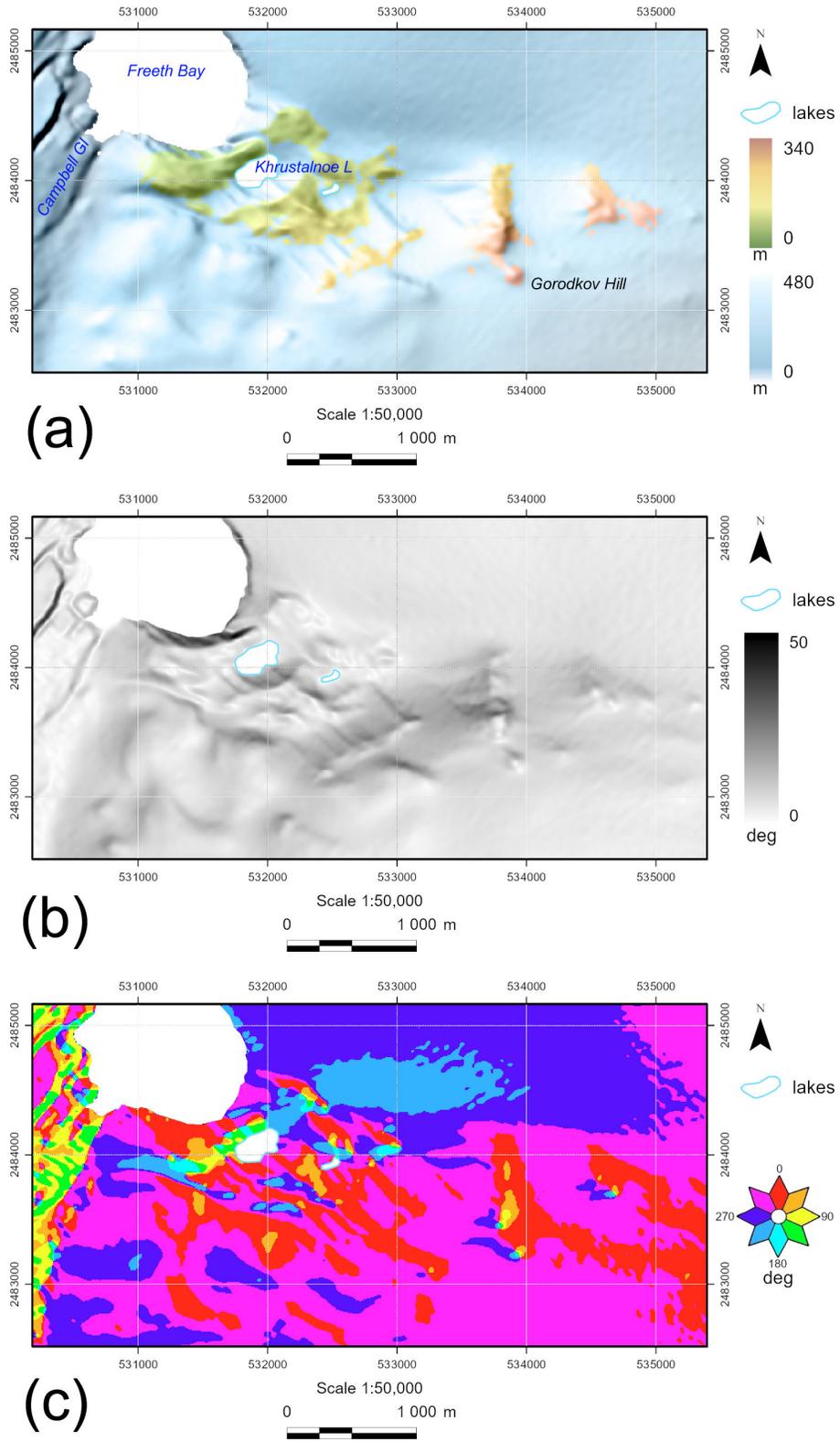

**Fig. 3** Konovalov Oasis: (a) Elevation. (b) Slope. (c) Aspect.

*(Continued)*

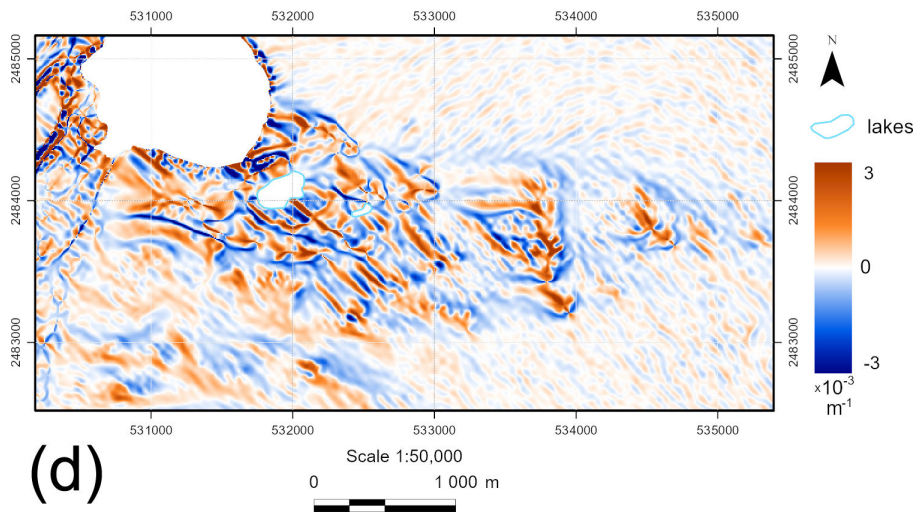

(d)

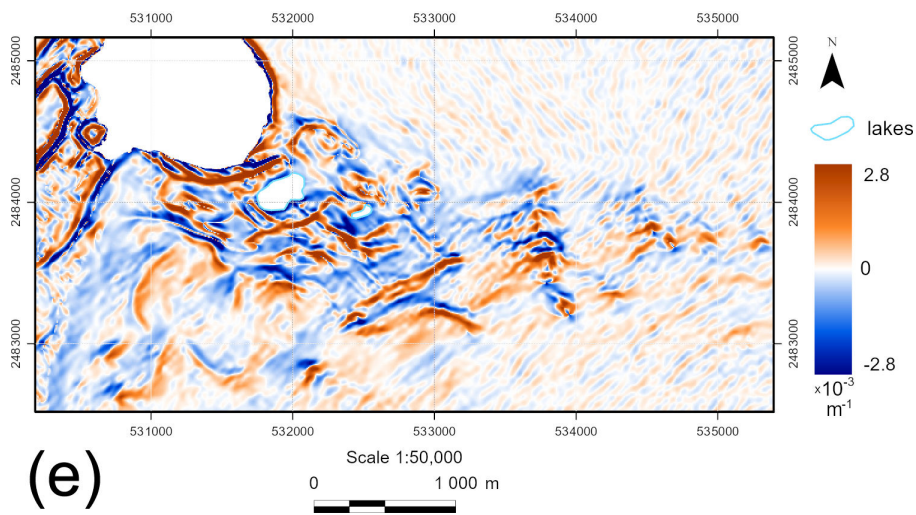

(e)

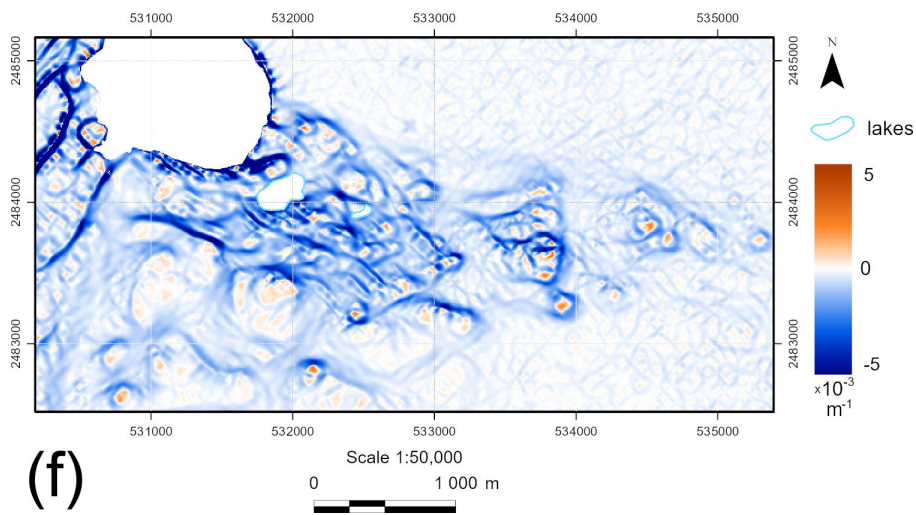

(f)

**Fig. 3, cont'd** Konovalov Oasis: (d) Horizontal curvature. (e) Vertical curvature. (f) Minimal curvature.

*(Continued)*

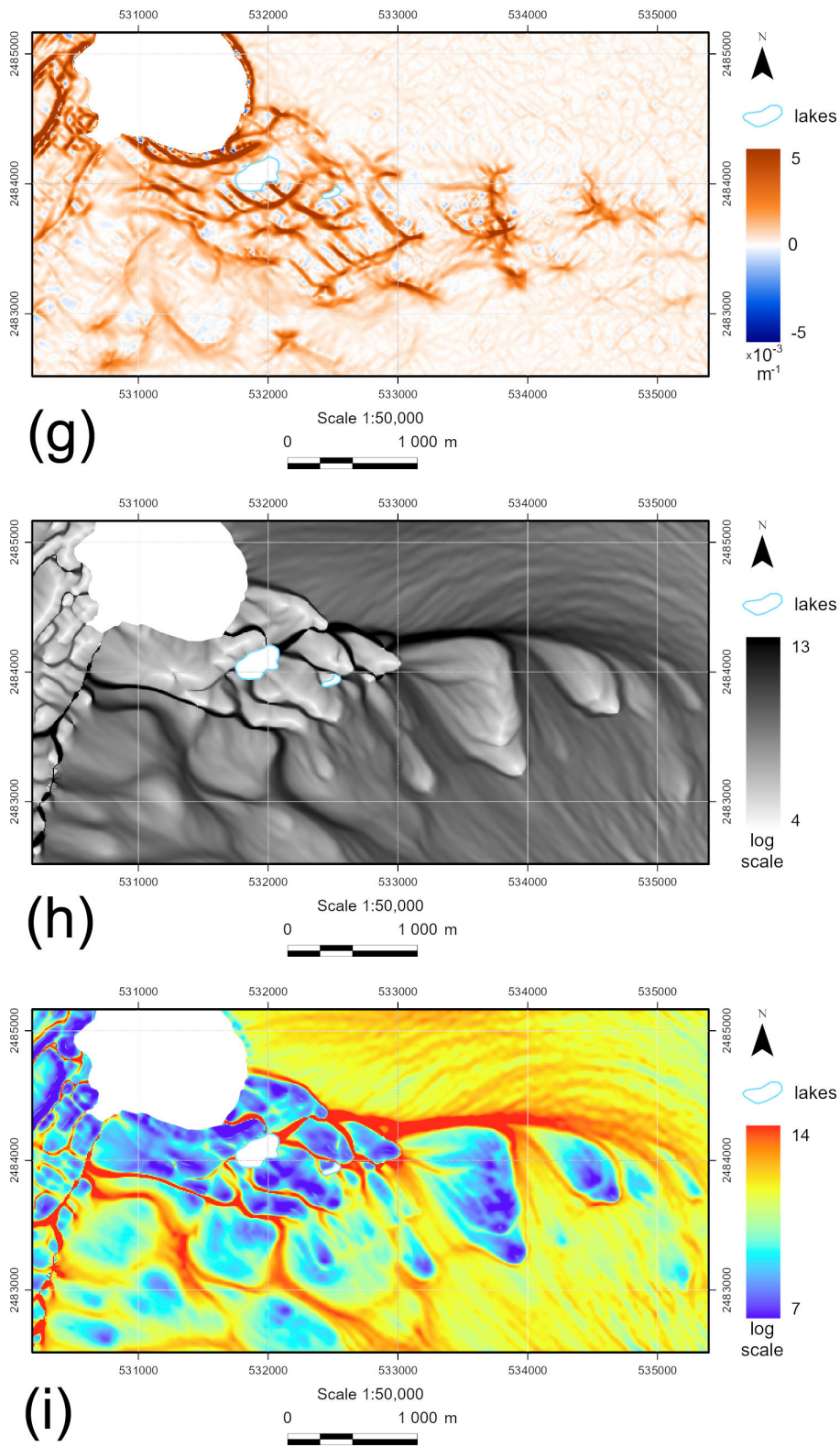

**Fig. 3, cont'd** Konovalov Oasis: (g) Maximal curvature. (h) Catchment area. (i) Topographic wetness index.

*(Continued)*

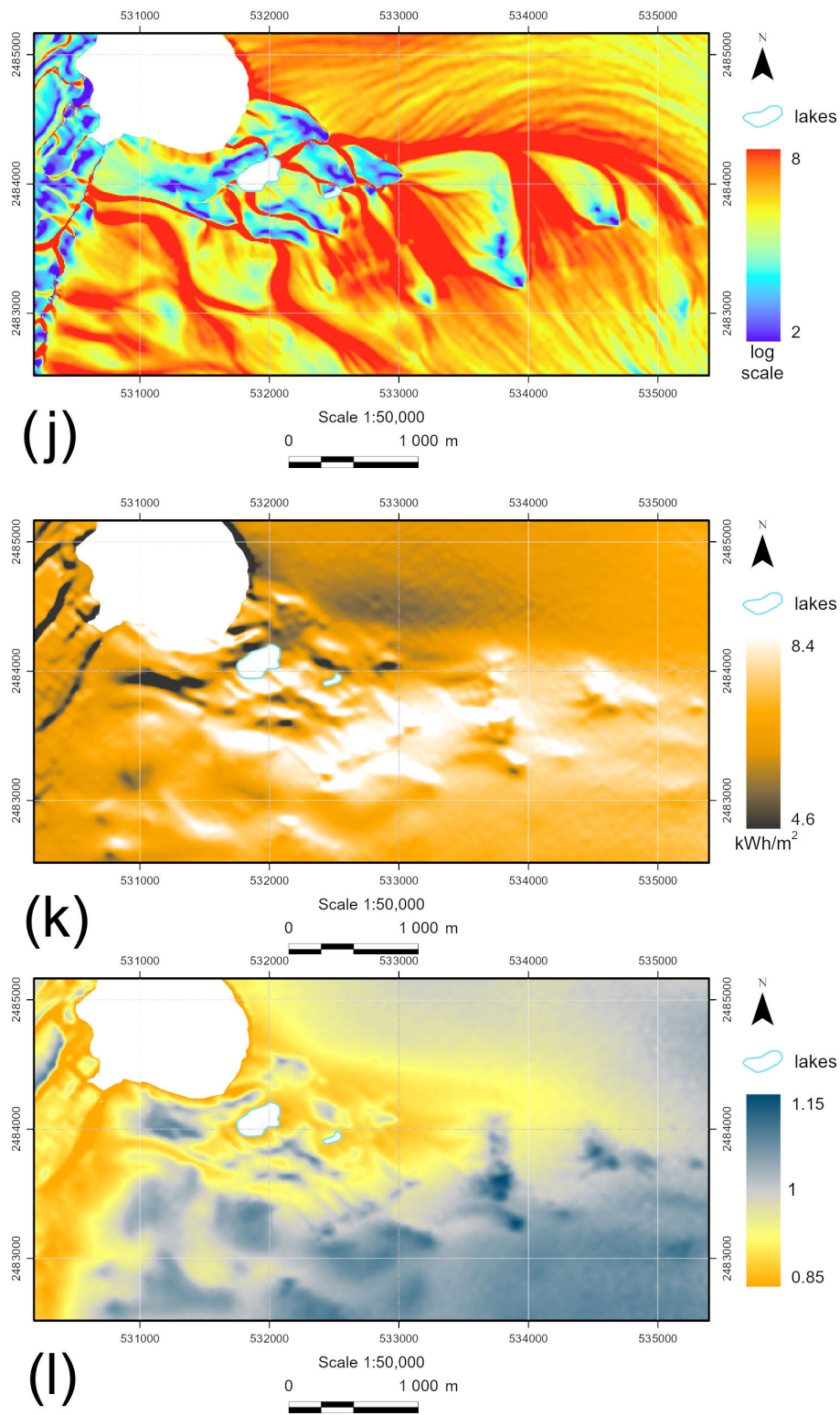

**Fig. 3, cont'd** Konovalov Oasis: ( j) Stream power index. (k) Total insolation. (l) Wind exposition index.

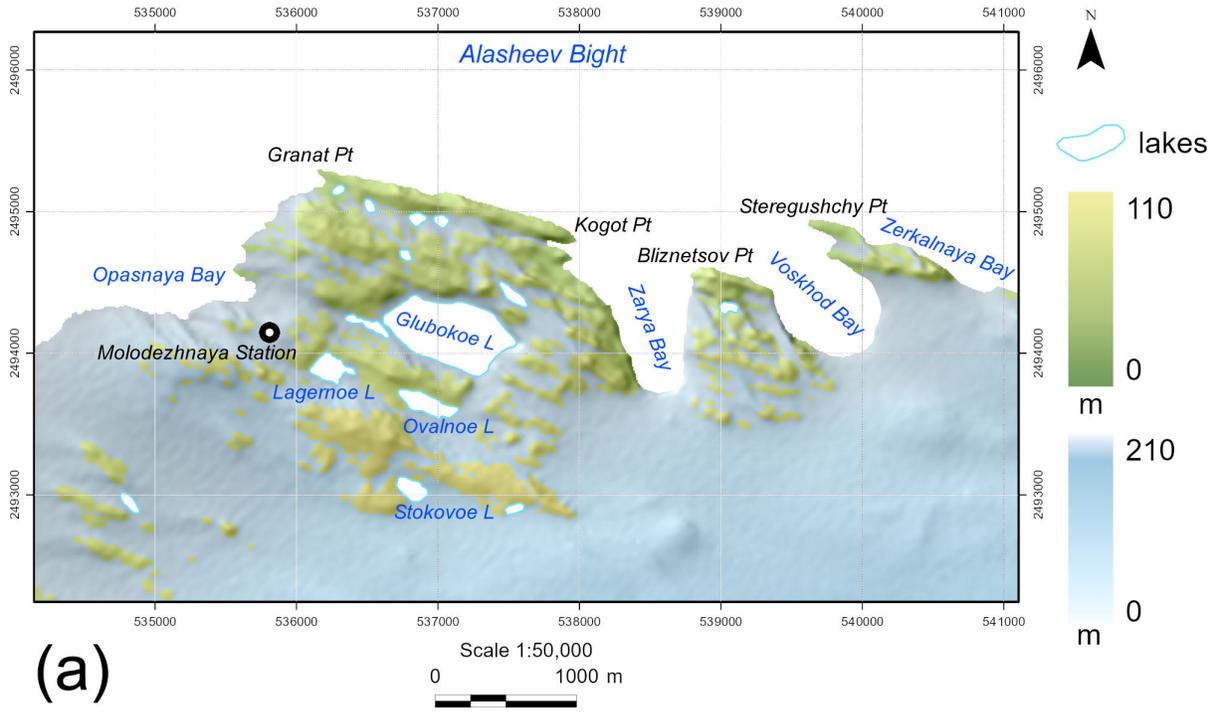

(a)

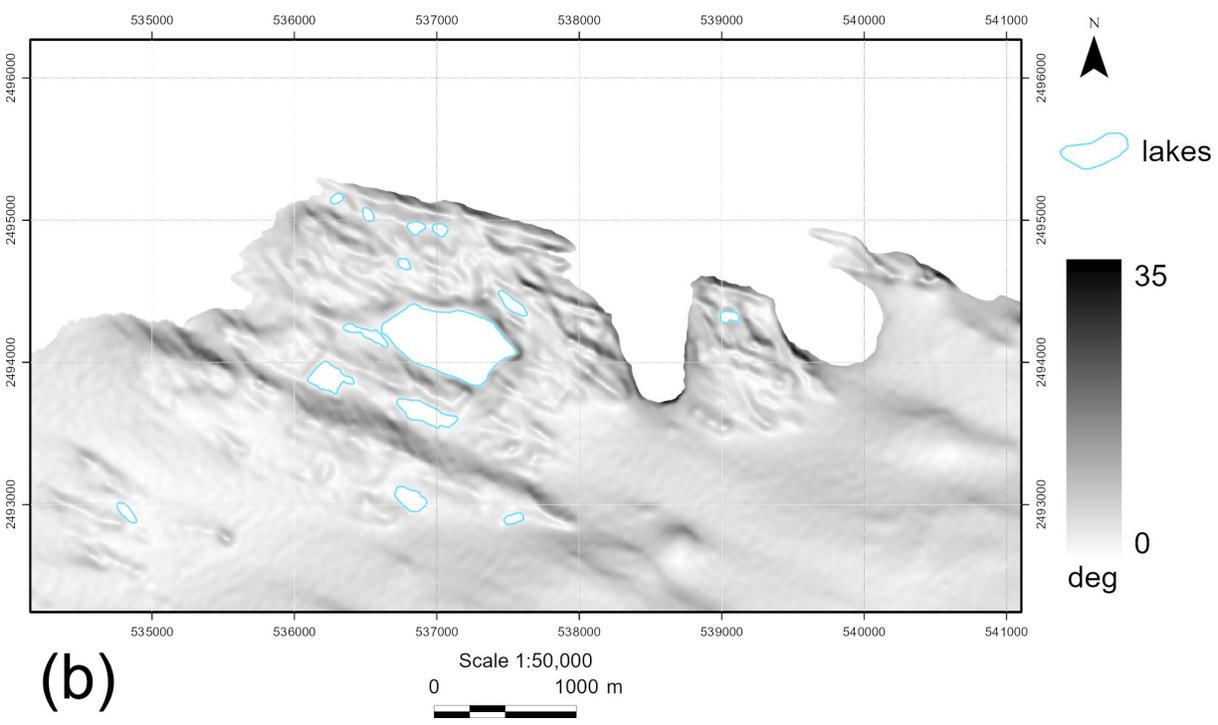

(b)

**Fig. 4** Molodezhny Oasis, Thala Hills: (a) Elevation. (b) Slope.

*(Continued)*

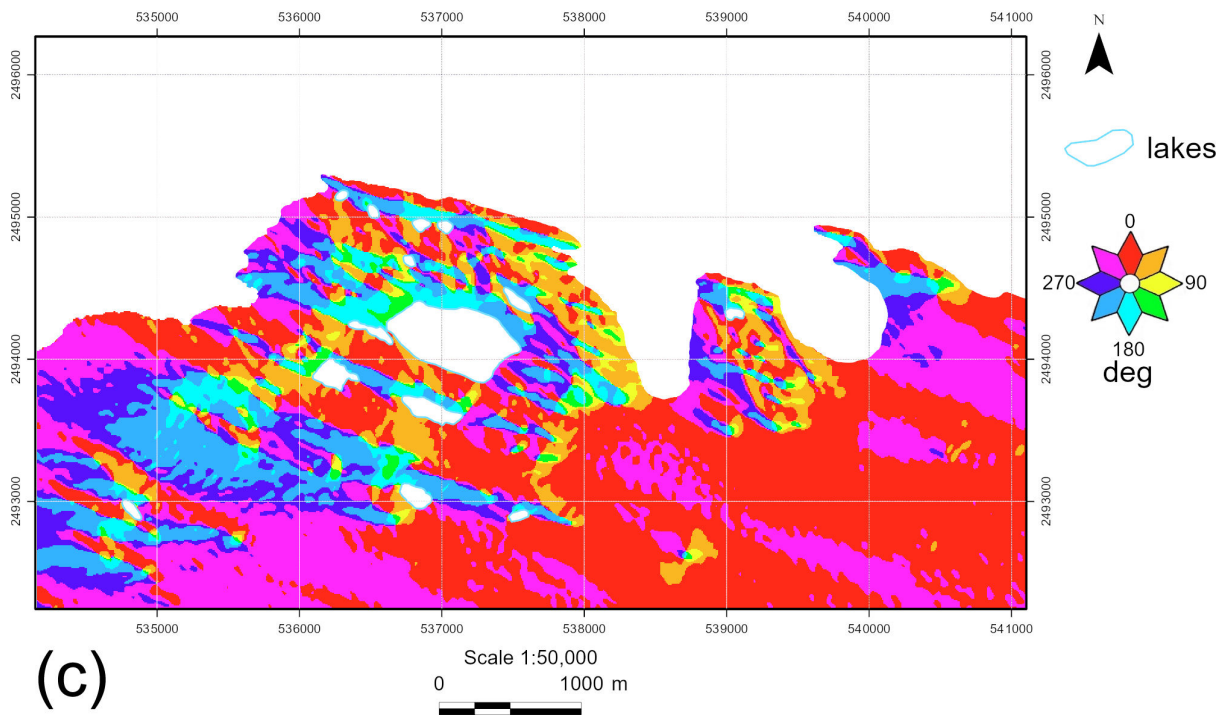

(c)

Scale 1:50,000

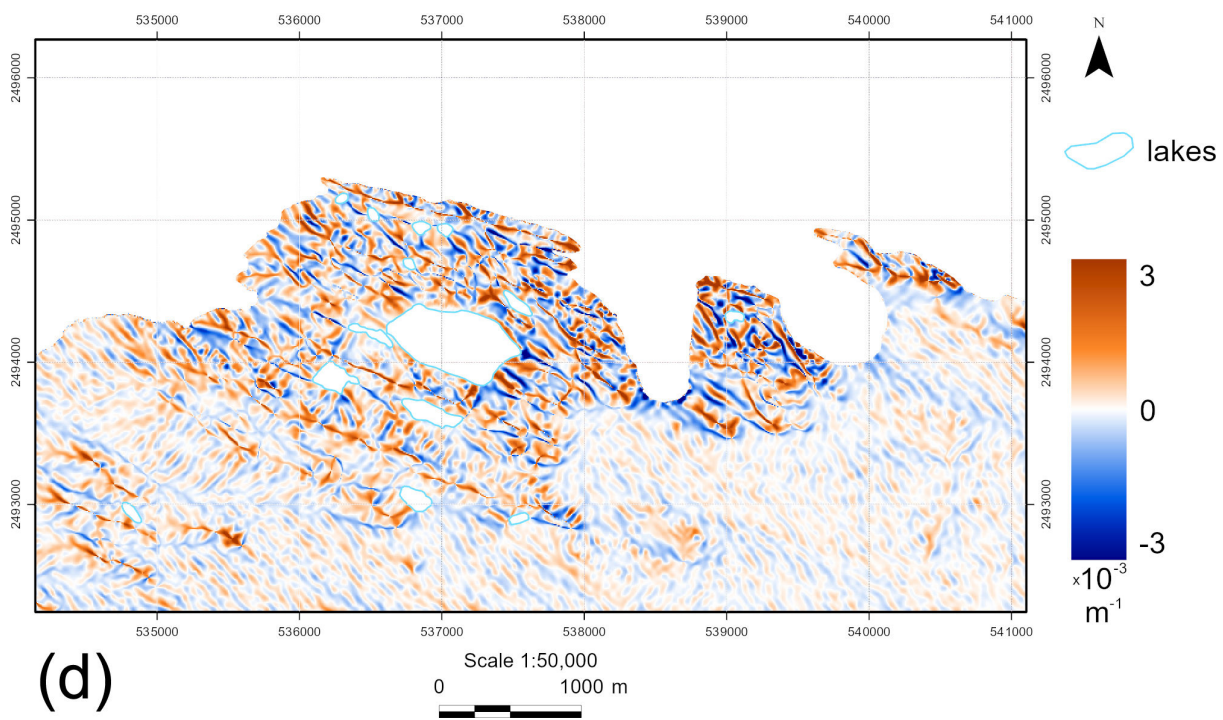

(d)

Scale 1:50,000

**Fig. 4, cont'd** Molodezhny Oasis, Thala Hills: (c) Aspect. (d) Horizontal curvature.

*(Continued)*

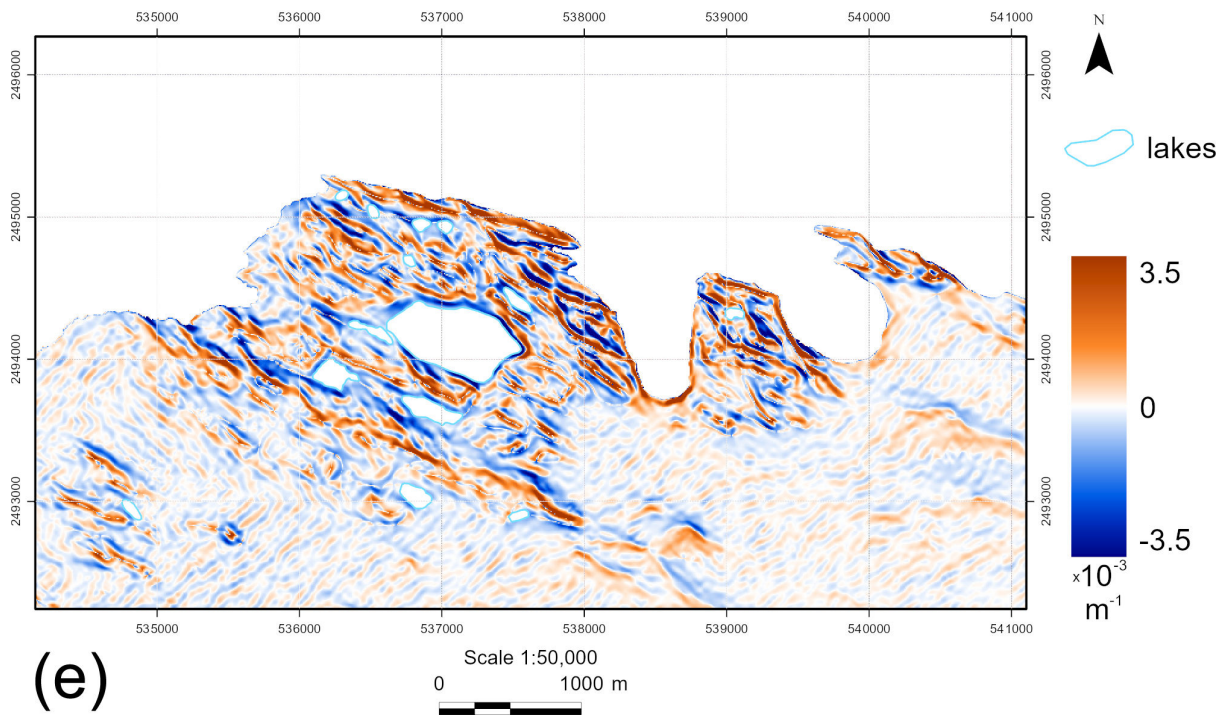

(e)

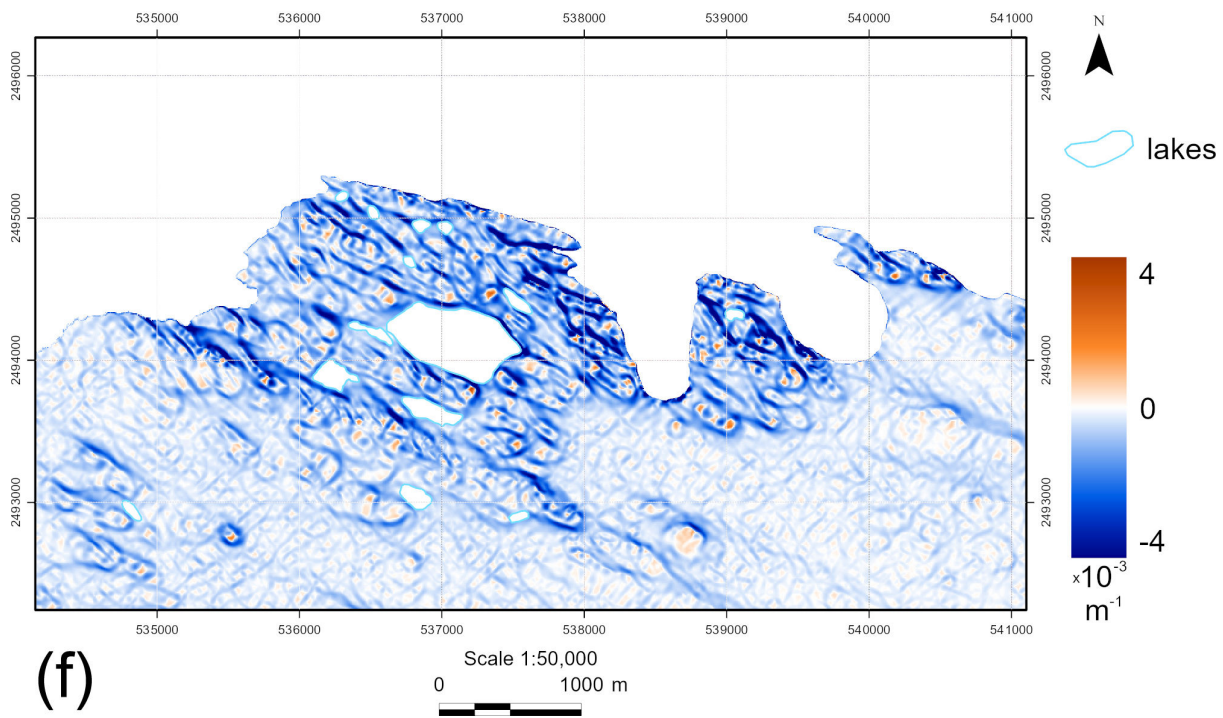

(f)

**Fig. 4, cont'd** Molodezhny Oasis, Thala Hills: (e) Vertical curvature. (f) Minimal curvature.

*(Continued)*

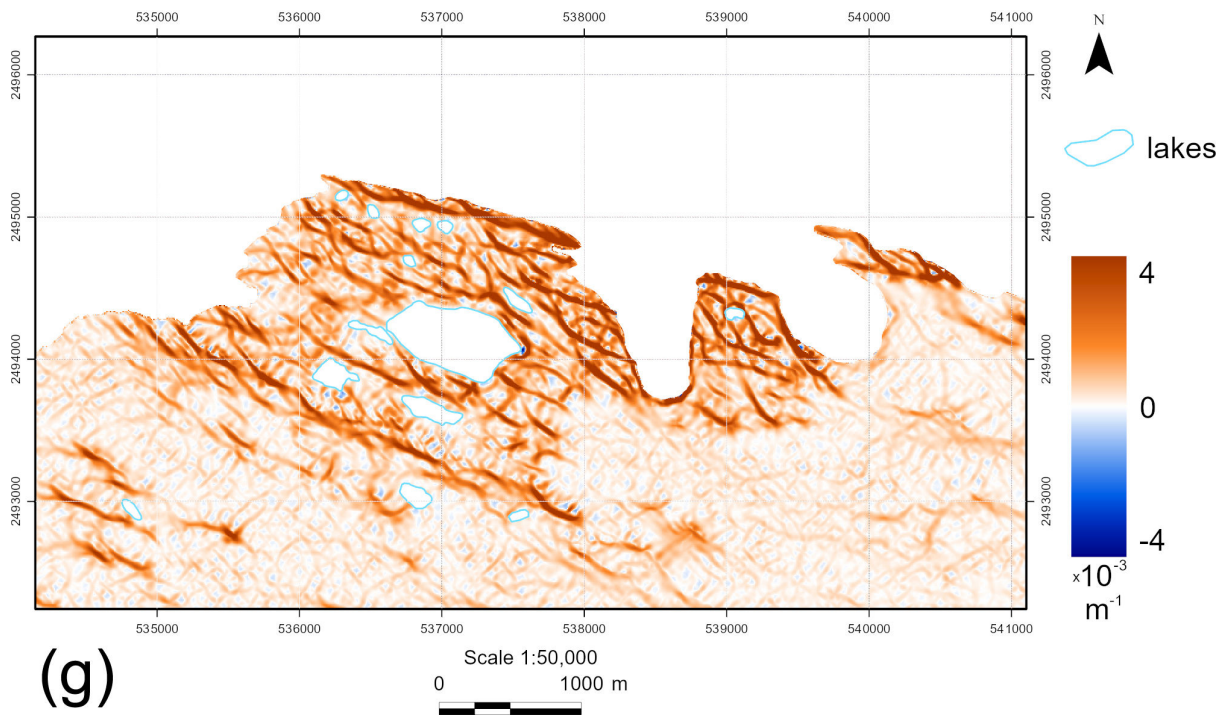

(g)

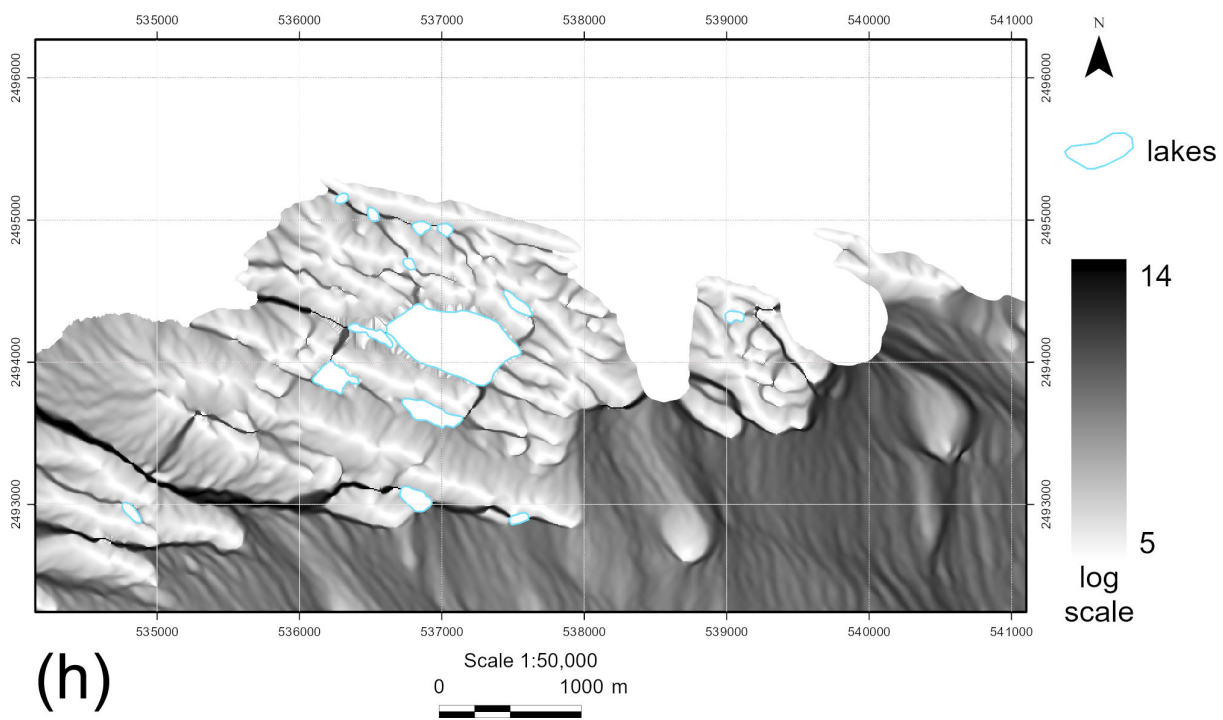

(h)

**Fig. 4, cont'd** Molodezhny Oasis, Thala Hills: (g) Maximal curvature. (h) Catchment area.

*(Continued)*

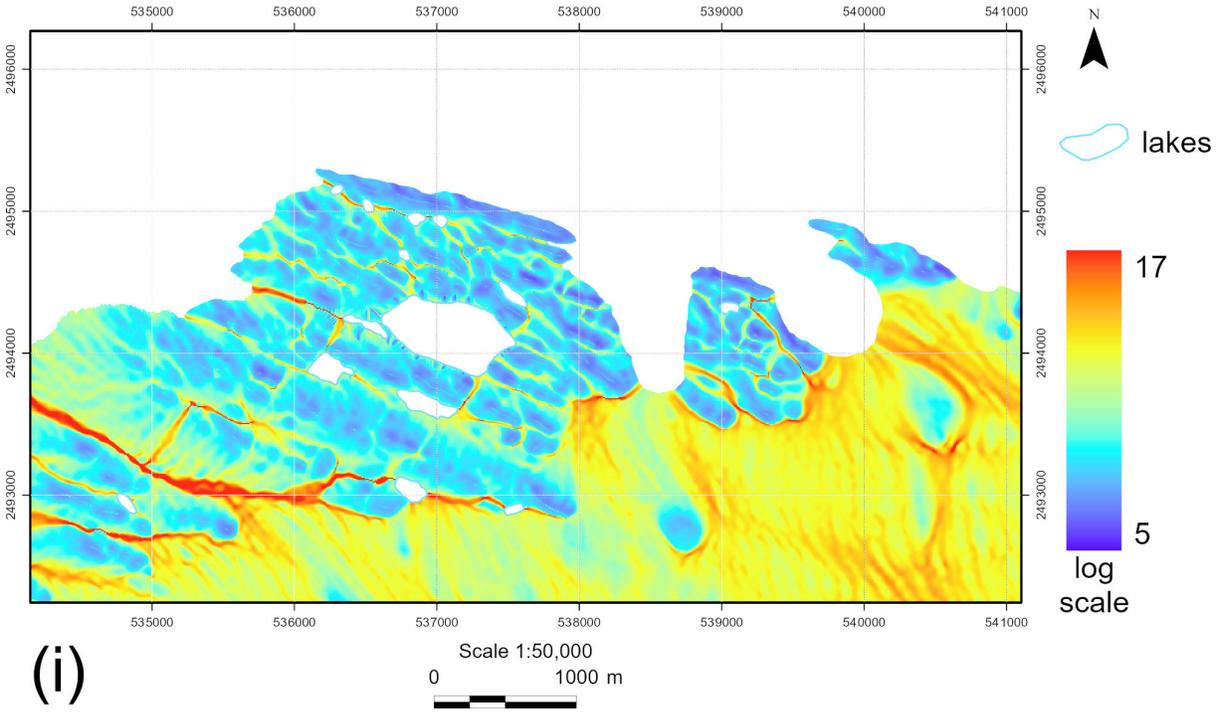

**(i)**

Scale 1:50,000

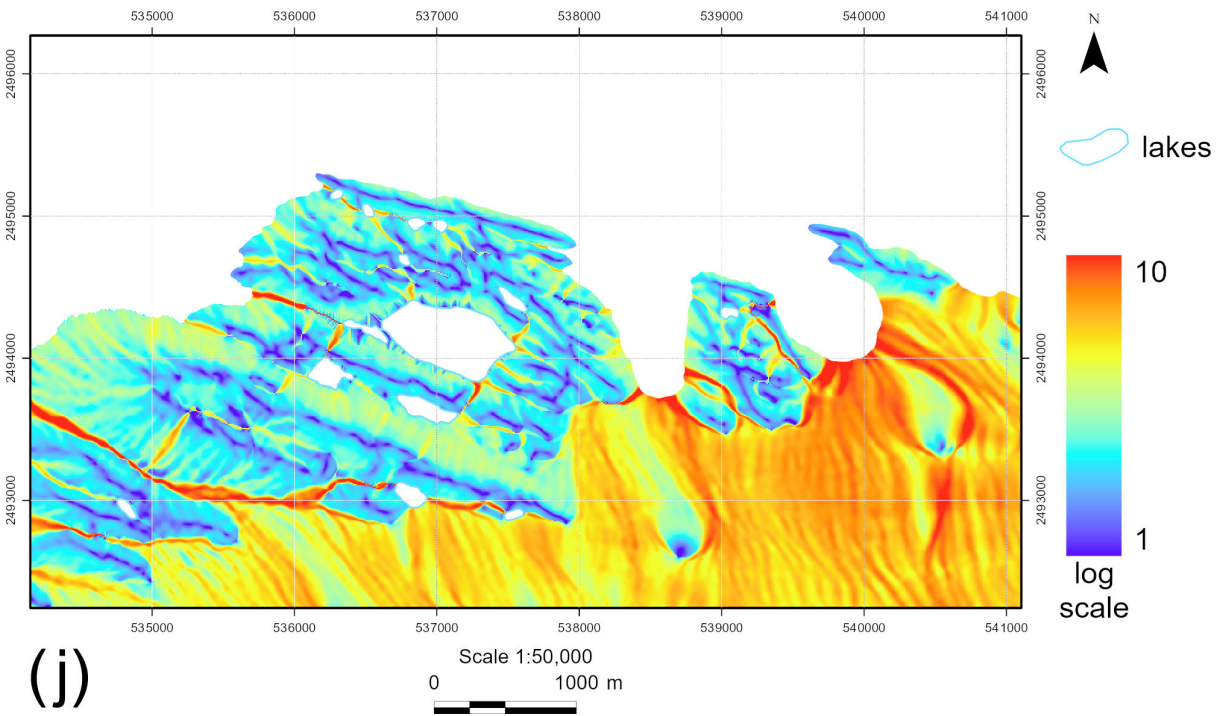

**(j)**

Scale 1:50,000

**Fig. 4, cont'd** Molodezhny Oasis, Thala Hills: (i) Topographic wetness index. (j) Stream power index.

*(Continued)*

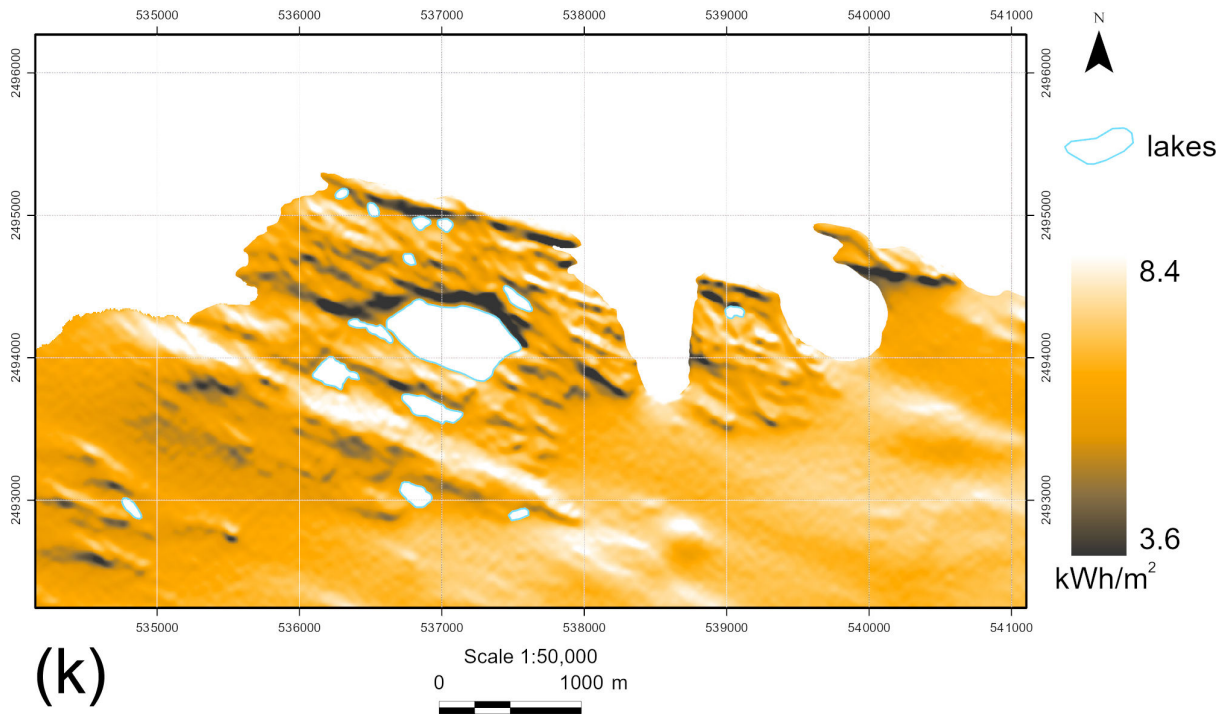

(k)

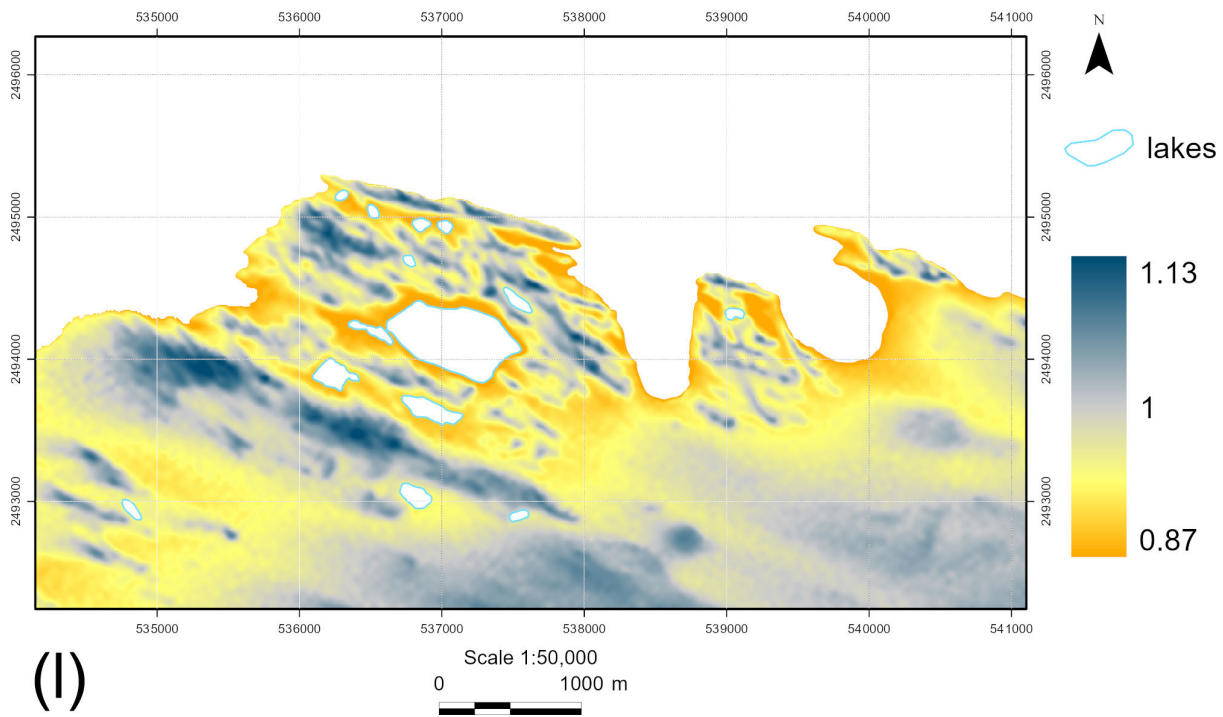

(l)

**Fig. 4, cont'd** Molodezhny Oasis, Thala Hills: (k) Total insolation. (l) Wind exposition index.

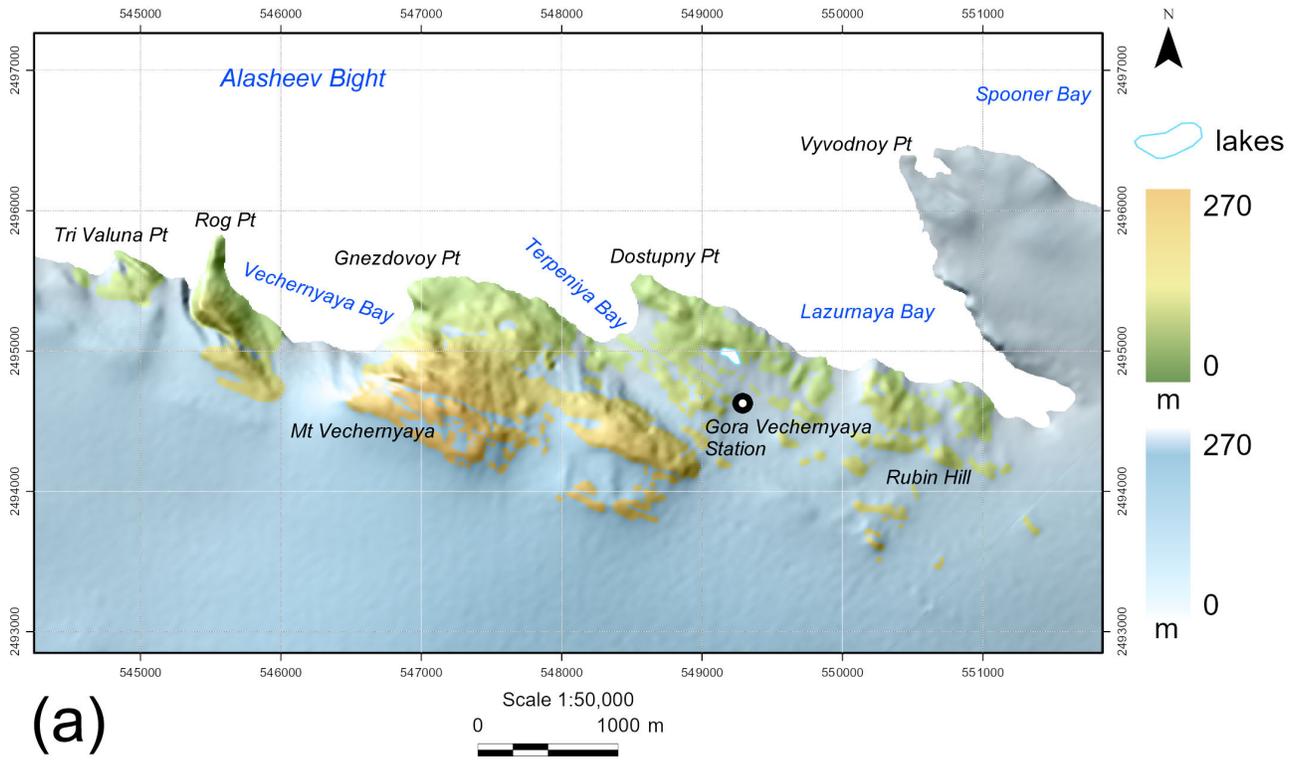

(a)

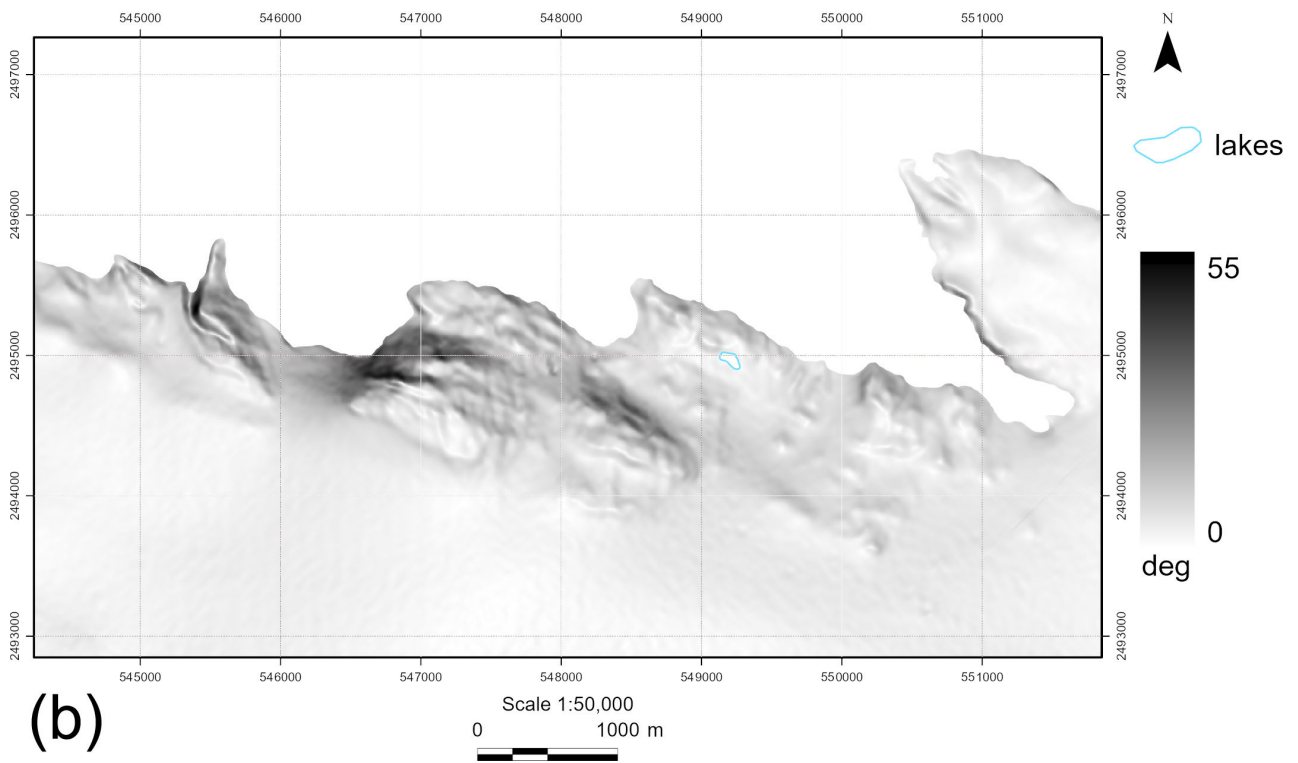

(b)

**Fig. 5, cont'd** Vecherny Oasis, Thala Hills: (a) Elevation. (b) Slope.

*(Continued)*

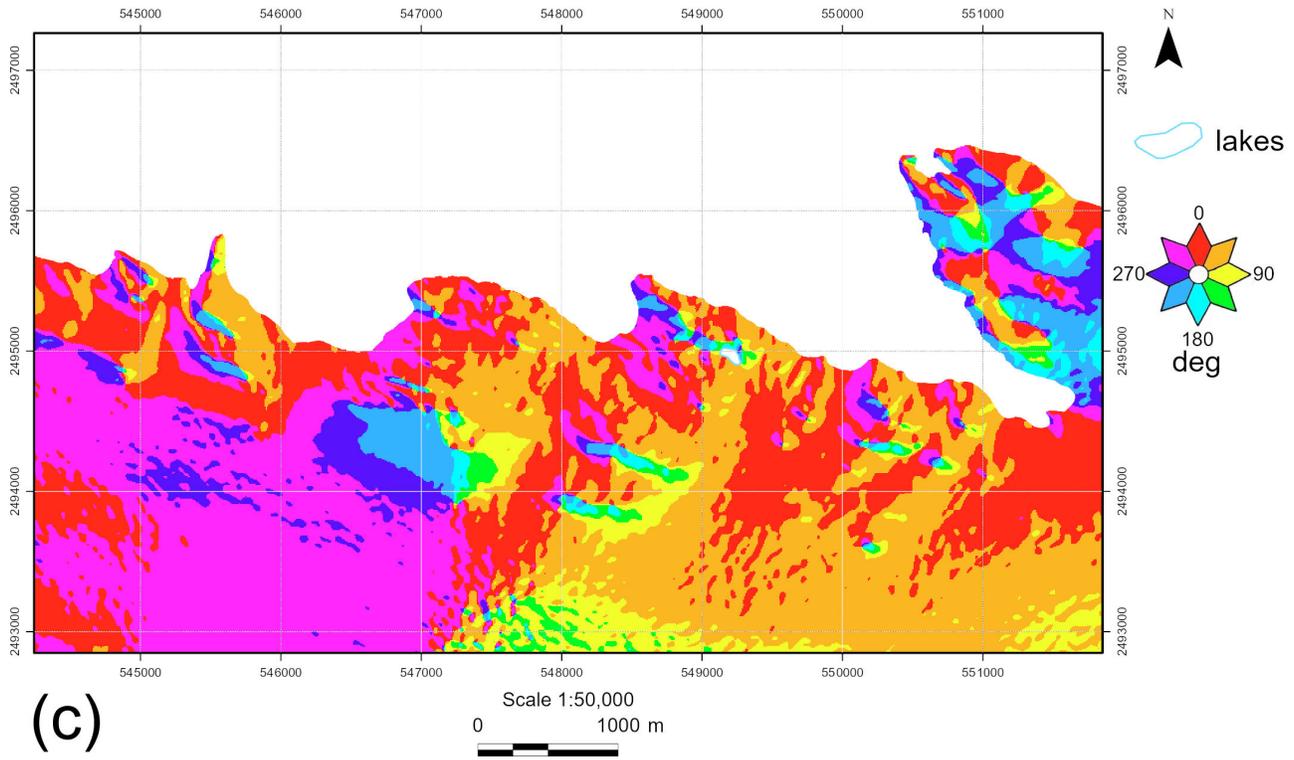

(c)

Scale 1:50,000

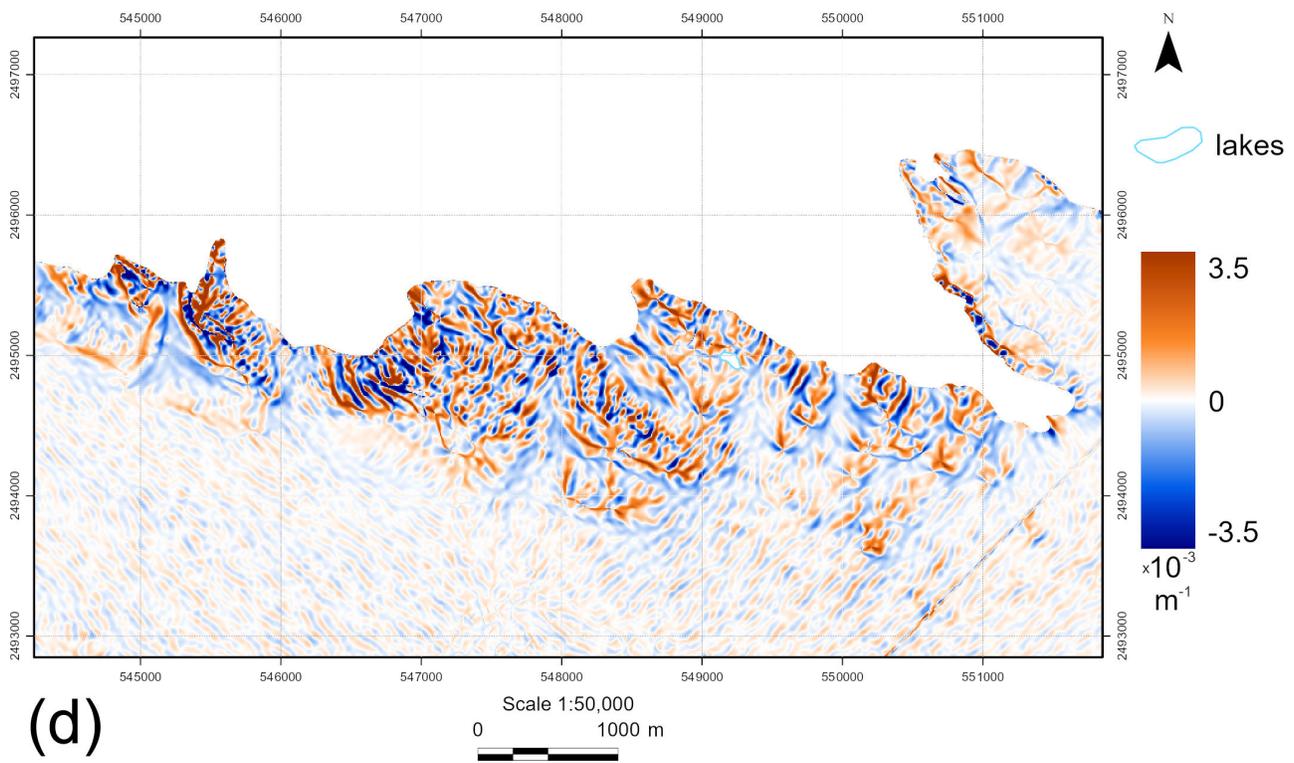

(d)

Scale 1:50,000

**Fig. 5, cont'd** Vecherny Oasis, Thala Hills: (c) Aspect. (d) Horizontal curvature.

*(Continued)*

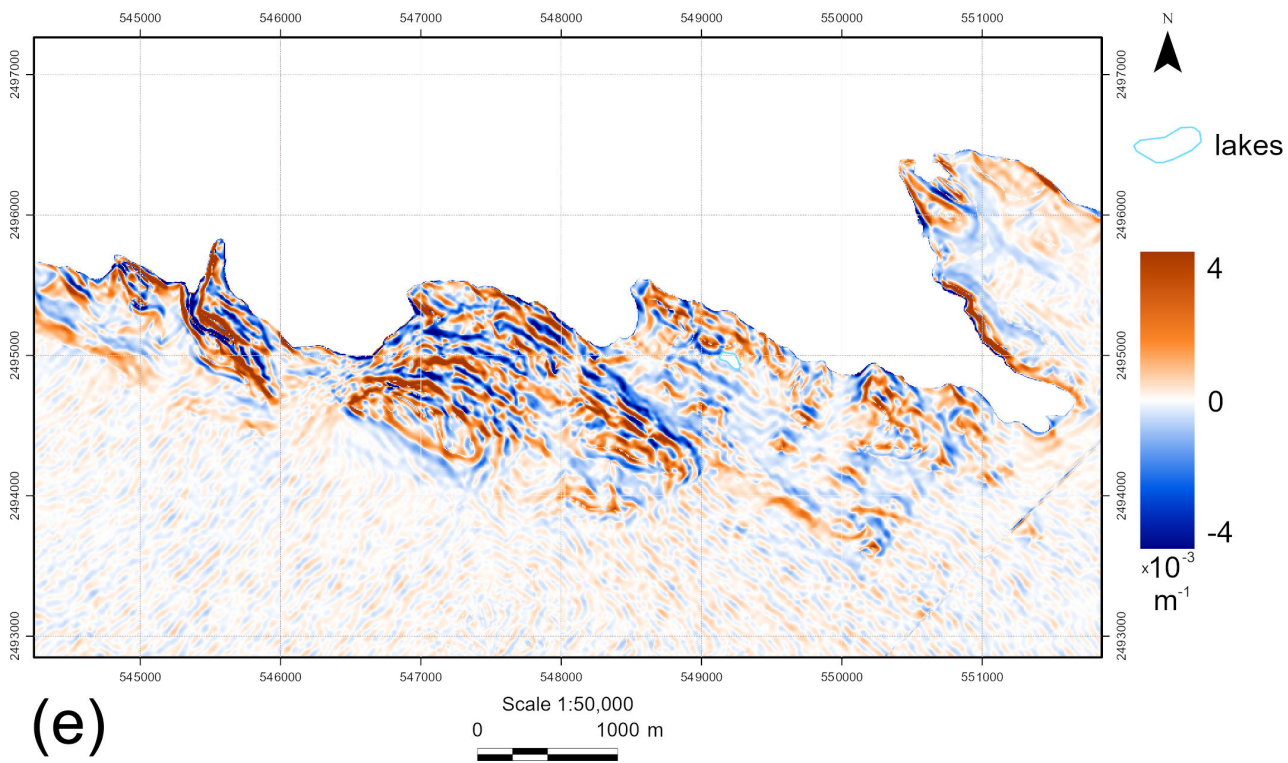

(e)

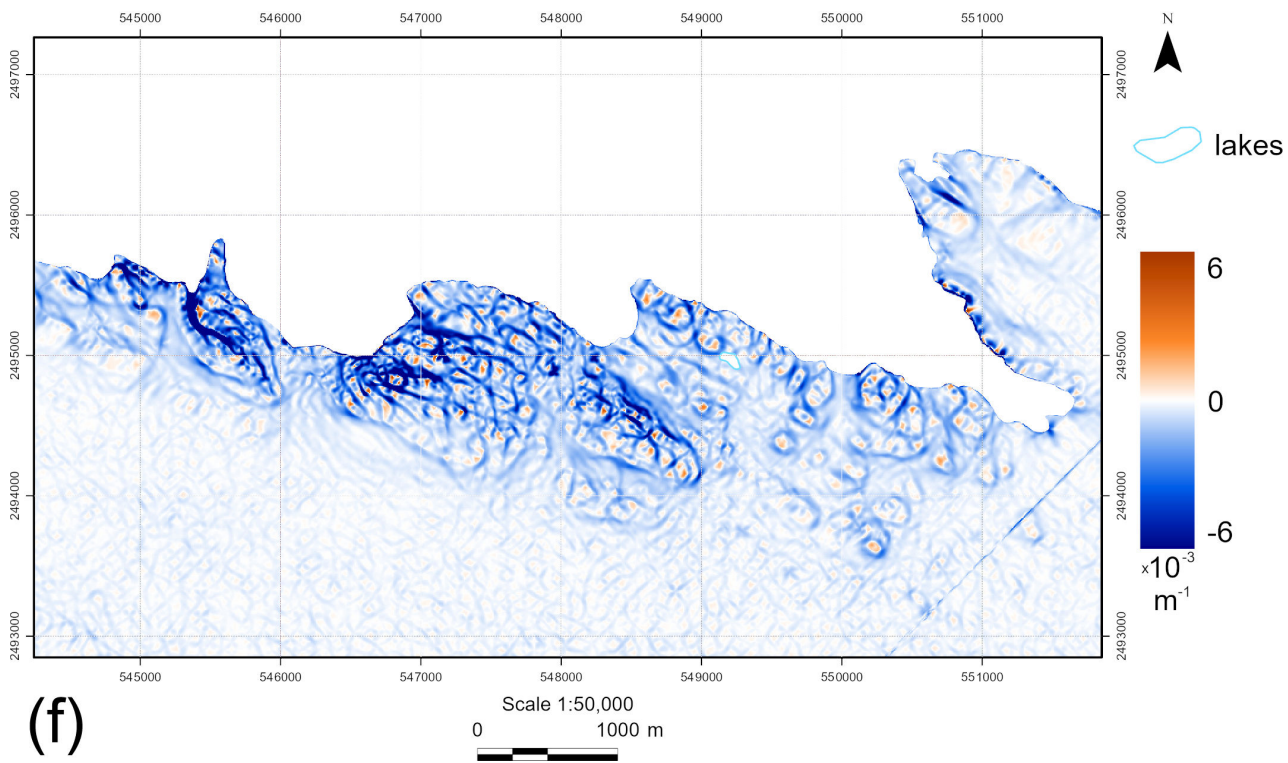

(f)

**Fig. 5, cont'd** Vecherny Oasis, Thala Hills: (e) Vertical curvature. (f) Minimal curvature.

*(Continued)*

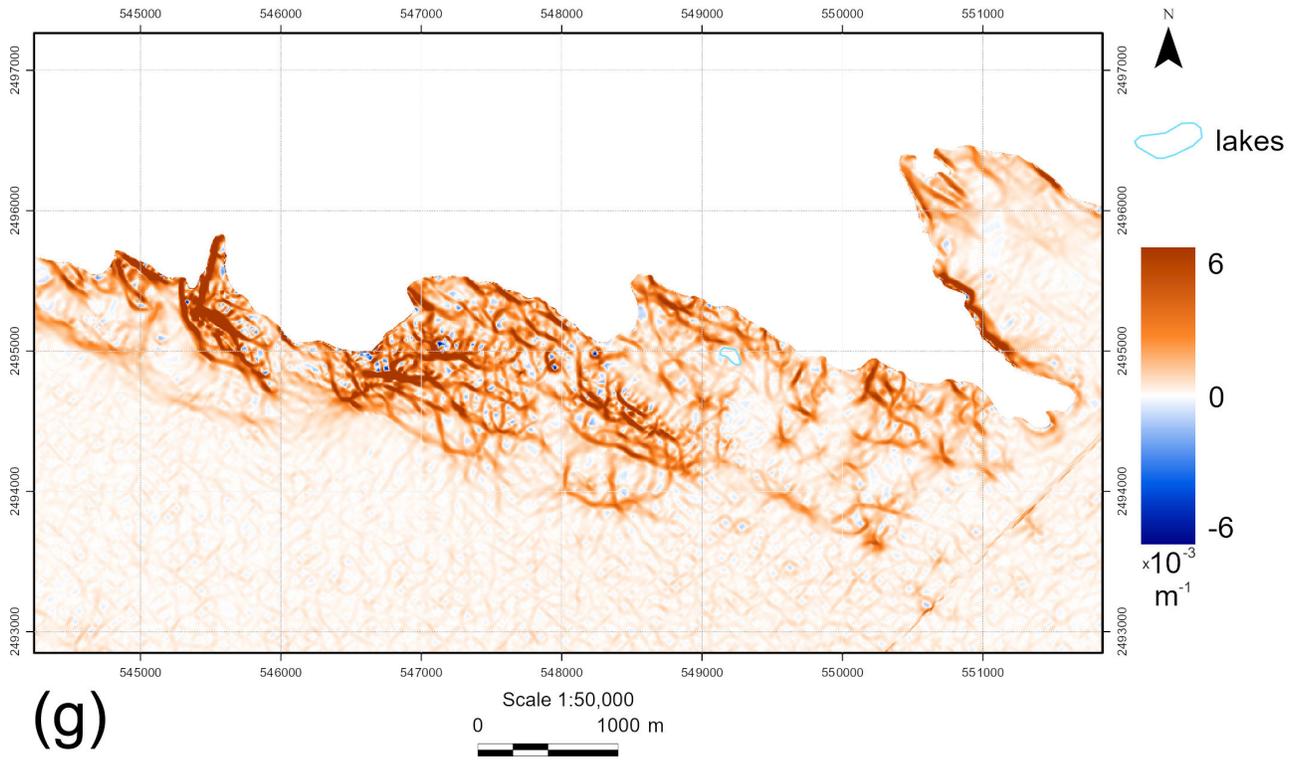

(g)

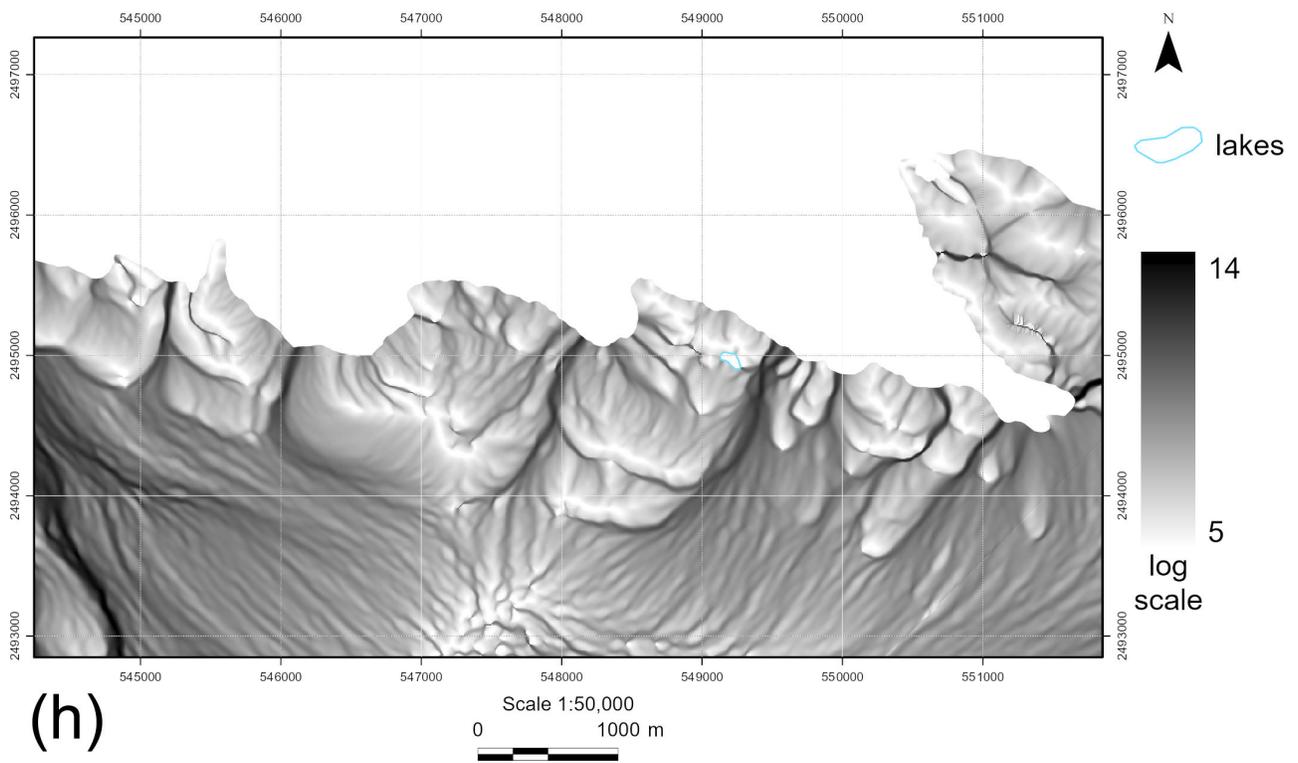

(h)

**Fig. 5, cont'd** Vecherny Oasis, Thala Hills: (g) Maximal curvature. (h) Catchment area.

*(Continued)*

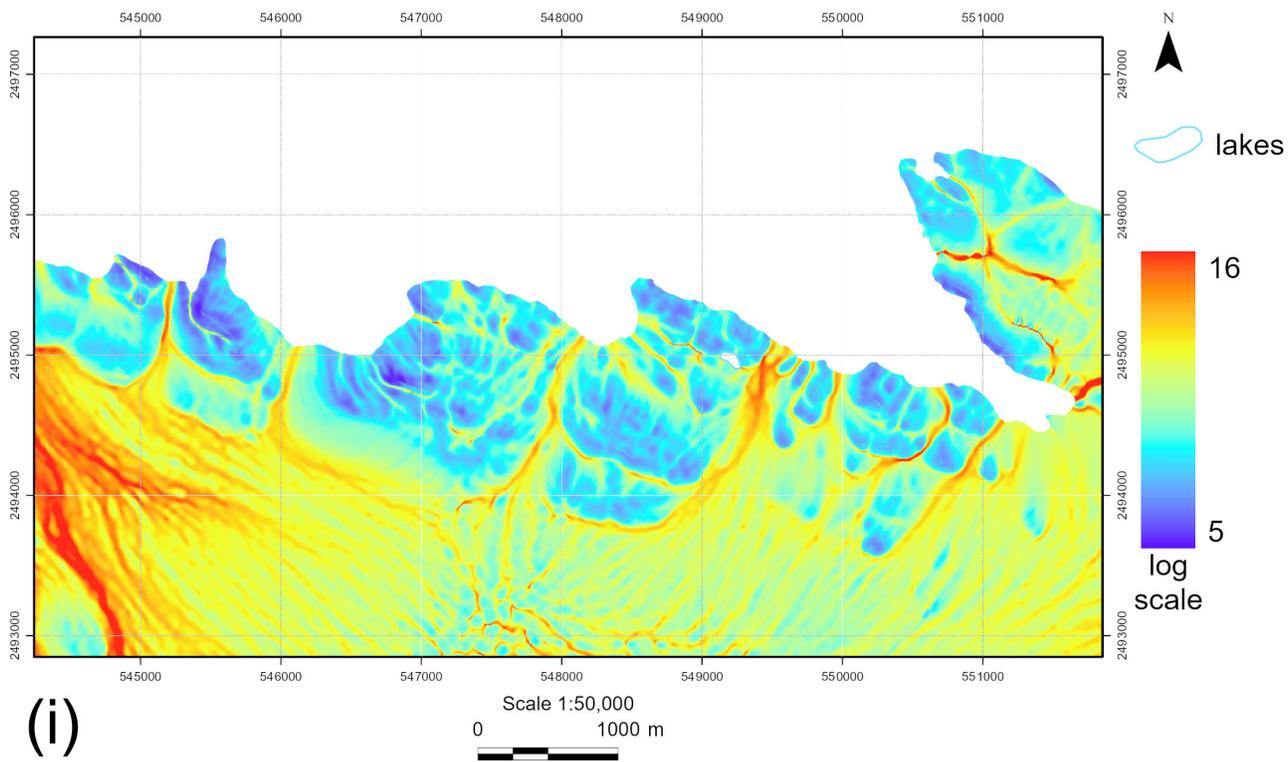

(i)

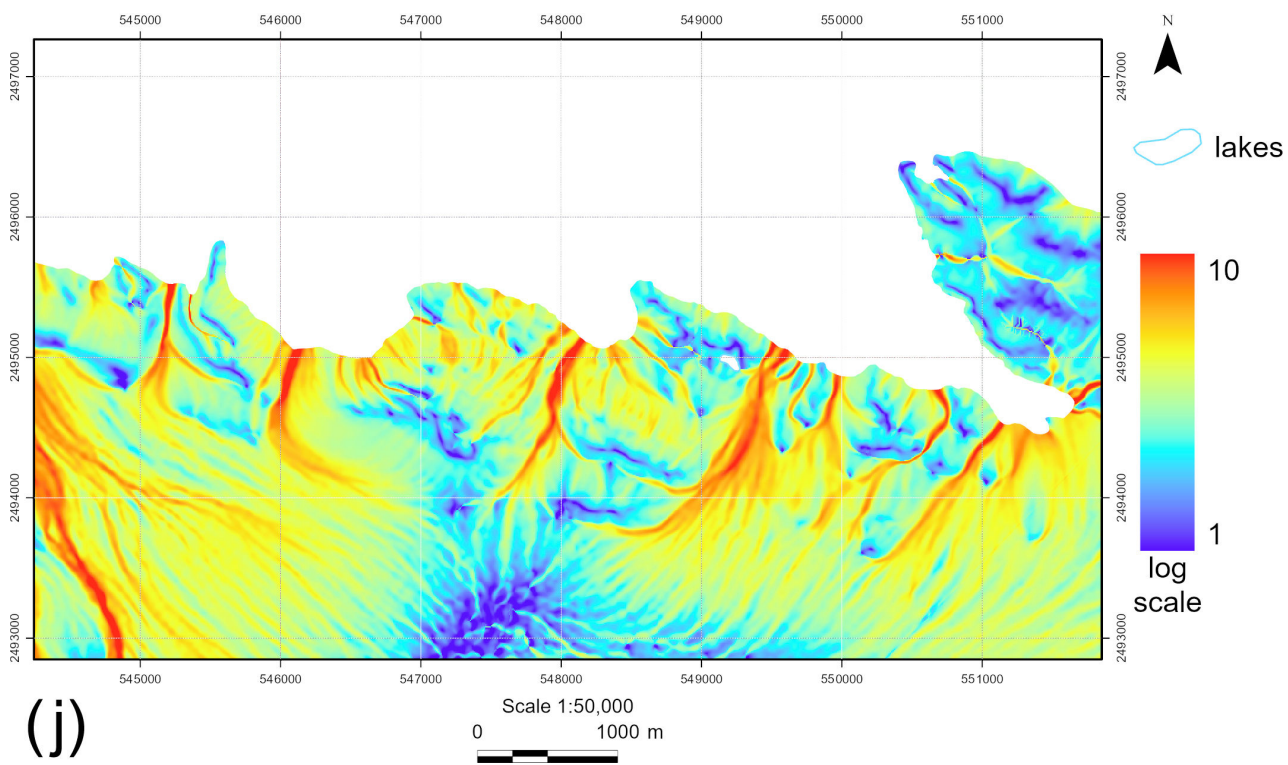

(j)

**Fig. 5, cont'd** Vecherny Oasis, Thala Hills: (i) Topographic wetness index. ( j) Stream power index.

*(Continued)*

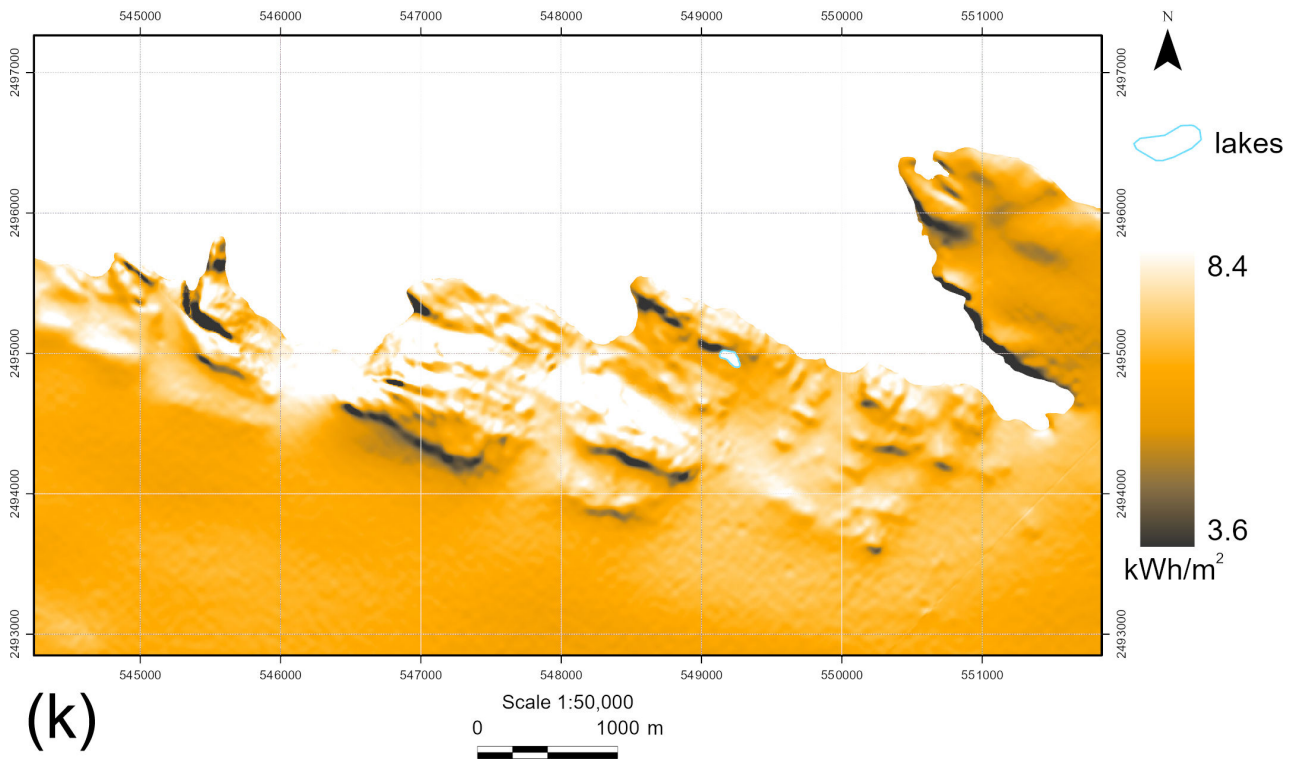

(k)

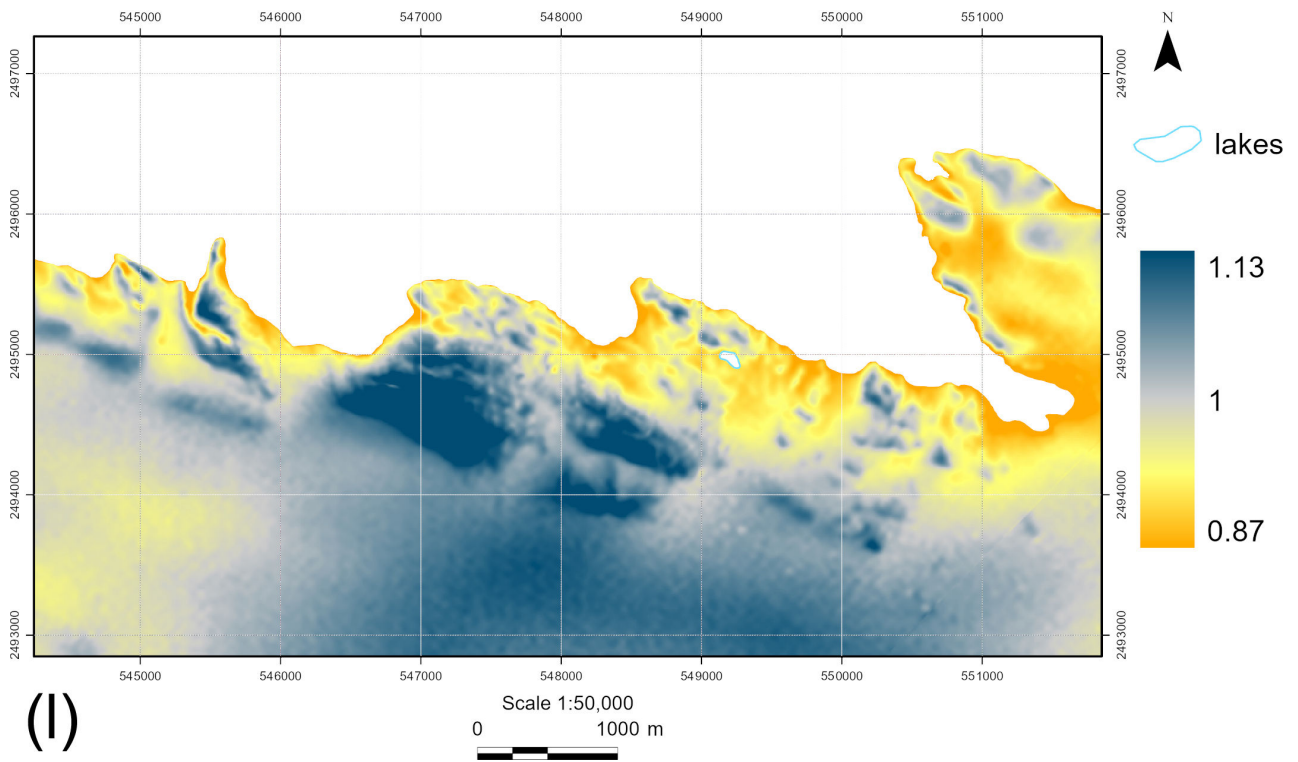

(l)

**Fig. 5, cont'd** Vecherny Oasis, Thala Hills: (k) Total insolation. (l) Wind exposition index.

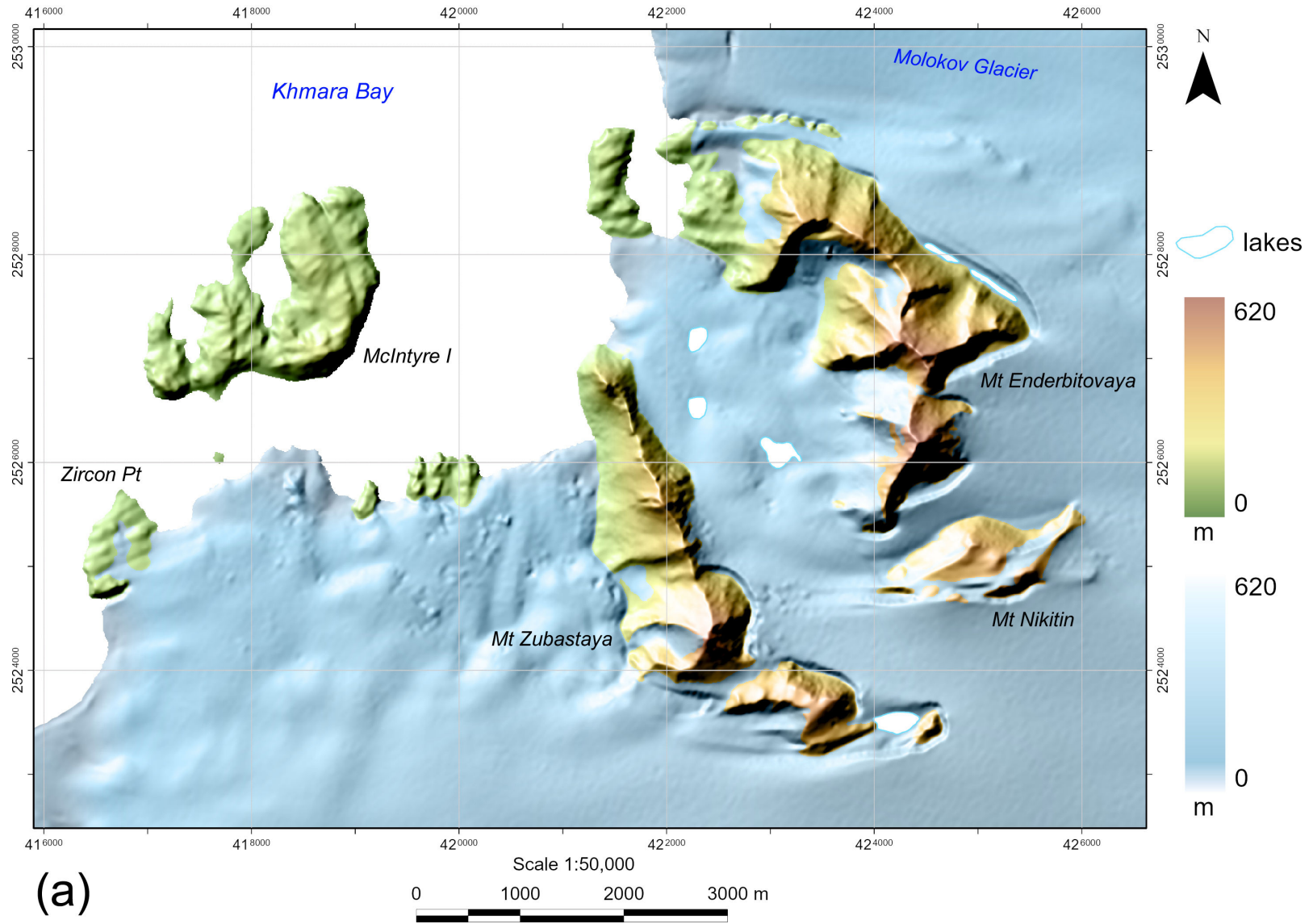

**Fig. 6** Fyfe Hills: (a) Elevation.

*(Continued)*

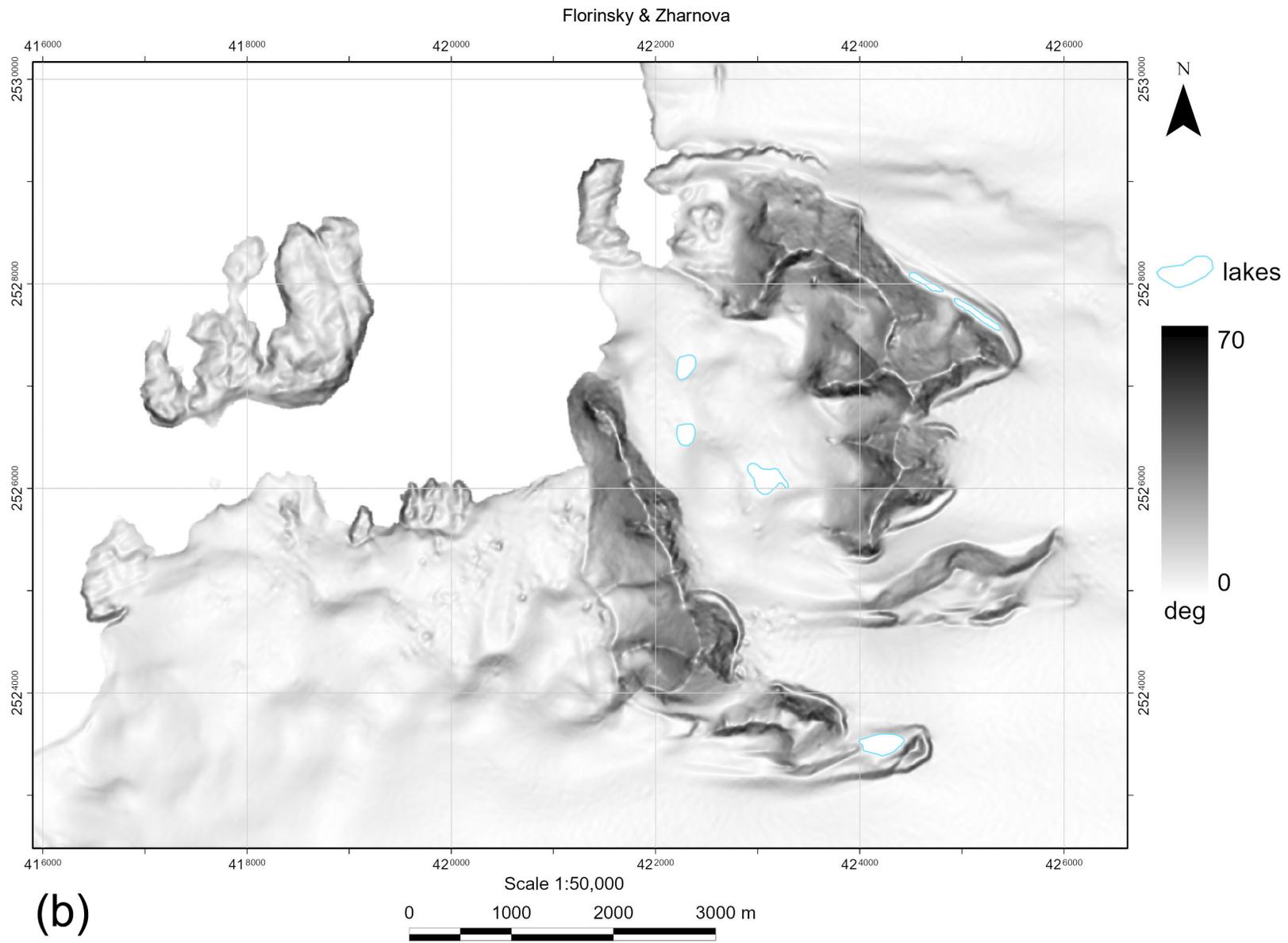

Scale 1:50,000

(b)

**Fig. 6, cont'd** Fyfe Hills: (b) Slope.

*(Continued)*

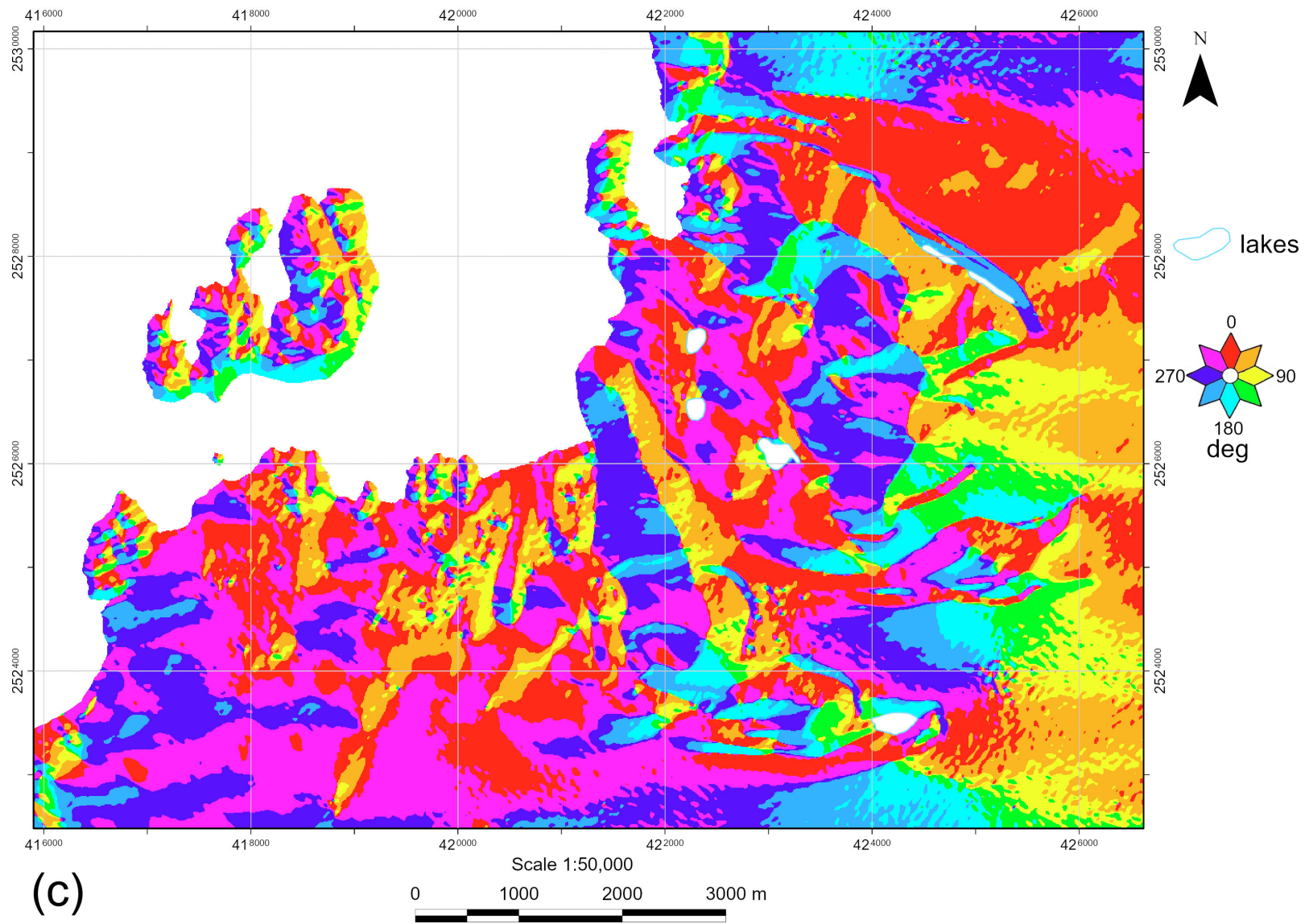

Scale 1:50,000

0       1000      2000      3000 m

(c)

**Fig. 6, cont'd** Fyfe Hills: (c) Aspect.

*(Continued)*

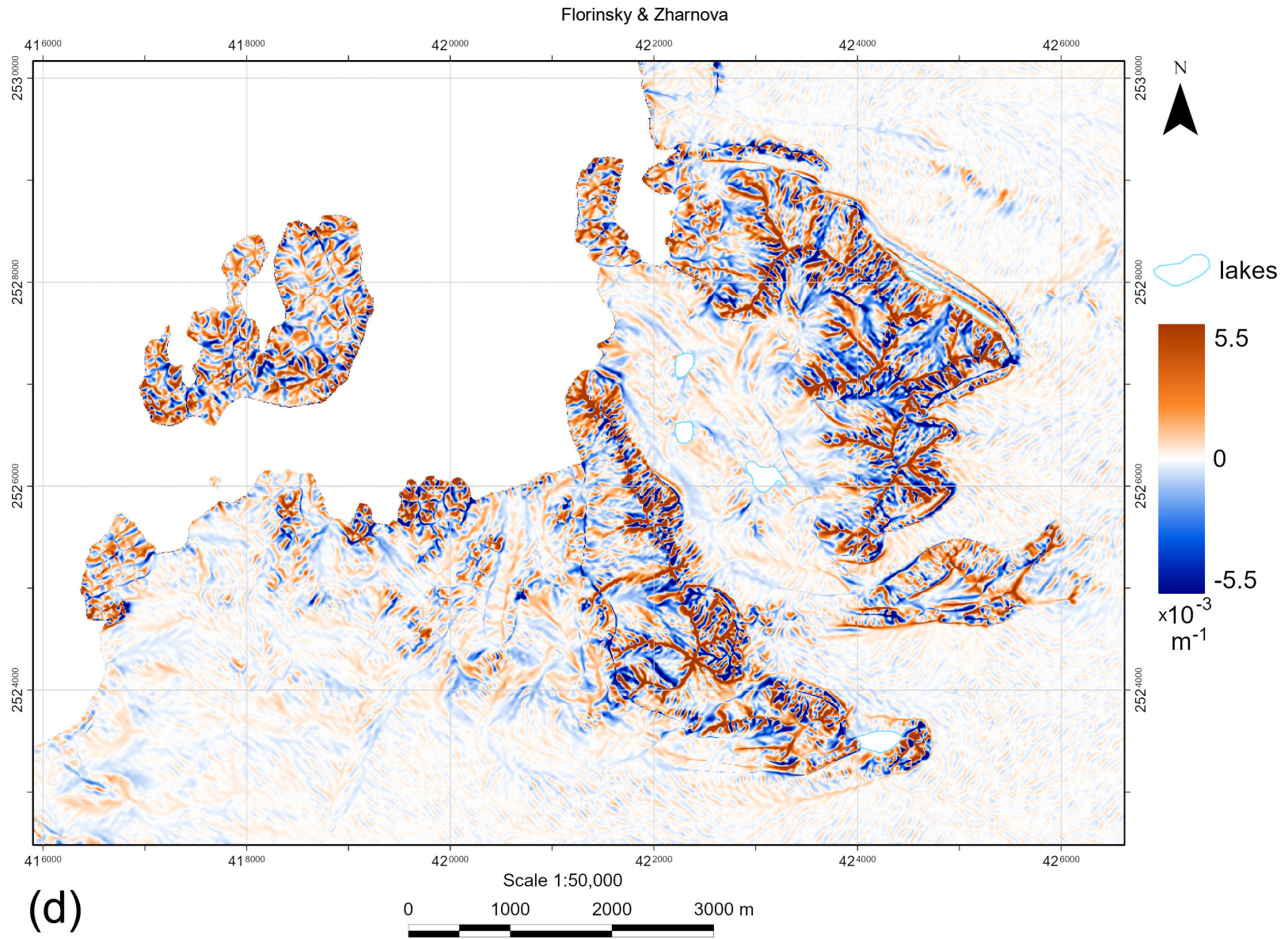

(d)

Scale 1:50,000

**Fig. 6, cont'd** Fyfe Hills: (d) Horizontal curvature.

*(Continued)*

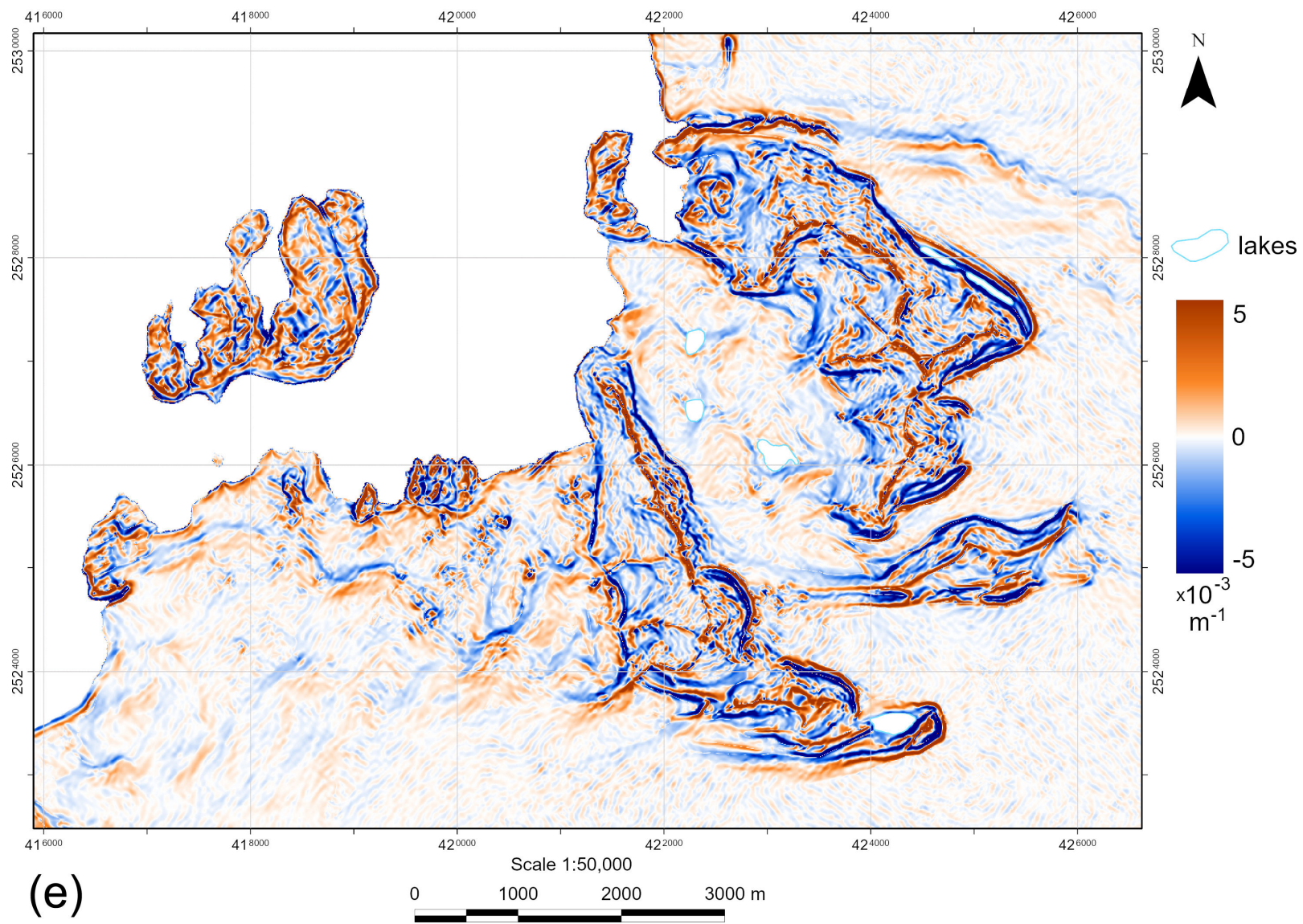

Scale 1:50,000

**Fig. 6, cont'd** Fyfe Hills: (e) Vertical curvature.

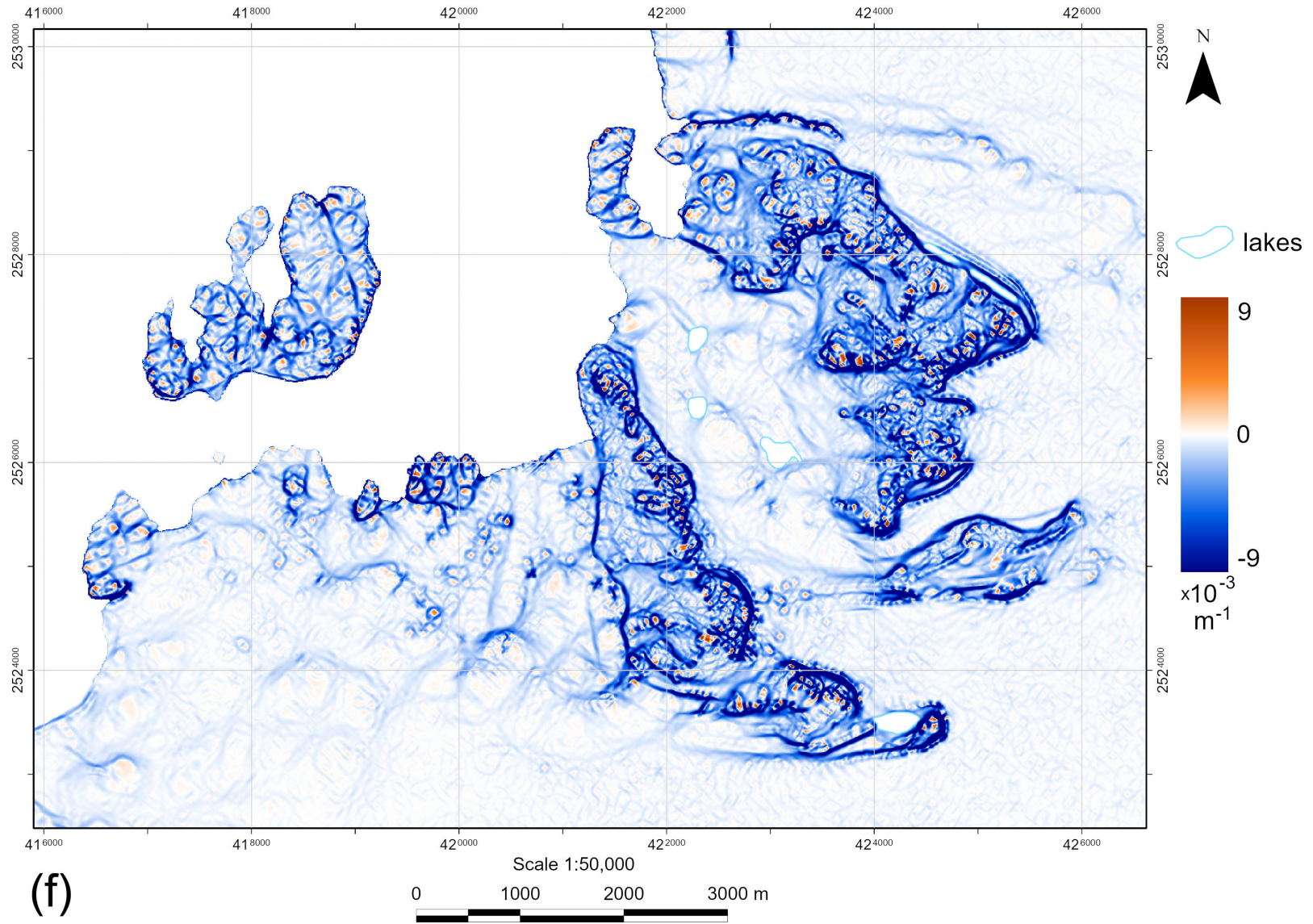

(f)

Scale 1:50,000

Fig. 6, cont'd Fyfe Hills: (f) Minimal curvature.

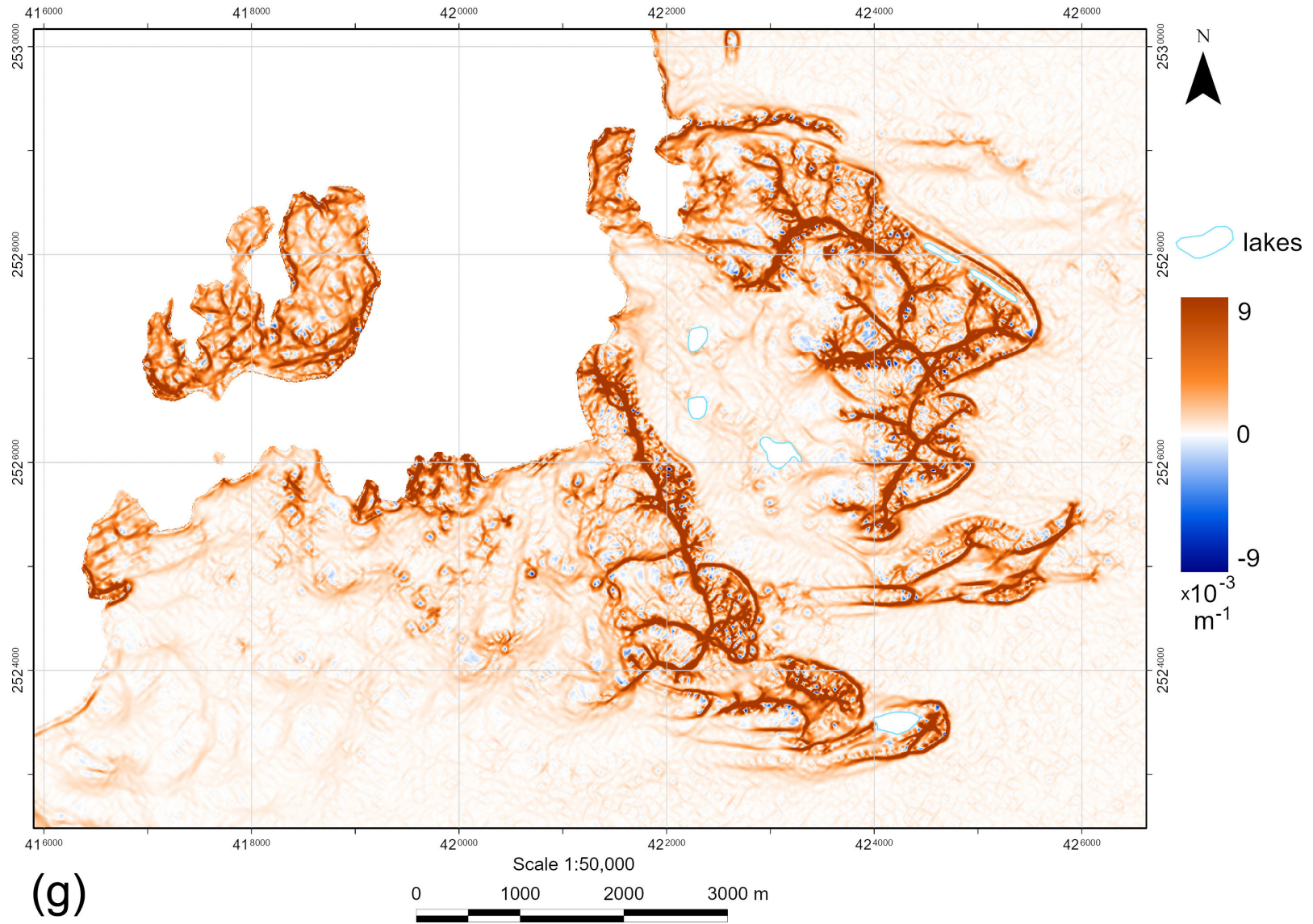

**Fig. 6, cont'd** Fyfe Hills: (g) Maximal curvature.

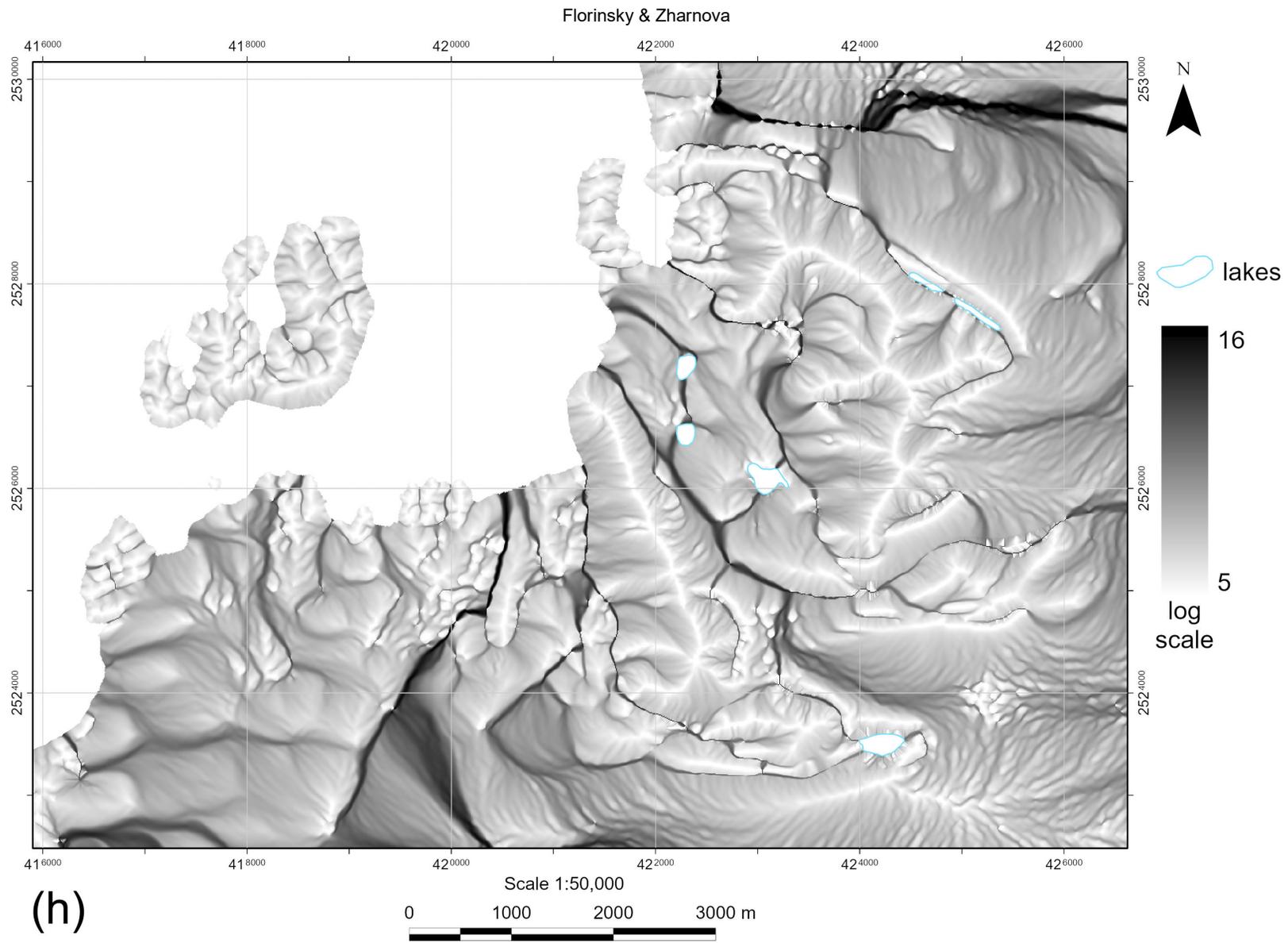

(h)

Scale 1:50,000

Fig. 6, cont'd Fyfe Hills: (h) Catchment area.

*(Continued)*

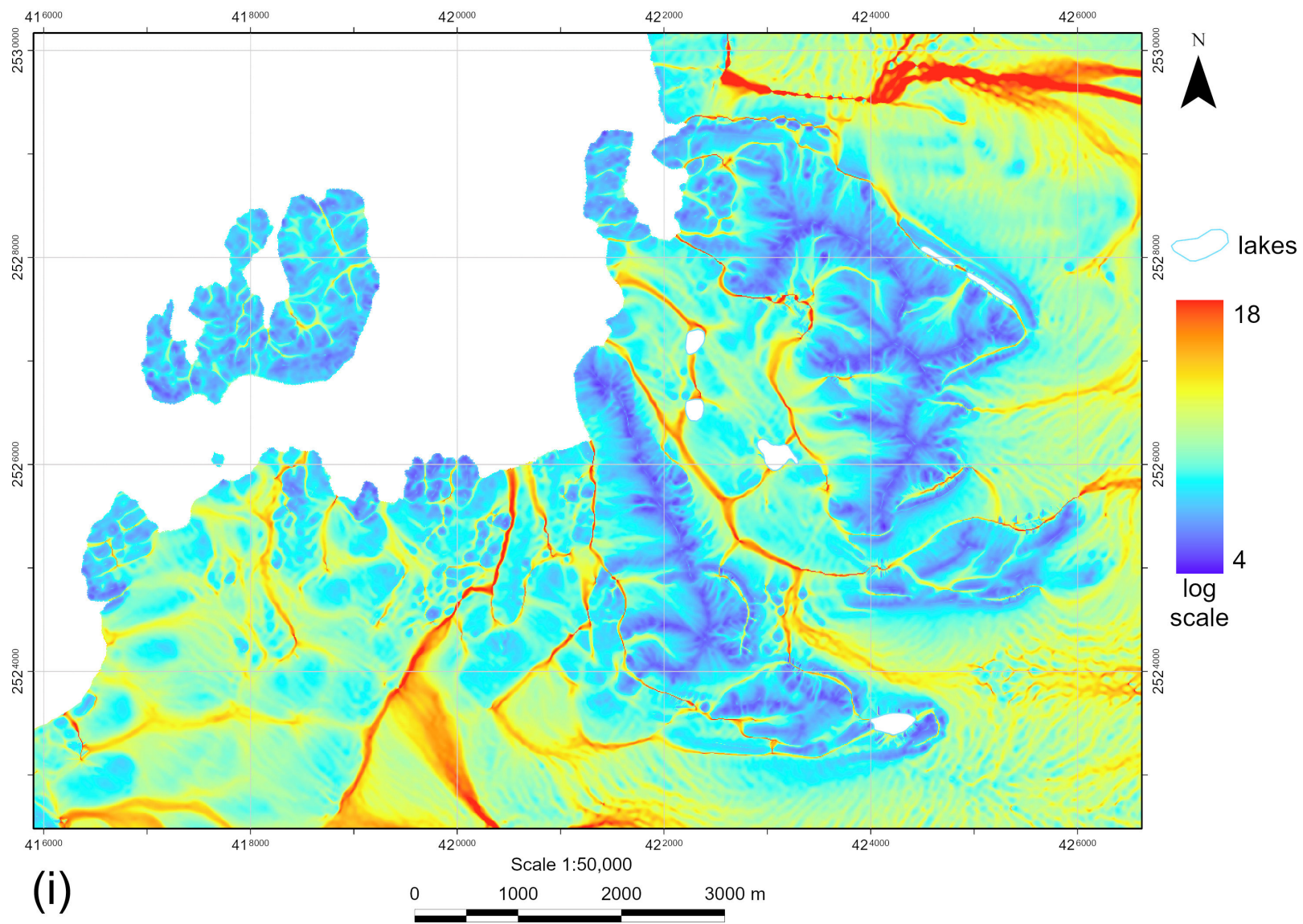

(i)

Scale 1:50,000

0    1000    2000    3000 m

**Fig. 6, cont'd** Fyfe Hills: (i) Topographic wetness index.

*(Continued)*

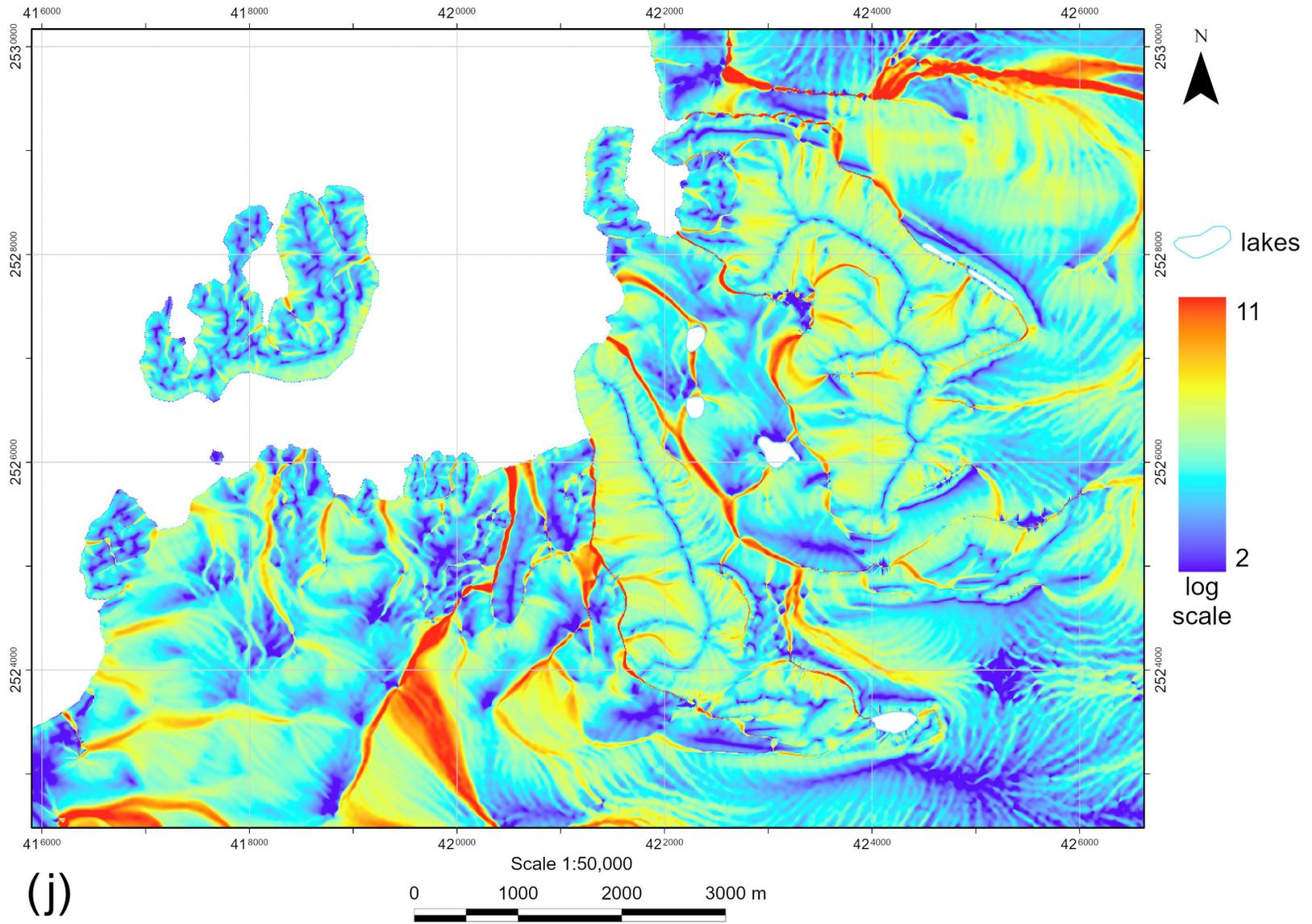

Scale 1:50,000

(j)

0    1000    2000    3000 m

**Fig. 6, cont'd** Fyfe Hills: ( j) Stream power index.

*(Continued)*

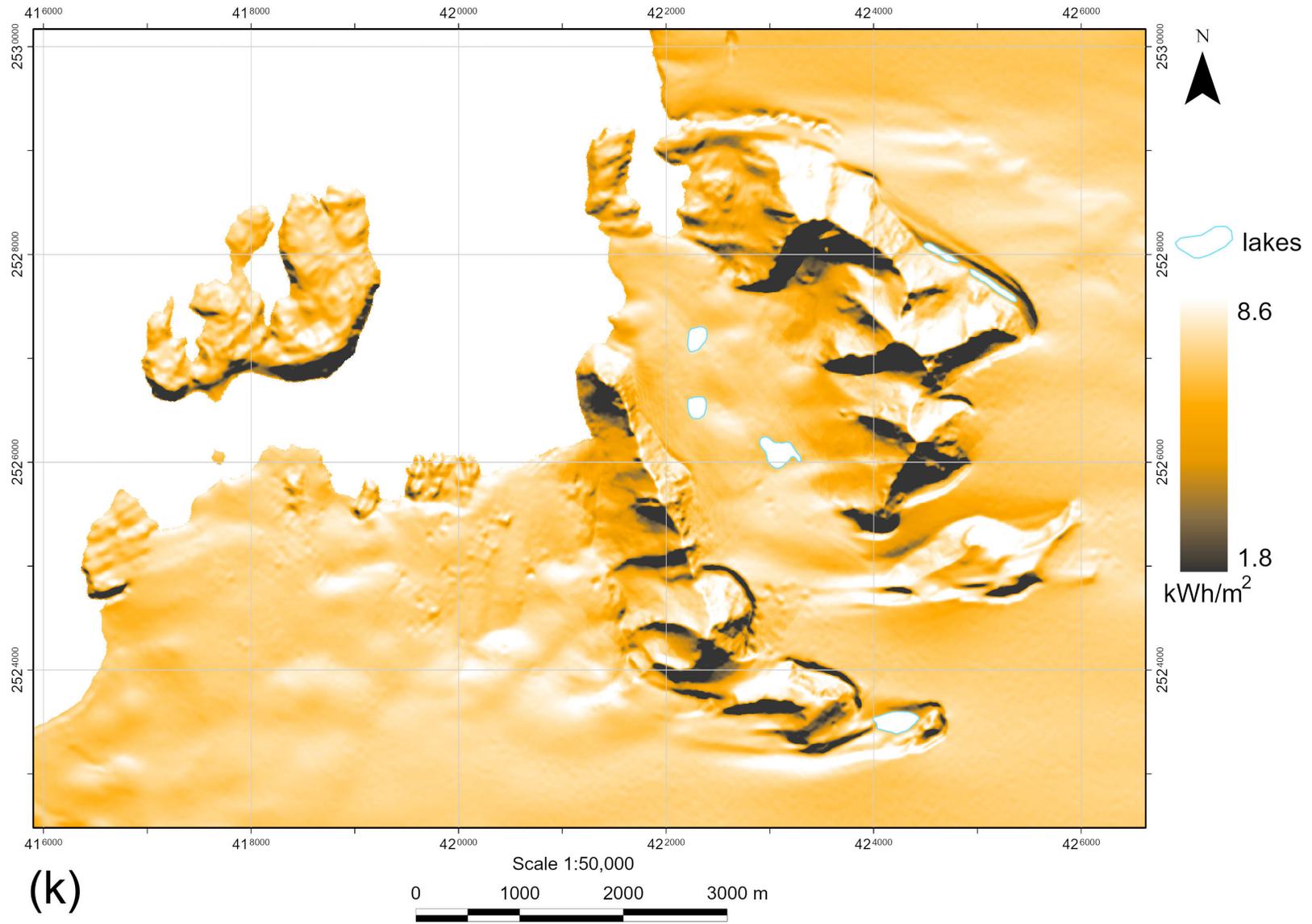

Scale 1:50,000

0    1000    2000    3000 m

**Fig. 6, cont'd** Fyfe Hills: (k) Total insolation.

(k)

*(Continued)*

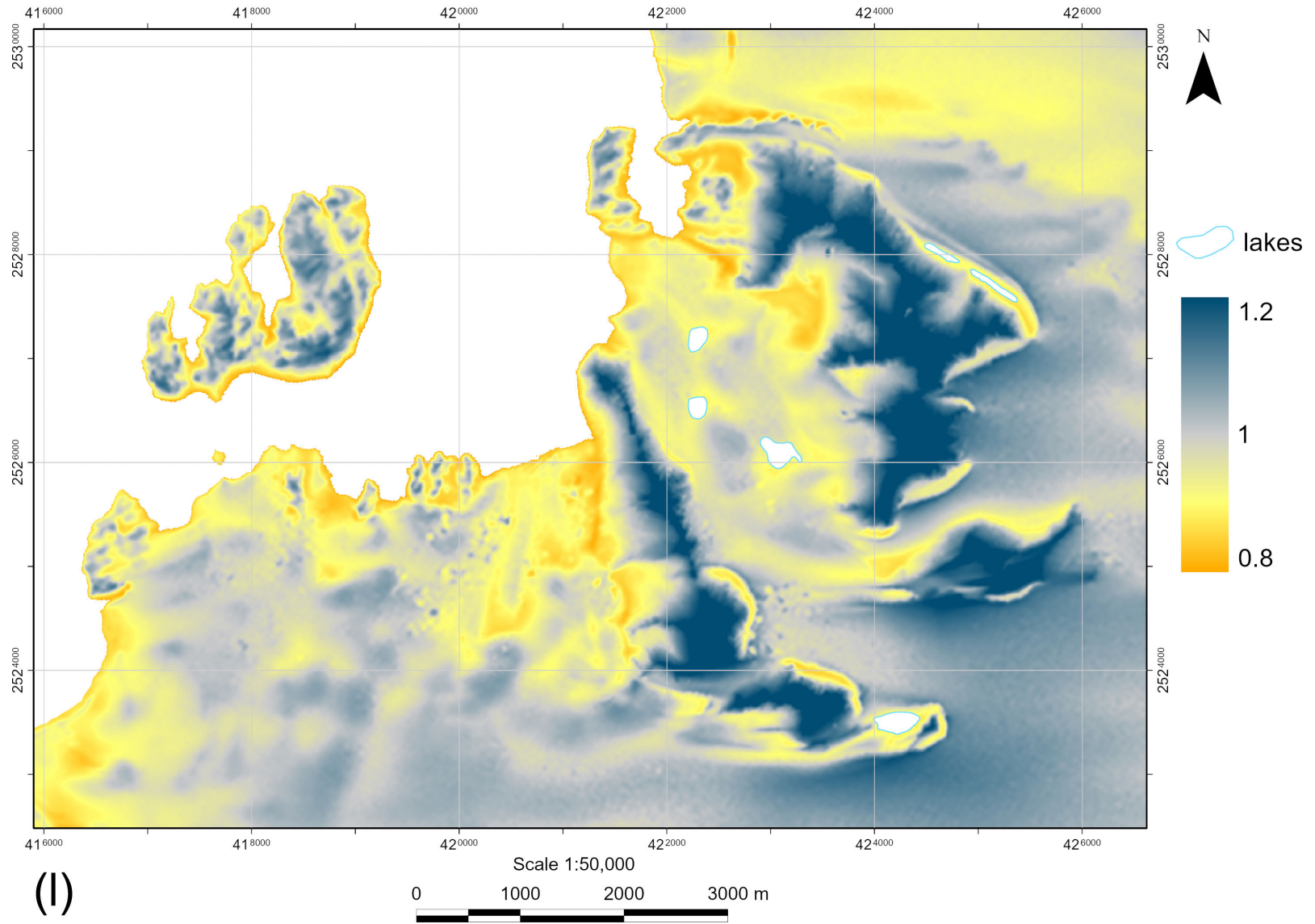

Scale 1:50,000

(l)

**Fig. 6, cont'd** Fyfe Hills: (l) Wind exposition index.

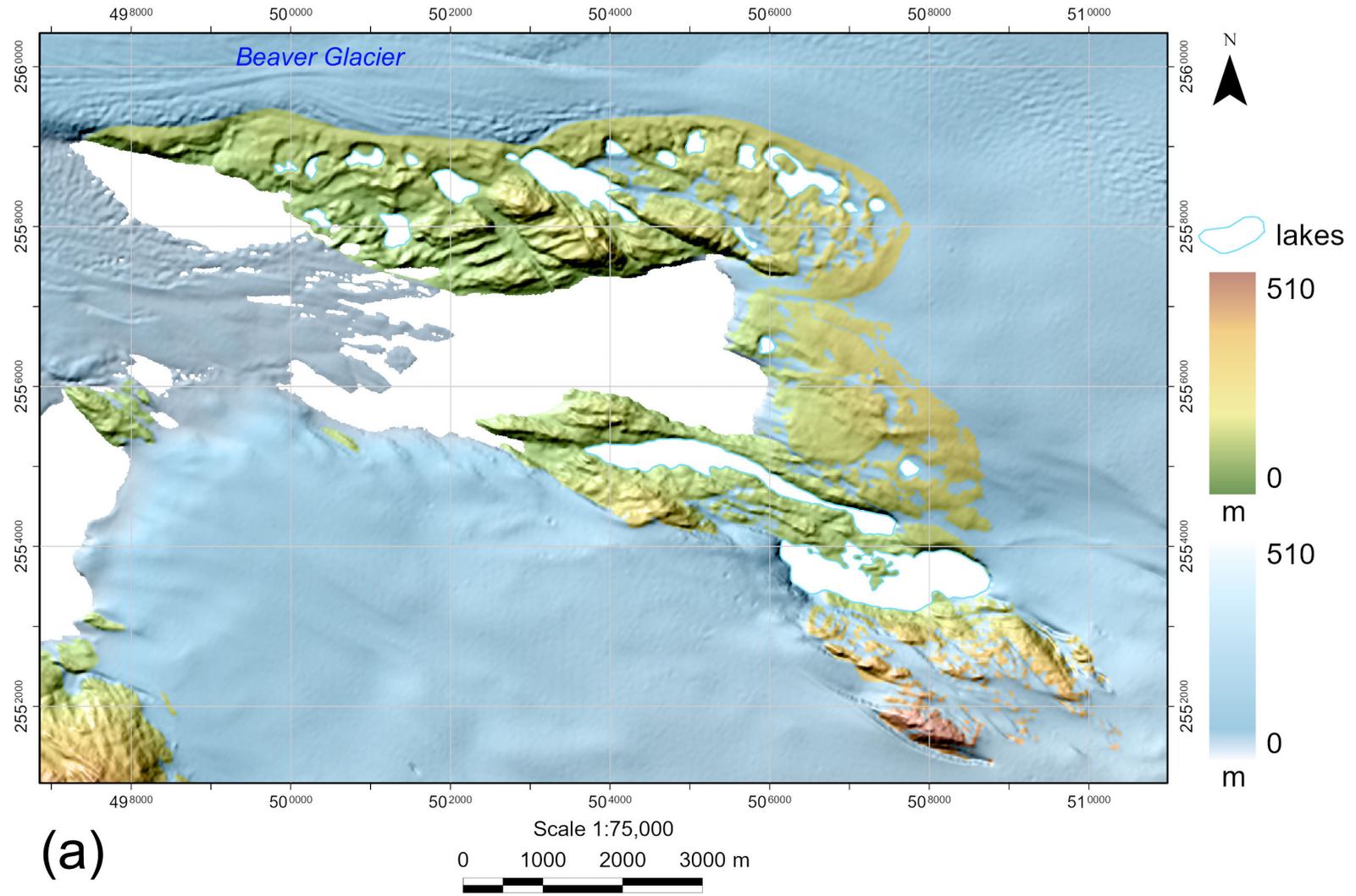

**Fig. 7** Howard Hills: (a) Elevation.

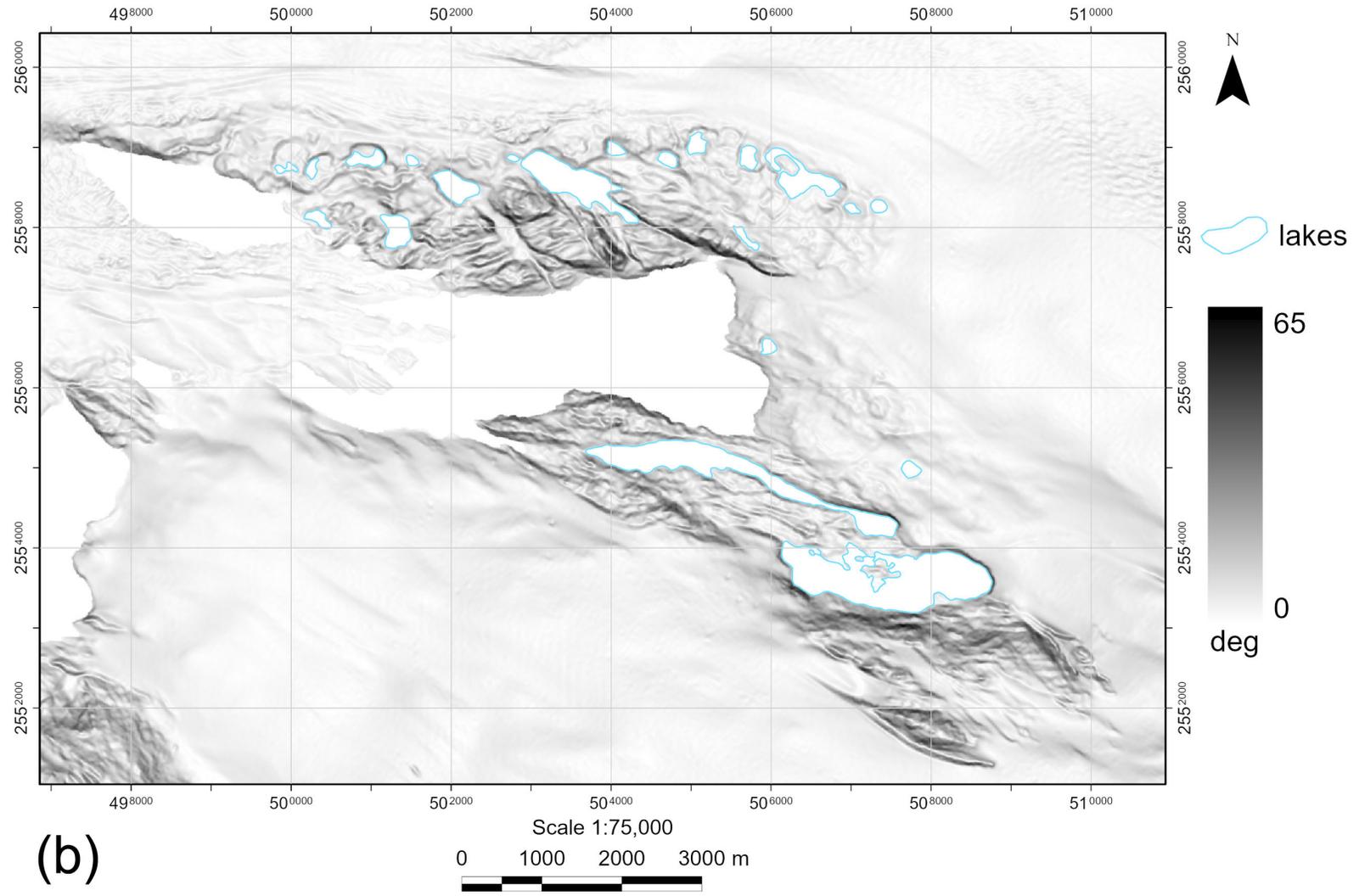

**(b)**

**Fig. 7, cont'd** Howard Hills: (b) Slope.

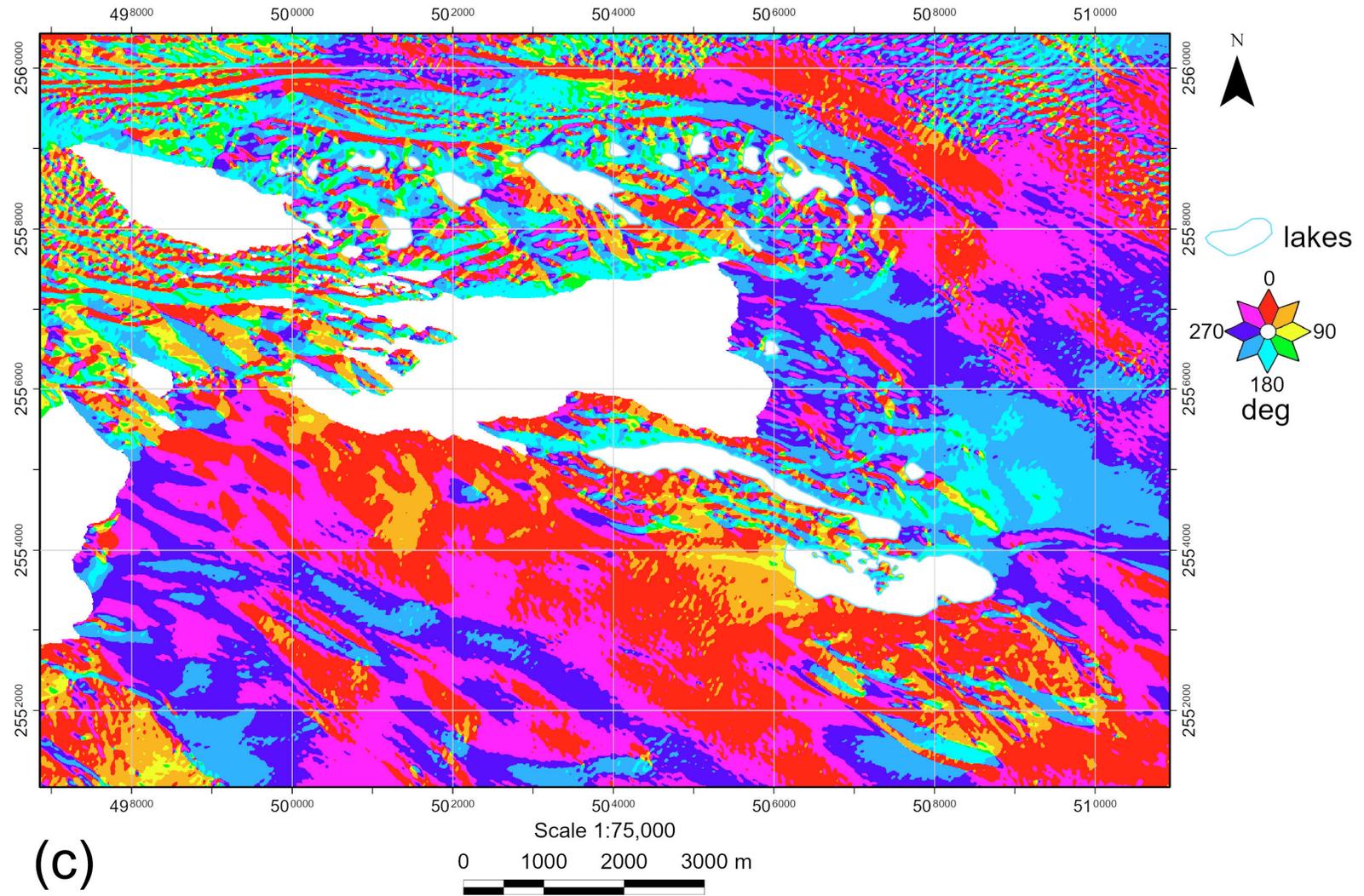

Scale 1:75,000

0  1000  2000  3000 m

**Fig. 7, cont'd** Howard Hills: (c) Aspect.

*(Continued)*

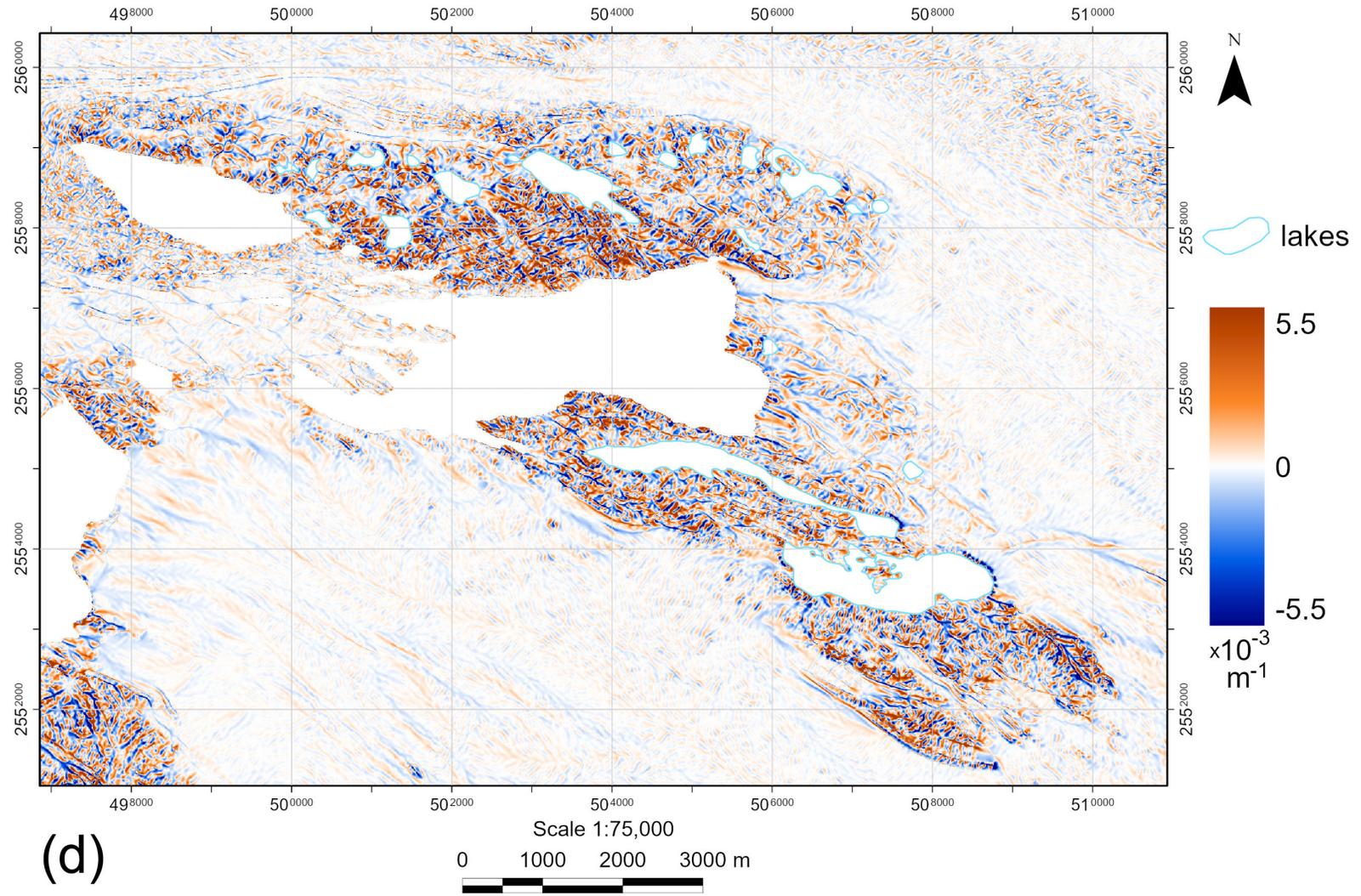

(d)

Scale 1:75,000

Fig. 7, cont'd Howard Hills: (d) Horizontal curvature.

*(Continued)*

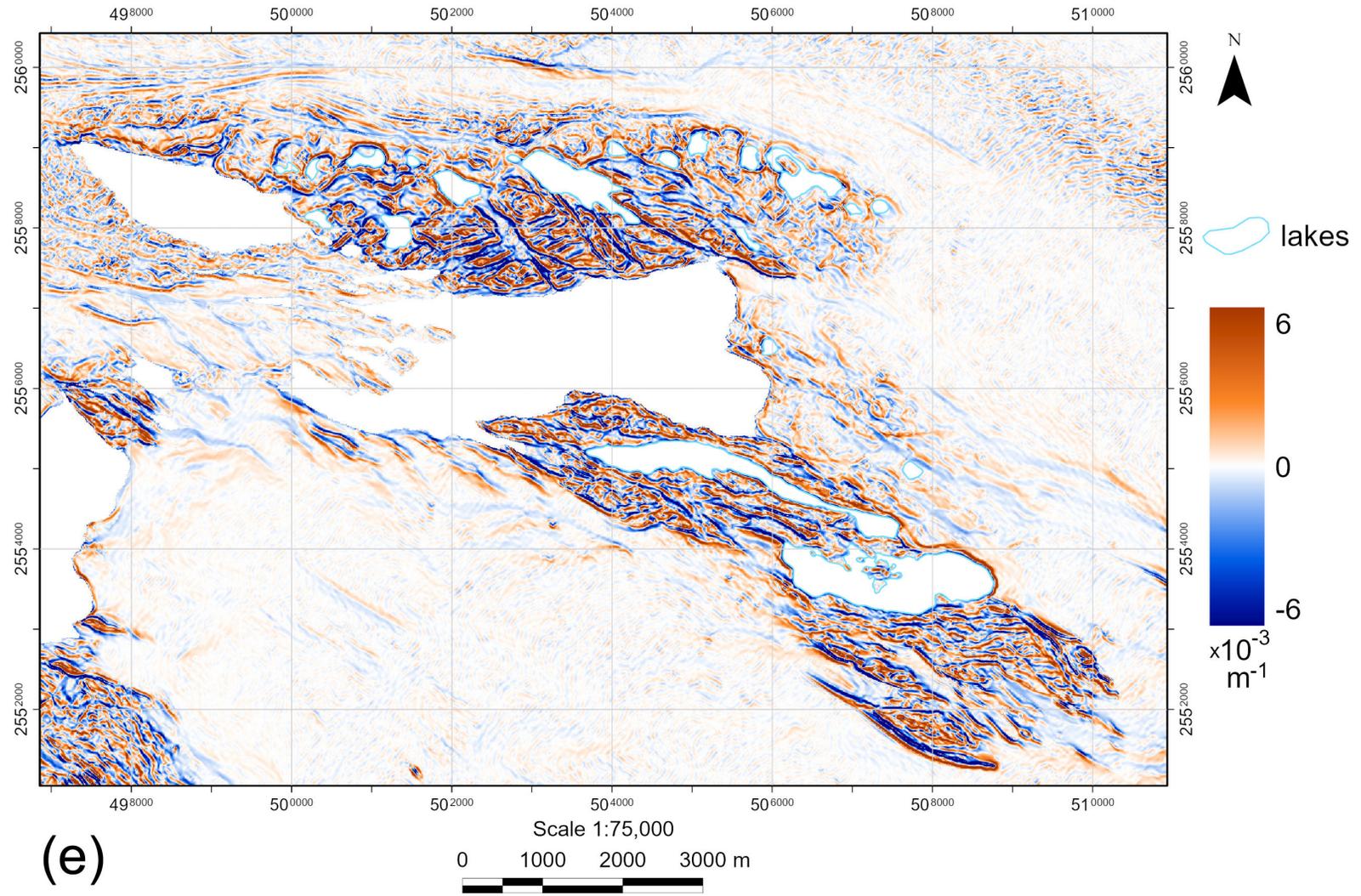

(e)

**Fig. 7, cont'd** Howard Hills: (e) Vertical curvature.

*(Continued)*

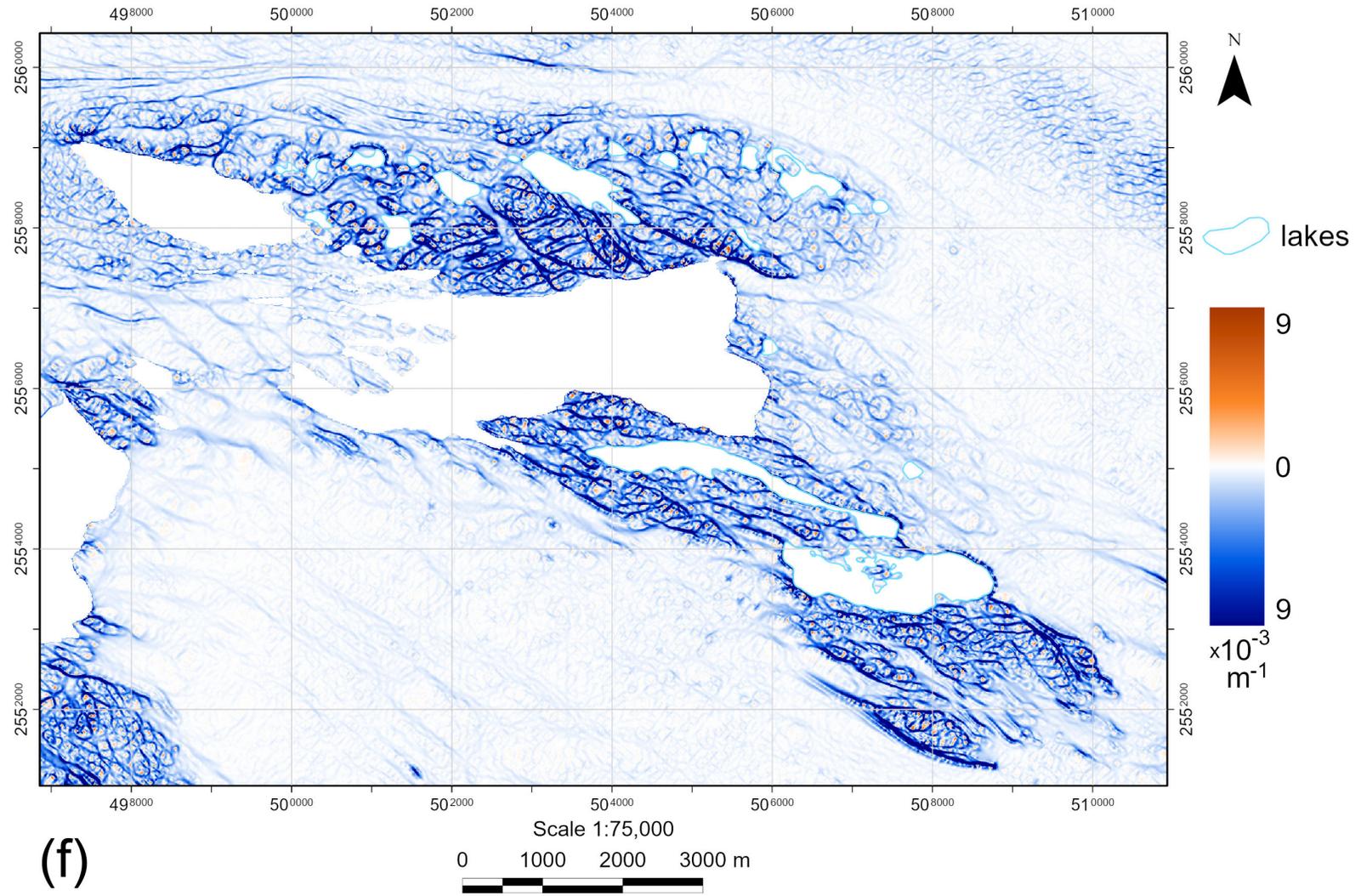

**(f)**

Scale 1:75,000

**Fig. 7, cont'd** Howard Hills: (f) Minimal curvature.

*(Continued)*

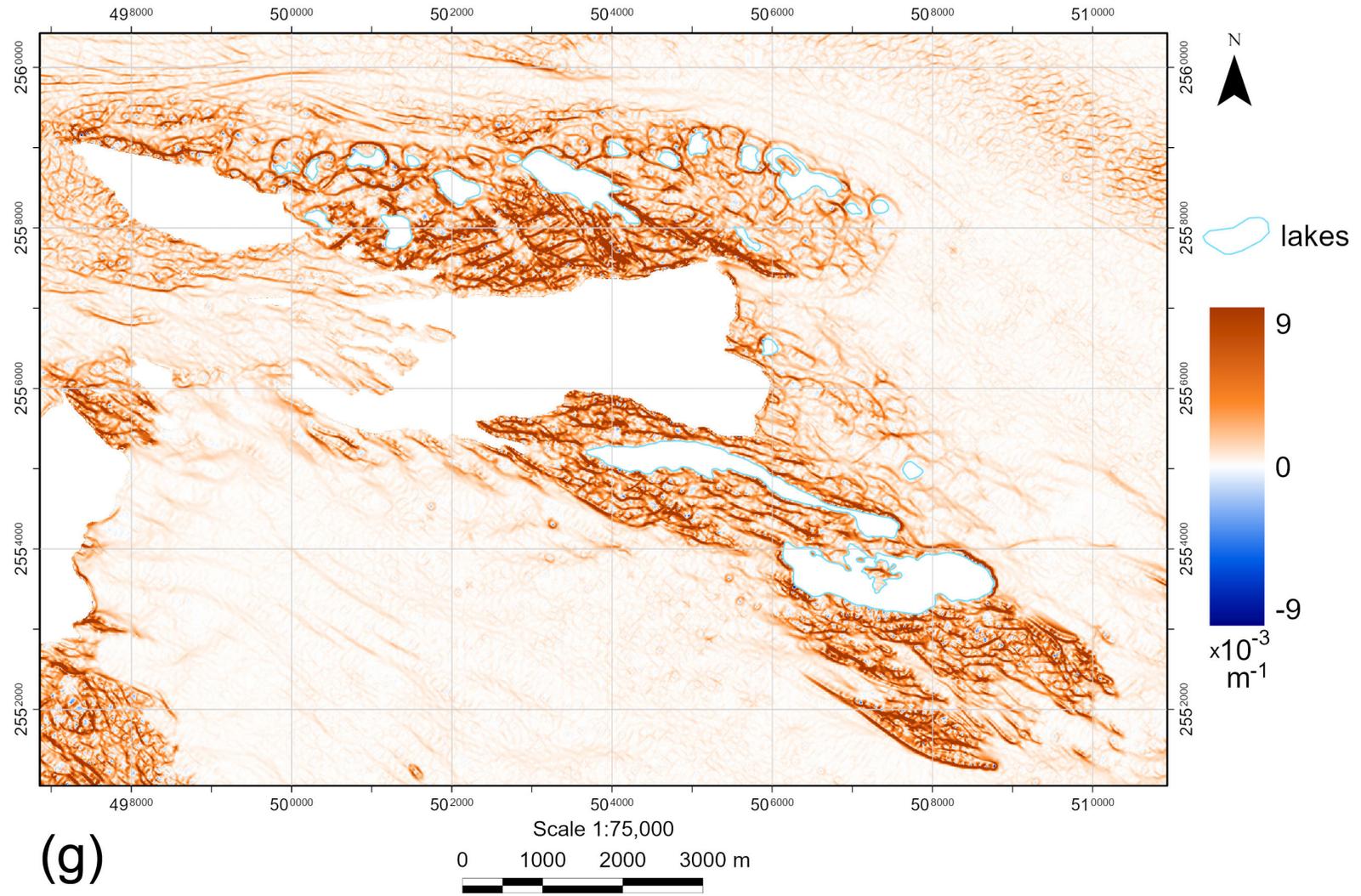

**(g)**

Scale 1:75,000

**Fig. 7, cont'd** Howard Hills: (g) Maximal curvature.

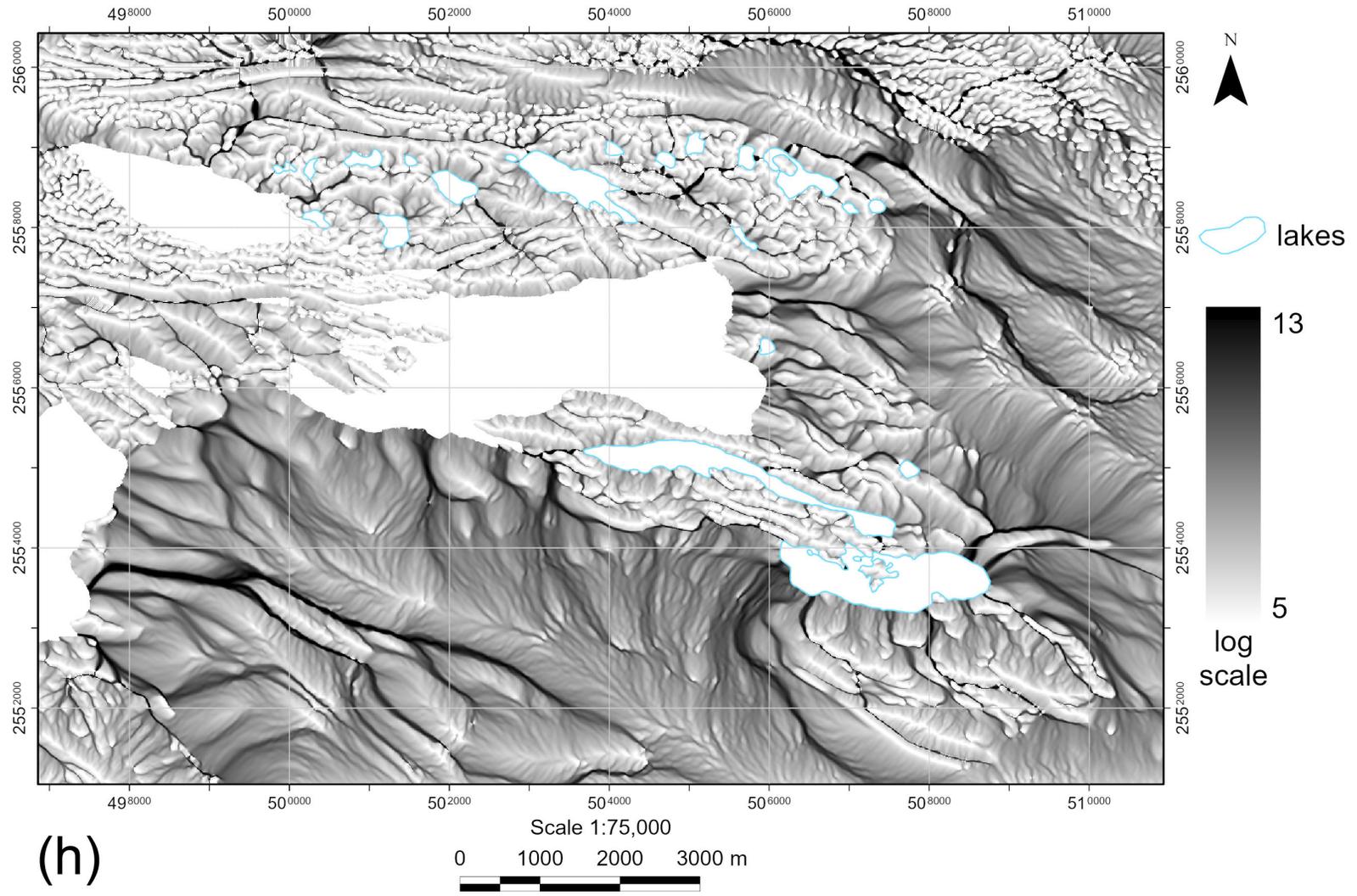

**Fig. 7, cont'd** Howard Hills: (h) Catchment area.

*(Continued)*

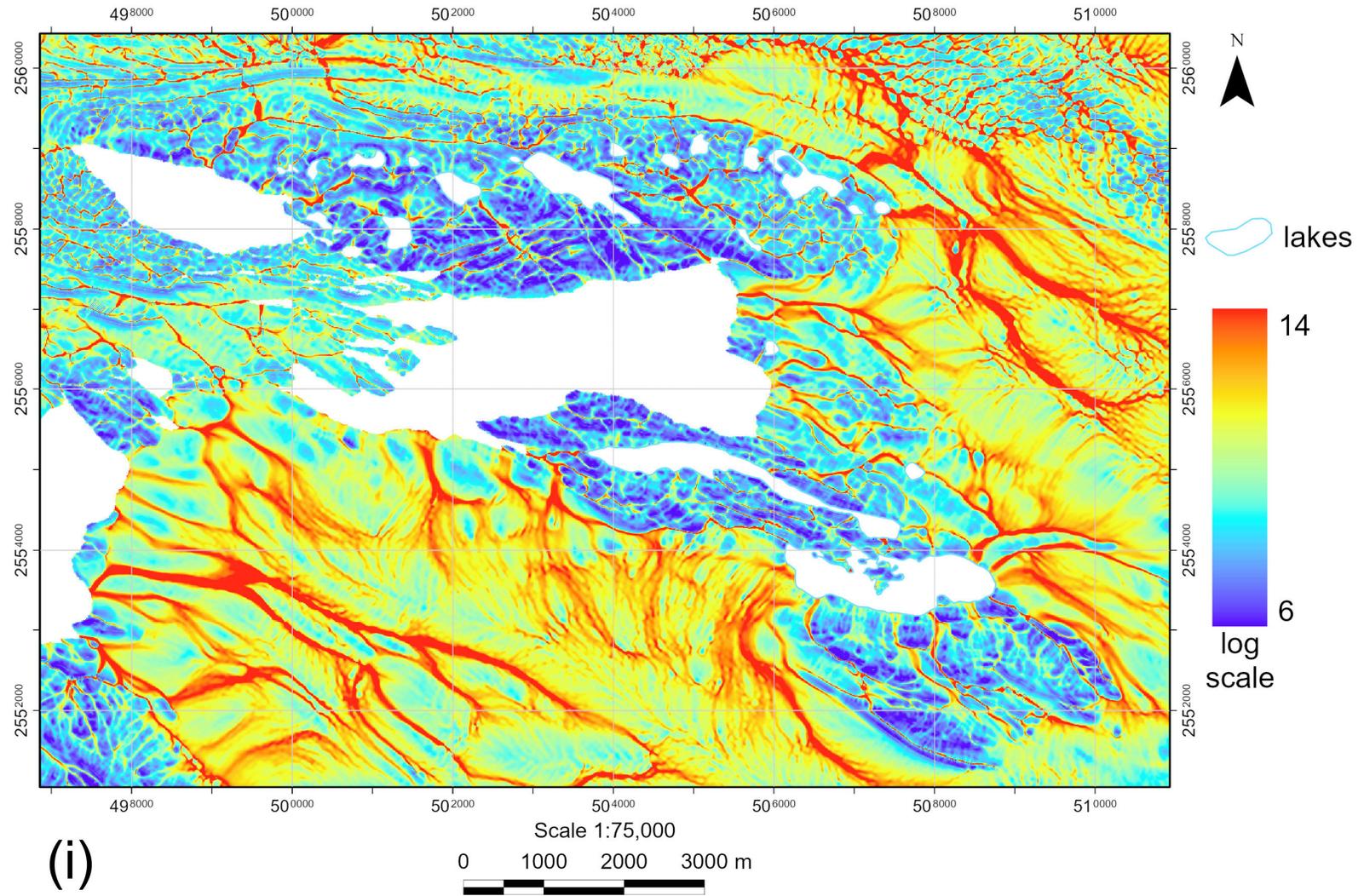

(i)

Scale 1:75,000

**Fig. 7, cont'd** Howard Hills: (i) Topographic wetness index.

*(Continued)*

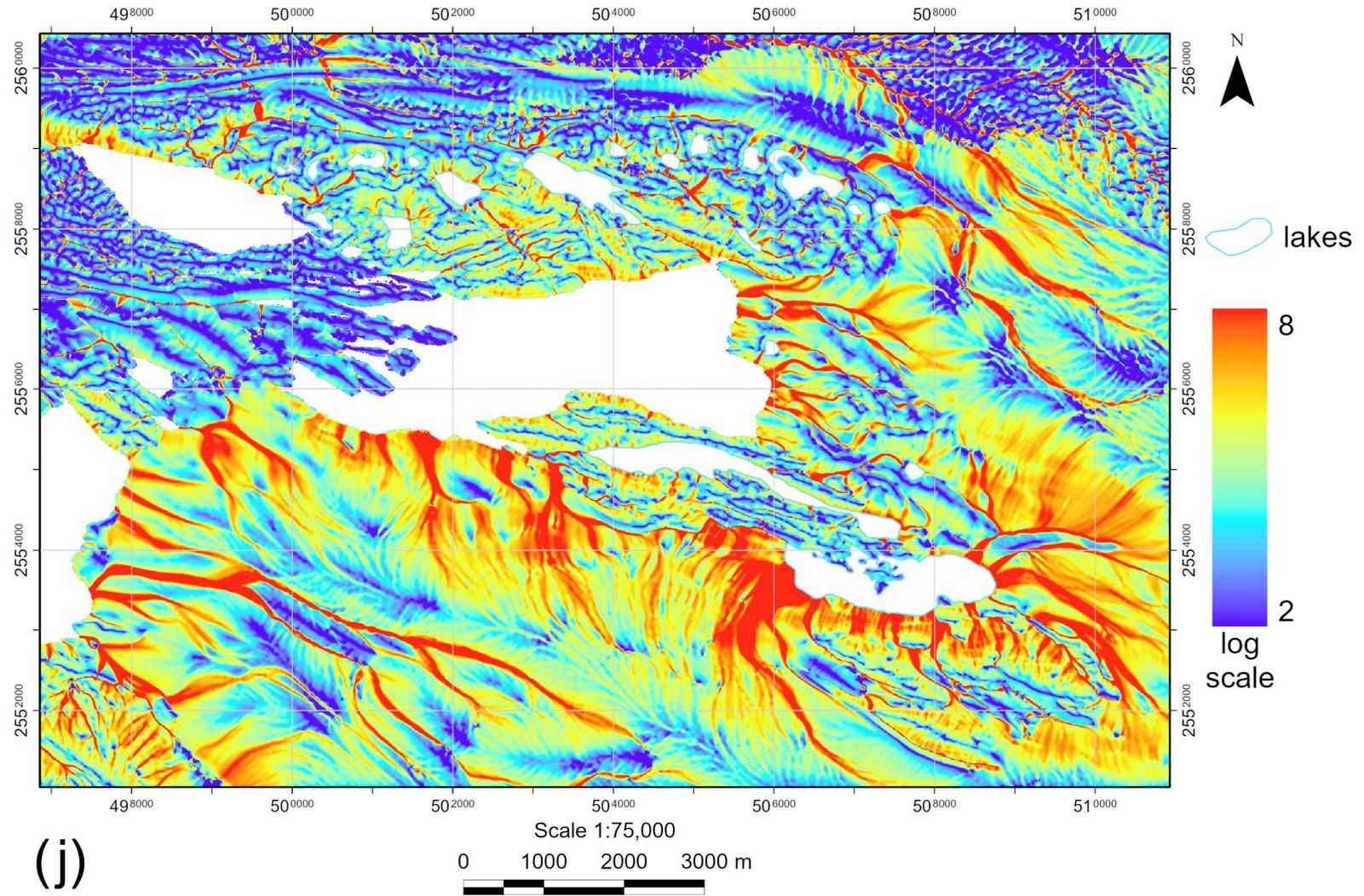

**(j)**

Scale 1:75,000

0   1000   2000   3000 m

**Fig. 7, cont'd** Howard Hills: ( j) Stream power index.

*(Continued)*

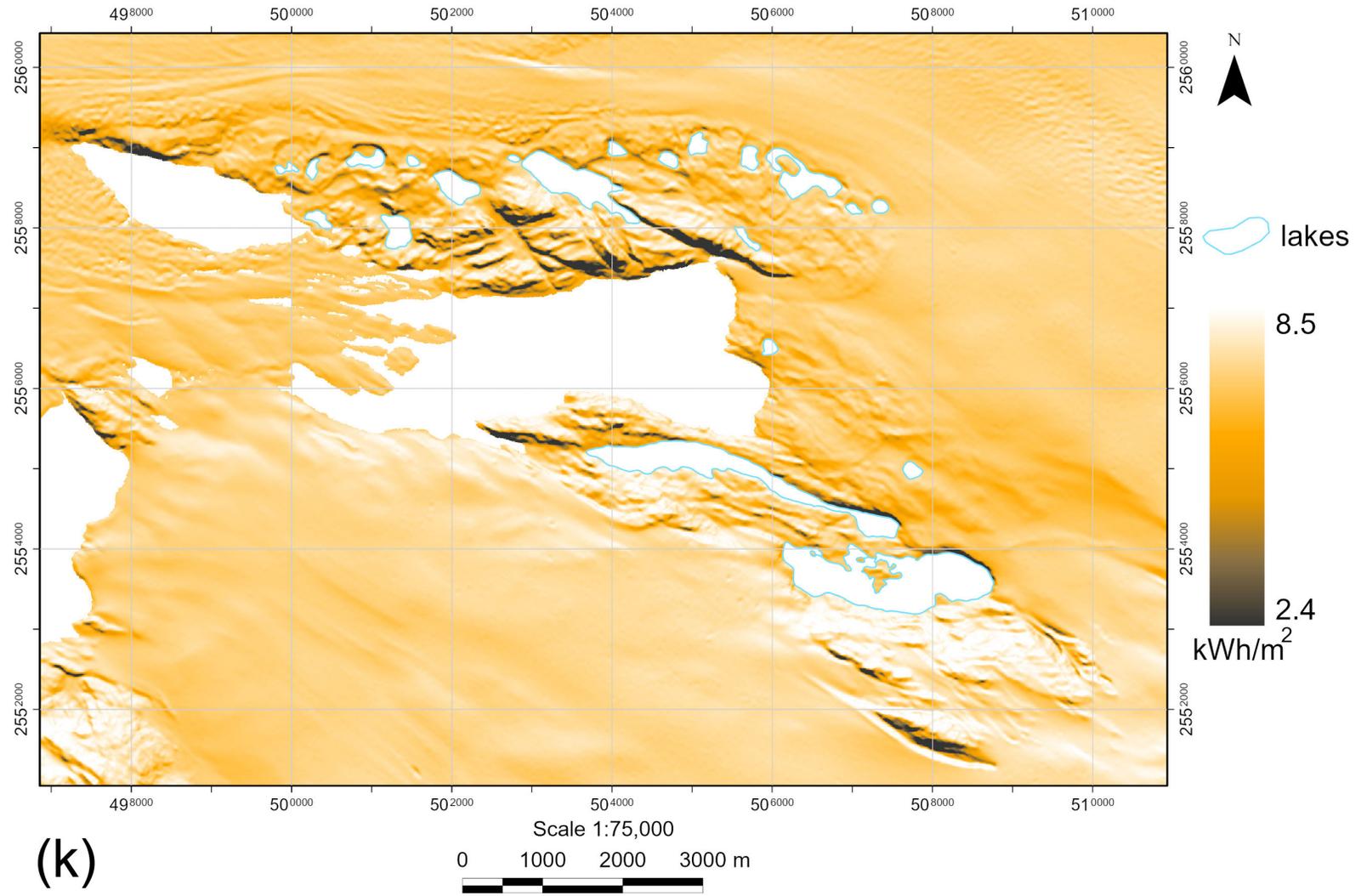

Scale 1:75,000

(k)

**Fig. 7, cont'd** Howard Hills: (k) Total insolation.

*(Continued)*

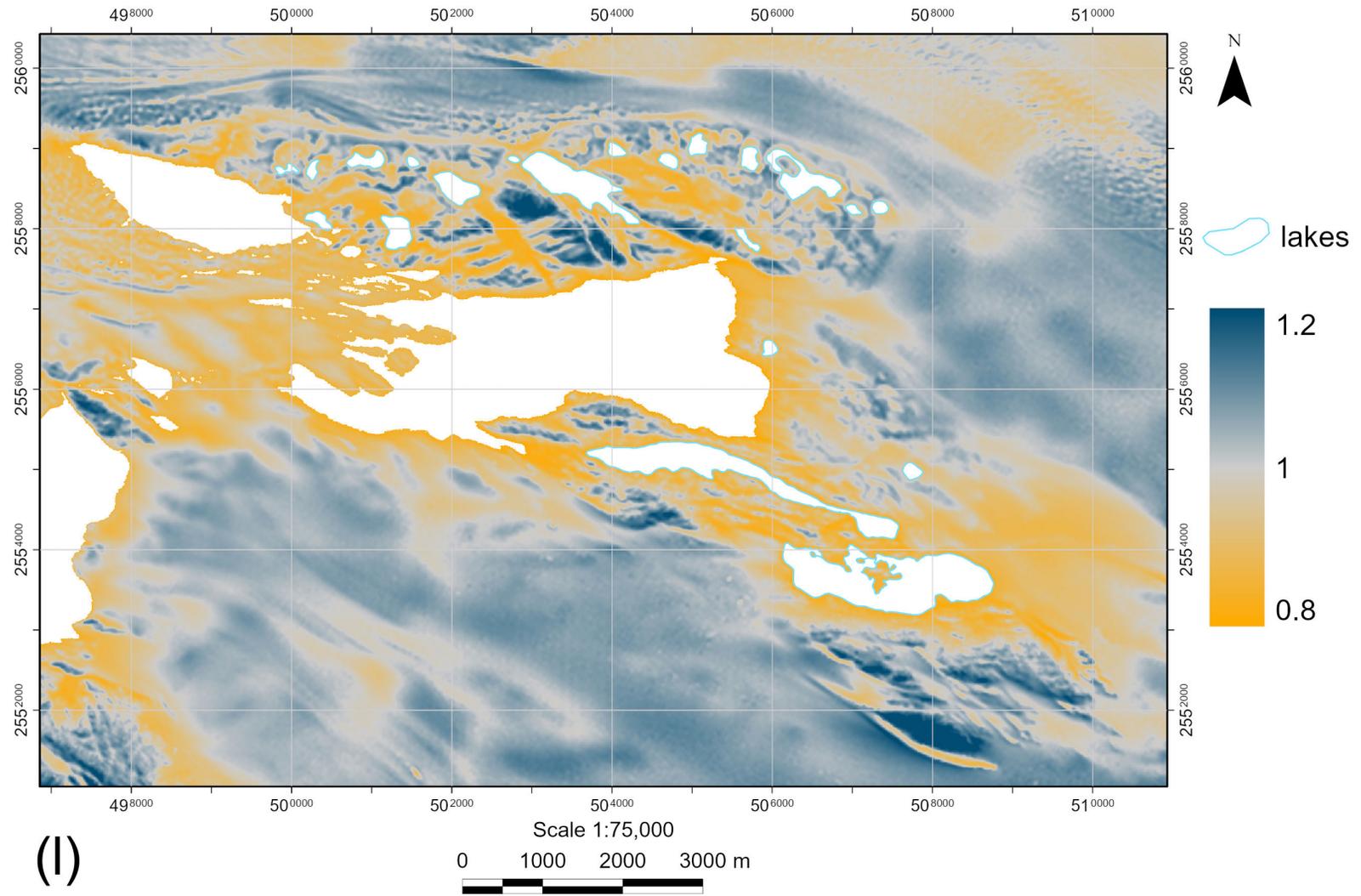

Scale 1:75,000

0    1000    2000    3000 m

**(l)**

**Fig. 7, cont'd** Howard Hills: (l) Wind exposition index.

## 5 Discussion

It is known[3] that topography controls the thermal, wind, and hydrological regimes of slopes, influencing therefore the distribution and properties of soils and vegetation (Huggett and Cheesman, 2002; Florinsky, 2025a).

The thermal regime of slopes depends in part on the incidence of solar rays to the land surface, so it depends on both $G$ and $A$. *TIns* directly considers this incidence and better describes the thermal regime (Böhner and Antonić, 2009). Information on the differentiation of slopes by insolation level is critical to predict the spatial distribution of primitive soils and lower plants in ice-free areas. To refine such a prediction, one can use *TWI* digital models describing topographic prerequisites of water migration and accumulation.

A further refinement of such a prediction can be done with *WEx* data. *WEx* digital models are utilized to identify areas affected by and protected from the wind impact (Böhner and Antonić, 2009; Florinsky, 2025a, chap. 2). In periglacial Antarctic landscapes, one of the main meteorological factors determining the microclimate is katabatic wind. Thus, *WEx* maps may be of great importance for modeling the distribution of primitive soil and lower plants, predicting the differentiation of snow accumulation in ice-free areas, and determining the optimal location of buildings and infrastructure at polar stations.

$k_h$ maps display the distribution of convergence and divergence areas of surface flows. Geomorphologically, these are spurs of valleys and crests, respectively. The combination of convergence and divergence areas creates an image of the flow structures (Florinsky, 2025a, chap. 2). $k_v$ maps show the distribution of relative deceleration and acceleration areas of surface flows. Geomorphologically, these maps represent cliffs, scarps, terrace edges, and other similar landforms or their elements with sharp bends in the slope profile (Florinsky, 2025a, chap. 2). In this regard, $k_h$ and $k_v$ digital models may be useful in geomorphological and hydrological studies of the ice-free areas.

Combination of $k_h$ and $k_v$ digital models allows revealing relative accumulation zones of surface flows (Florinsky, 2025a, chap. 2). These zones, marked by both $k_h < 0$ and $k_v < 0$, coincide with the fault intersection sites and are characterized by increased rock fragmentation and permeability. Within these zones, one can observe an interaction and exchange between two types of substance flows: (a) lateral, gravity-driven substance flows moved along the land surface and in the near-surface layer, such as water, dissolved and suspended substances, and (b) vertical, upward substance flows, such as fluids, groundwater of different mineralization and temperature (Florinsky, 2025a, chap. 15). Maps of accumulation zones may be useful for geochemical studies in the ice-free areas.

$k_{max}$ and $k_{min}$ maps are informative in terms of structural geology because they clearly display elongated linear landforms (Florinsky, 2017, 2025a). In Antarctica, such lineaments can be interpreted as a reflection of the local fault and fracture network, which topographic manifestation has been amplified by erosional, exaration, and nival processes (Florinsky, 2023b; Florinsky and Zharnova, 2025a). Thus, researchers are able to use $k_{max}$ and $k_{min}$ models for compiling lineament maps and comparing them with other geological data.

*CA* maps can be used to identify the fine flow structure of drainage basins and then to incorporate this information into geochemical and hydrological analysis. *TWI* digital models can be applied for prediction of the ground moisture content in the ice-free areas as well as for forecasting the spatial distribution of snow puddles on adjacent glaciers in summer. *SPI* data can be useful for prediction of slope erosion in the ice-free areas as well as erosion of snow cover and ice by meltwater flows on adjacent glaciers in summer.

---

[3] This one-page explanation of the usefulness and possible application of morphometric models and maps in Antarctic research has earlier been published in our papers on ice-free geomorphometry of Antarctica (Florinsky, 2025b; Florinsky and Zharnova, 2025b, 2025c, 2025d).

## 6 Conclusions

We performed geomorphometric modeling of the five key coastal oases of Enderby Land, East Antarctica. For the first time, we created the series of morphometric models and maps for the Konovalov Oasis, Thala Hills (Molodezhny and Vecherny Oases), Fyfe Hills, and Howard Hills. In total, we derived 60 maps in 1:50,000 and 1:75,000 scales.

The obtained maps rigorously, quantitatively, and reproducibly describe the coastal oases of Enderby Land. New morphometric data can be used in further geological, geomorphological, glaciological, hydrological, and ecological studies of this Antarctic region.

The study was performed within the framework of the project for creating a physical geographical thematic scientific reference geomorphometric atlas of ice-free areas of Antarctica (Florinsky, 2024, 2025b).

## References

Alexandrov, M.V., 1971. Some physical and geographical features of the Molodezhnaya Station area and some issues of its mapping. *Proceedings of the Soviet Antarctic Expedition* 47: 190–216 (in Russian).

Alexandrov, M.V., 1972. Landscape surveys in Antarctica (works at the Fyfe Hills). *Information Bulletin of the Soviet Antarctic Expedition* 84: 15–20 (in Russian).

Alexandrov, M.V., 1985. *Landscape Structure and Mapping of the Enderby Land Oases*. Hydrometeoizdat, Leningrad, USSR, 152 p. (in Russian).

Alexandrov, M.V., Simonov, I.M., 1971. Intra-landscape division of lowland oases in East Antarctica (exemplified by the Molodezhny Oasis). *Proceedings of the Arctic and Antarctic Research Institute* 304: 210–228 (in Russian).

Australian Antarctic Division, 2023. *1:1,000,000 Antarctic Topographic Series*, sheets SQ 37–38, SQ 39–40. Australian Antarctic Division, Kingston, TAS.

Bakaev, V.G., Tolstikov, E.I. (Eds.), 1966. *Atlas of Antarctica*, vol. I. Central Board of Geodesy and Cartography, Moscow–Leningrad, USSR, 241 p. (in Russian).

Beyer, L., Bölter, M. (Eds.), 2002. *Geoecology of Antarctic Ice-Free Coastal Landscapes*. Springer, Berlin, Germany, 429 p. doi:10.1007/978-3-642-56318-8.

Black, L.P., James, P.R., Harley, S.L., 1983. The geochronology, structure and metamorphism of early Archaean rocks at Fyfe Hills, Enderby Land, Antarctica. *Precambrian Research* 21: 197–222. doi:10.1016/0301-9268(83)90041-4.

Böhner, J., 2004 Regionalisierung bodenrelevanter Klimaparameter für das Niedersächsische Landesamt für Bodenforschung (NLfB) und die Bundesanstalt für Geowissenschaften und Rohstoffe (BGR). *Arbeitshefte Boden* 4: 17–66.

Böhner, J., Antonić, O., 2009. Land-surface parameters specific to topo-climatology. In: Hengl, T., Reuter, H.I. (Eds.), *Geomorphometry: Concepts, Software, Applications*. Elsevier, Amsterdam, the Netherlands, pp. 195–226. doi:10.1016/S0166-2481(08)00008-1.

Conrad, O., Bechtel, B., Bock, M., Dietrich, H., Fischer, E., Gerlitz, L. et al., 2015. System for Automated Geoscientific Analyses (SAGA) v. 2.1.4. *Geoscientific Model Development* 8: 1991–2007. doi:10.5194/gmd-8-1991-2015.

Division of National Mapping, 1962–1964. *Australian Antarctic Territory, Enderby Land, 1:250,000 Series*, sheets SQ 38–39/14 Tange Promontory, SQ 38–39/15 Simpson Peak, SQ 38–39/16. Division of National Mapping, Canberra, Australia.

Donidze, G.I. (Ed.), 1987. *Dictionary of Antarctic Geographic Names*. Central Scientific Research Institute for Geodesy, Aerophotosurveying, and Cartography, Moscow, USSR, 407 p. (in Russian).

ESA, 2025. *Sentinel-2*. European Space Agency. https://sentiwiki.copernicus.eu/web/sentinel-2 (accessed 29 Aug. 2025).

ESRI, 2015–2024. *ArcGIS Pro*. Environmental Systems Research Institute. https://www.esri.com/en-us/arcgis/products/arcgis-pro/overview (accessed 29 Aug. 2025).

Evans, I.S., 1972. General geomorphometry, derivations of altitude, and descriptive statistics. In: Chorley, R.J., (Ed.), *Spatial Analysis in Geomorphology*. Methuen, London, UK, pp. 17–90.

Evans, I.S., 1980. An integrated system of terrain analysis and slope mapping. *Zeitschrift für*

*Geomorphologie* Suppl. 36: 274–295.

Florinsky, I.V., 2017. An illustrated introduction to general geomorphometry. *Progress in Physical Geography* 41: 723–752. doi:10.1177/0309133317733667.

Florinsky, I.V., 2023a. Geomorphometric modeling and mapping of Antarctic oases. *arXiv*:2305.07523 [physics.geo-ph], 84 p. doi:10.48550/arXiv.2305.07523.

Florinsky, I.V., 2023b. Larsemann Hills: geomorphometric modeling and mapping. *Polar Science* 38: 100969. doi:10.1016/j.polar.2023.100969.

Florinsky, I.V., 2024. A project of a geomorphometric atlas of ice-free Antarctic territories. *InterCarto InterGIS* 30(2): 53–79 (in Russian, with English abstract). doi:10.35595/2414-9179-2024-2-30-53-79.

Florinsky, I.V., 2025a. *Digital Terrain Analysis*, 3rd rev. enl. edn. Academic Press, London, UK, 455 p. doi:10.1016/C2023-0-51092-4.

Florinsky, I.V., 2025b. Geomorphometric atlas of ice-free Antarctic areas: problem statement, concept, and key principles. *arXiv*:2508.02846 [physics.geo-ph], 33 p. doi:10.48550/arXiv.2508.02846.

Florinsky, I.V., Zharnova, S.O., 2025a. Geomorphometry of the Bunger Hills, East Antarctica. *Advances in Polar Science* 36: 95–112. doi:10.12429/j.advps.2024.0042.

Florinsky, I.V., Zharnova, S.O., 2025b. Ice-free geomorphometry of Queen Maud Land, East Antarctica: 1. Sôya Coast. *arXiv*:2508.10462 [physics.geo-ph], 89 p. doi:10.48550/arXiv.2508.10462.

Florinsky, I.V., Zharnova, S.O., 2025c. Ice-free geomorphometry of Queen Maud Land, East Antarctica: 2. Prince Olav Coast. *Research Square Preprints,* rs-7428671, 57 p. doi:10.21203/rs.3.rs-7428671.

Florinsky, I.V., Zharnova, S.O., 2025d. Ice-free geomorphometry of Enderby Land, East Antarctica: 1. Mountainous areas. *Geomorphometric Modeling Group, IMPB KIAM RAS, Preprint GMG-0825-1*. Zenodo, 203 p., doi:10.5281/zenodo.17009437.

Grew, E.S., 1975. Geologic studies of Precambrian basement around Molodezhnaya Station, Enderby Land. *Antarctic Journal of the United States* 10: 245–248.

Grew, E.S., 1978. Precambrian basement at Molodezhnaya Station, East Antarctica. *Geological Society of America Bulletin* 89: 801–813. doi:10.1130/0016-7606(1978)89<801:PBAMSE>2.0.CO;2.

Hengl, T. and Reuter, H.I. (Eds.), 2009. *Geomorphometry: Concepts, Software, Applications*. Elsevier, Amsterdam, the Netherlands, 796 p.

Howat, I.M., Porter, C., Smith, B.E., Noh, M.-J., Morin, P., 2019. The Reference Elevation Model of Antarctica. *Cryosphere* 13: 665–674. doi:10.5194/tc-13-665-2019.

Howat, I., Porter, C., Noh, M.-J., Husby, E., Khuvis, S., Danish, E. et al., 2022. The Reference Elevation Model of Antarctica—Mosaics, Version 2. *Harvard Dataverse*, V1, doi:10.7910/DVN/EBW8UC.

Huggett, R.J., Cheesman, J., 2002. *Topography and the Environment*. Pearson Education, Harlow, UK, 274 p.

Ishizuka, H., 2008. An overview of geological studies of JARE in the Napier Complex, Enderby Land, East Antarctica. *Geological Society, London, Special Publications* 308: 121–138. doi:10.1144/SP308.5.

Kamenev, E.N., 1979. The oldest metamorphic strata of the Fyfe Hills (Antarctica). In: Avsyuk, G.A. (Ed.), *Antarctica. Reports of the Commission, No. 18*. Nauka, Moscow, pp. 11–19 (in Russian).

Kamenev, E.N., Hofmann, J., 1988. Geology and structural analysis of Precambrian metamorphic formations of the Thala Hills in the area of the Molodezhnaya Station (Enderby Land, East Antarctica). In: Avsyuk, G.A. (Ed.), *Antarctica. Reports of the Commission, No. 27*. Nauka, Moscow, pp. 47–56 (in Russian).

Klimov, L.V., Dukhanin, S.F., and Mitroshin, M.I., 1962. Geological research in the western part of Enderby Land. *Information Bulletin of the Soviet Antarctic Expedition* 37: 5–7 (in Russian).

Korotkevich, E.S., 1972. *Polar Deserts*. Hydrometeoizdat, Leningrad, USSR, 419 p. (in Russian).

Korotkevich, E.S., Fomchenko, V.D., Friedman, B.S. (Eds.), 2005. *Atlas of the Oceans: Antarctica*. Head Department of Navigation and Oceanography, Arctic and Antarctic Research Institute, St. Petersburg, Russia, 280 p. (in Russian, with English contents, summary, and index).

Kovaleva, O.V., 2014. Perfecting the representation of topography by hypsometric tints. New classification of hypsometric scales. *Geodesy and Cartography* 75(11): 21–29 (in Russian, with English abstract). doi:10.22389/0016-7126-2014-893-11-21-29.

Krylov, D.P., Hoernes, S., Bridgwater, D., 1998. Changes in the $^{18}O/^{16}O$ ratios of fluids as evidence for different metamorphic episodes in high grade gneisses from the Konovalov Mountains area (Rayner Complex, East Antarctica). *Chemical Geology* 147: 295–312. doi:10.1016/S0009-2541(98)00022-9.

Markov, K.K., Bardin, V.I., Lebedev, V.L., Orlov, A.I., Suetova, I.A., 1970. *The Geography of Antarctica*. Israel Program for Scientific Translations, Jerusalem, Israel, 370 p.

Minár, J., Krcho, J., Evans, I.S., 2016. Geomorphometry: quantitative land-surface analysis. In: Elias, S.A. (Ed.), *Reference Module in Earth Systems and Environmental Sciences*. Elsevier, Amsterdam, the Netherlands. doi:10.1016/B978-0-12-409548-9.10260-X.

Moore, I.D., Grayson, R.B., Ladson, A.R., 1991. Digital terrain modelling: a review of hydrological, geomorphological and biological applications. *Hydrological Processes* 5: 3–30. doi:10.1002/hyp.3360050103.

Myasnikov, O.V., 2011. Geological structure of the Vechernegorskaya territory (Western Enderby Land, Antarctica). In: Samodurov, V.P. (Ed.), *Actual Problems of Geology and Prospecting of Mineral Deposits: Proceedings of the 5th University Geological Readings, Minsk, Belarus, 8–9 April 2011*. Belarus State University, Minsk, Belarus, pp. 17–20 (in Russian).

Myers, P.B.J.. MacNamara, E.E., 1970. Rocks of coastal Enderby Land near Molodezhnaya station, Antarctica. Antarctic Journal of the United States 5: 158–159.

Patterson, T., Jenny, B., 2011. The development and rationale of cross-blended hypsometric tints. *Cartographic Perspectives* 69: 31–45. doi:10.14714/CP69.20.

PGC, 2022–2024. *REMA Explorer*. Polar Geospatial Center, University of Minnesota, Saint Paul, MN. https://rema.apps.pgc.umn.edu (accessed 29 Aug. 2025).

Pickard, J. (Ed.), 1986. *Antarctic Oasis: Terrestrial Environment and History of the Vestfold Hills*. Academic Press, New York, NY, 367 p.

Sandiford, M., Wilson, C.J.L., 1984. The structural evolution of the Fyfe Hills–Khmara Bay region, Enderby Land, East Antarctica. *Australian Journal of Earth Sciences* 31: 403–426. doi:10.1080/08120098408729301.

Segal, D.B., 1982. Theoretical basis for differentiation of ferric-iron bearing minerals, using Landsat MSS data. In: *Proceedings of Symposium for Remote Sensing of Environment, 2nd Thematic Conference on Remote Sensing for Exploratory Geology, Fort Worth, TX*, pp. 949–951.

Shary, P.A., Sharaya, L.S., Mitusov, A.V., 2002. Fundamental quantitative methods of land surface analysis. *Geoderma* 107: 1–32. doi:10.1016/S0016-7061(01)00136-7.

Sheraton, J.W., 1985. *Geology of Enderby Land and Western Kemp Land, Australian Antarctic Territory*, scale 1:500 000. Bureau of Mineral Resources, Geology and Geophysics, Canberra, Australia, 1 chart.

Simonov, I.M., 1968. Physical and geographical characteristics of the Molodezhny Oasis. *Proceedings of the Soviet Antarctic Expedition*, 38: 6–21 (in Russian).

Simonov, I.M., 1971. *Oases of East Antarctica*. Hydrometeoizdat, Leningrad, USSR, 176 p. (in Russian).

Sobotovich, E.V., Kamenev, V.N., Kornanstyy, A.A., Rudnik, V.A., 1976. The oldest rocks of Antarctica (Enderby Land). *International Geology Review* 18: 371–388. doi:10.1080/00206817609471218.

Sokratova, I.N., 2010. *Antarctic Oases: History and Results of Investigations*. Arctic and Antarctic Research Institute, St. Petersburg, Russia, 274 p. (in Russian).

Wilson, J.,. Gallant, J.C. (Eds.), 2000. *Terrain Analysis: Principles and Applications*. Wiley, New York, NY, 479 p.

Xu, H., 2006. Modification of normalised difference water index (NDWI) to enhance open water features in remotely sensed imagery. *International Journal of Remote Sensing* 27: 3025–3033. doi:10.1080/01431160600589179.

Yoshimura, Y., Motoyoshi, Y., Grew, E.S., Miyamoto, T., Carson, C.J., Dunkley, D.J., 2000. Ultrahigh-temperature metamorphic rocks from Howard Hills in the Napier Complex, East Antarctica. *Polar Geoscience* 13: 60–85.